\documentclass[12pt,fleqn, twoside]{book}
\long\def\nop#1{}
\usepackage{epsf,psfig,pstricks,multicol,color}
\usepackage{amssymb,amsmath,wasysym,delarray,array}
\usepackage{caption}
\usepackage{fancyhdr}
\usepackage{exscale}
\usepackage{graphicx}
\renewcommand{\chaptermark}[1]%
        {\markboth{#1}{}}
\renewcommand{\sectionmark}[1]%
        {\markright{\thesection \ #1}}
\begin{document}


\begin{titlepage}
\begin{center}

\hrule\vspace{0.5cm}

{\LARGE\bf Bosonization of dimerized spinless-fermion and Hubbard chains}

\vspace{0.5cm}
\hrule
\vspace{3.5cm}

Zur Erlangung des akademischen Grades eines \\
Doktors der Naturwissenschaften\\ der Mathematisch-Naturwissenschaftlichen
Fakult\"at\\ der Universit\"at Augsburg 
\vspace{1.0cm}\\
angenommene \vspace{1.0cm}\\

\large{Dissertation}\vspace{2.5cm}\\
von  \vspace{0.5cm}\\
\large{\bfseries{M. Sc. ~Carmen Mocanu}}\vspace{0.5cm}\\

\end{center}
\vspace{1.0cm}
\small{Erstgutachter: Prof. Dr. Ulrich Eckern\\
Zweitgutachter: Prof. Dr. Gert-Ludwig Ingold\\
Tag der m\"undlichen Pr\"ufung: 11 Februar  2005}
\end{titlepage}
\vspace{4cm}

\thispagestyle{empty}

\tableofcontents
%
%

\chapter{Introduction}

The  interplay between electron-electron interaction and
electron-phonon coupling in strongly correlated fermionic systems
has been intensively studied in the past  and is 
still  a major  challenge for a realistic description 
and a better understanding of many materials.

In the last decades, 
particular  attention has
been paid  to 
 one-dimensional fermionic systems, 
 from both the theoretical and the experimental point of view. 
 Starting from the features of one-dimensional electron systems, the initial
 goal was to understand correlation effects in higher dimensions, too.
Furthermore, one-dimensional models are attractive from a theoretical point of
view: there exists, in fact, a variety of methods that work exclusively in one
dimension, 
 sometimes
even allowing for an exact solution. 
As a special feature of one dimension,
the Fermi liquid theory breaks
down and a new paradigm has to be introduced, 
namely the Luttinger liquid \cite{Tomonaga50,Luttinger63}.
The first version of the generic model  was proposed by Tomonaga, who showed that
the excitations of a one-dimensional electron gas with linear dispersion 
are  bosons,
even though the elementary constituents are fermions. The excitations involve
two particles, and the wave function of the two fermion states has bosonic
properties.
This model  proved to be of fundamental 
importance for
purely one- or quasi one-dimensional systems, e.g.\ quasi
one-dimensional organic conductors \cite{Luther74,Heeger77,Jerome82,Heeger88,Keiss92} 
or spin-Peierls compounds \cite{Hase93,Isobe96} 
where correlations  are known to play an important role. In the last years, 
evidence for  Luttinger liquid  behavior has been found  
in semiconductor quantum
wires \cite{Tarucha95} and  carbon nanotubes \cite{Bockrath99}. 

The concept of a Luttinger liquid is intimately connected with the bosonization
technique whose origins date back  to the seminal
paper by Tomonaga \cite{Tomonaga50} in 1950. During the following decades the method
was worked out and successfully applied to one-dimensional electron and spin
systems \cite{Haldane81,Schulz,Delft98, Schulz00}.  Despite its long history 
there are
still some subtle points in the bosonization formalism which are not taken into
consideration in the majority of the literature. One of these issues is the proper
treatment of the so-called Klein factors which have to be introduced in order to
preserve the anti-commutation property of the fermionic fields during the
bosonization procedure. 
Often,
the existence of Klein factors has been ignored in the literature:
this may be justified for infinite systems \cite{Schulz00},
but in general they  have to be treated carefully as pointed
out, for example, in the context of impurity models and two-leg ladders
\cite{Delft98,Kotliar96,Schonhammer01,Schonhammer02,Marston2002}.
The importance 
 of Klein factors has also  been  emphasized  by Sch\"onhammer 
\cite{Schonhammer01,Schonhammer02}, who  noted that
the common practice of neglecting them may lead to erroneous results when
nonlinear terms are  considered.
Such nonlinear terms arise in the presence of perturbations 
like impurity scattering or a modulation of the hopping.
For the system with a perturbation an exact solution 
is known
only in some special cases \cite{baxter1982}.
In general one has to resort to approximative methods like
renormalization group calculations \cite{Solyom}.
Another more intuitive method is the self-consistent harmonic approximation (SCHA)
where the nonlinear terms are replaced by a harmonic potential
with parameters to be determined  self-consistently 
according to a variational principle for the
energy or the free energy. 
The SCHA has been successfully applied to various nonlinear models
\cite{Coleman75,Nakano81,Fukuyama,Gogolin93,Rojas96,Cosima98,Gouvea99}.

In this work we handle the Klein factors in a systematic way,
both in the thermodynamic limit and for finite systems.
We develop  an extension of the
SCHA which
treats the bosonic fields and the Klein factors on equal footing.
As prototypical models  we
consider  one-dimensional dimerized spinless fermions and   Hubbard
models with periodic modulation of the hopping (Peierls-Hubbard model) and of
the chemical potential (ionic Hubbard model), respectively.
Note that, throughout this work, we consider exclusively ground-state properties,
i.e.\ the zero-temperature limit.

 We start with a short presentation of the bosonization
method in chapter \ref{bosonization}, where we focus in particular on the properties of the Klein factors.
The chapter ends with a description 
of the Luttinger model, which, in certain cases, can be  solved exactly. 
In the next  two chapters 
we present a method for treating the Klein factors in 
bosonized Hamiltonians with nonlinear terms, and apply this method to spinless
fermions in chapter \ref{spinless}, and to the Hubbard model in chapter 
\ref{hub}.
  
In chapter \ref{vanadium} we consider a model of one-dimensional
interacting electrons coupled  to  a three-dimensional  weakly  correlated
conduction band through a Coulomb  interaction. This model is motivated 
by materials like VO$_2$, where at the Fermi energy one encounters
both  a band with one-dimensional dispersion and bands with  three-dimensional
character. 
We calculate the
spin and the charge susceptibility in order to arrive  at a description in terms
of coupled one-dimensional chains with intra- and inter chain coupling in the
spin and charge channel mediated through the  $3d$
environment.

In chapter \ref{summary} we summarize our results and give an outlook to some 
still open  problems. Some technical details are described in six appendices.

\chapter{Bosonization}
\label{bosonization}
\section{Bosonization prerequisite}
\label{prerequisite}
 Fermion systems in one dimension have features quite distinct from those in
 higher dimensions. In the early work of Mattis and Lieb \cite{mattis65} and 
 Bychkov et al.\ \cite{Bychkov66} 
 it has been shown that Landau type quasi-particles do not exist and the 
 Fermi liquid theory  breaks down  in these systems. As a
 consequence a  new concept  has to be  introduced to describe
  one-dimensional interacting electrons,  
 known as the 
 Tomonaga-Luttinger liquid. Corresponding models were first introduced 
 by Tomonaga \cite{Tomonaga50} and Luttinger \cite{Luttinger63} 
 and have been  continuously  
 improved in order  to obtain an accurate  characterization  of real systems 
 \cite{mattis65,Luther74,Solyom,Haldane81,Schulz,Delft98,
 Schulz00,Schonhammer01}.
 The theory of the Luttinger liquid is closely related to the
 bosonization technique, which has been  
 successfully applied to strongly
 correlated one-dimensional models. The basic idea behind
 bosonization is that particle-hole excitations have a bosonic character and are
 well defined  quasi-particles at low energies.
 It consists essentially in a systematic
 mapping of a  fermionic system (states, operators, Hamiltonians) into a bosonic
 one. It turns out that the bosonic language is often more suited for the understanding
 of the physics of the system, sometimes even allowing for an exact solution.

 A feature specific to one dimension and central to the whole development is the reduction 
 of the Fermi sphere  to two disconnected Fermi points $\pm k_F$. For energies 
 close to the  Fermi level, the dispersion relation can be
 linearized  around the Fermi points. Following closely  the notation  of von 
 Delft and Schoeller \cite{Delft98}, we  
 consider a Hamiltonian  with linear dispersion 
\begin{eqnarray}
\label{kin-lin}
   H_0
   &=&\sum_{k<0,\sigma}
      \left[\epsilon_F- v_F(k+ k_F)\right]c^+_{k\sigma}c_{k\sigma}
     + \sum_{k>0,\sigma }
      \left[\epsilon_F+ v_F(k- k_F)\right]c^+_{k\sigma}c_{k\sigma},
\end{eqnarray}
where  $\epsilon_F$ is the Fermi energy and $c^+_{k\sigma}$  
($c_{k\sigma}$)  creates  (annihilates) an
electron with momentum $k$ and spin direction $\sigma=\uparrow,\downarrow$; 
we take $\hbar =1$ here and in the following. Since we are interested in low
energy excitations, but not in the total ground state enery, 
we can extend the momentum domain of the first 
sum from $k\in (-\infty,0)$ to 
$k\in (-\infty,\infty)$, and the momentum domain
    of the second sum  from $k\in (0,\infty)$ to 
 $k\in (-\infty,\infty)$. In the following we will refer to 
 these sums as left and right branches and we
 introduce  the notion of left and  right movers. 
 As a result of the extension of the spectrum to minus infinity,
the Hilbert space of the model is not the usual electron 
Hilbert space, but has been
expanded to include a sea of unphysical  states as well. 
This second, unphysical set of
fermions  requires high energies for their excitation and   will not affect the low
energy properties.
In order to avoid the
divergences associated with them, the normal
ordering 
\begin{equation}
\label{normal-ordering}
:A: \;=A-~_0\langle 0|A| 0\rangle_0
\end{equation}
is introduced, where the reference state $| 0\rangle_0$ is the filled non-interacting Fermi sea.
\begin{figure}
   \centerline{\psfig{figure=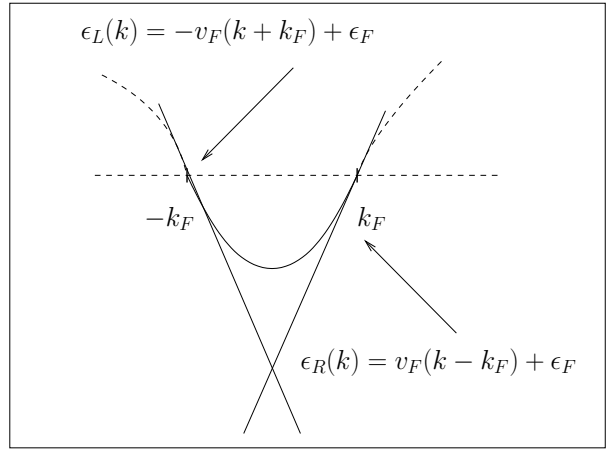,width=8cm,height=6cm}}
   \caption{Linearization of the spectrum around the Fermi points $\pm k_F$. 
   The  momentum domain of the left branch $k\in (-\infty,0)$ and 
    of the right branch $k\in (0,\infty)$
    was extended to 
 $k\in (-\infty,\infty)$.}
\end{figure} 

The physical fermion fields,  defined as
\begin{eqnarray}
   \psi_{phys}(x)&=&\frac{1}{\sqrt L}
   \sum_{\sigma}\sum_{k=-\infty}^{\infty}\textrm{e}^{ikx}c_{k\sigma},
\end{eqnarray}
are separated in left
and right movers with respect to the $\pm k_F$ points as follows:
\begin{eqnarray}
   \psi_{phys}(x)
  &\approx &\frac{1}{\sqrt L}
   \sum_{\sigma}\sum_{k>0}^{\infty}\left[
   \textrm{e}^{ikx}c_{k\sigma}+\textrm{e}^{-ikx}c_{-k\sigma}\right]
   \nonumber\\
   &=&\frac{1}{\sqrt L}\sum_{\sigma}\sum_{k>-k_F}^{\infty}
       [\textrm{e}^{-ik_Fx}\textrm{e}^{-ikx}c_{-k-k_F\sigma}+
       \textrm{e}^{ik_Fx}\textrm{e}^{ikx}c_{k+k_F\sigma}]. 
\end{eqnarray}
 For an interacting model, 
the separation into left and right moving fermions allows the  sorting of  various
scattering processes according to their initial and final states 
\cite{Solyom} (``g-ology"). 
Using the notation 
$\psi_{kL\sigma}=c_{-k-k_F\sigma}$ for the left movers and 
$\psi_{kR\sigma}=c_{k+k_F\sigma}$ for the right movers, 
the corresponding left and right fermion fields are defined by 
 \begin{equation}
  \label{fermionfield}
    \psi_{L\sigma}(x)=\frac{1}{\sqrt L}\sum\limits_{k=-\infty}^{\infty}
       \textrm{e}^{-ikx}\psi_{kL\sigma},
    \end{equation}
   \begin{equation}\label{fermionfield1}
   \psi_ {R\sigma}(x)=\frac{1}{\sqrt L}\sum\limits_{k=-\infty}^{\infty}
           \textrm{e}^{ikx}\psi_{kR\sigma},
 \end{equation}	   
where again we extended the 
 domain of momentum $k\in (-k_F,\infty)$ of each  
branch  to  $k\in (-\infty,\infty)$. 	   
These fermion fields satisfy the usual anti-commutation relations
\begin{equation}
   \{\psi_{\gamma}(x),\psi^+_{\gamma'}(x')\}=
   \delta(x-x')\delta_{\gamma\gamma'},
\end{equation}
where $\gamma,\gamma' \in \{R\uparrow, R\downarrow, L\uparrow, L\downarrow\}$.

The density operator for particles with spin $\sigma$ is generally
\begin{equation}
\label{rhosigma}
\rho_{\sigma} (q)=\sum\limits_{k}c^+_{k+q\sigma}c_{k\sigma}.
\end{equation} 
When the momentum $q$ is small, it may be written as a sum of  left and right movers 
\begin{eqnarray}
\rho_{\sigma} (q)&\approx& 
\sum_{k>0 }c^+_{k+q\sigma}c_{k\sigma}
+\sum_{k<0 }c^+_{k+q\sigma}c_{k\sigma}\nonumber\\
&=& \sum_{k }\psi^+_{k+q R\sigma}\psi_{kR\sigma}+
   \sum_{k}\psi^+_{k-q L\sigma}\psi_{kL\sigma}=
     \rho_{R\sigma} (q)+\rho_{L\sigma} (q).
\end{eqnarray}
The left and  right density operators $\rho_{R\sigma} (q)$ and 
 $\rho_{L\sigma} (q)$
 obey the following commutation relations: 
\begin{equation}
\label{co1}
[\rho_{R\sigma} (q),\rho_{R \sigma'}(-q')] =
   -\frac{qL}{2\pi}\delta_{\sigma \sigma'}\delta_{q q'},
\end{equation}
\begin{equation}
\label{co2}
[\rho_{L\sigma} (q),\rho_{L \sigma'}(-q')] = 
   \; \frac{qL}{2\pi} \delta_{\sigma \sigma'}\delta_{q q'},
\end{equation}
\begin{equation}
\label{co3}
[\rho_{L\sigma} (q),\rho_{R \sigma'}( q')]\;\; \;=\; \;0.  
\end{equation}
In evaluating the commutators (\ref{co1})-(\ref{co3}) we use  Eq.\ ({\ref{normal-ordering}}) for the 
 density operators\\ 
 $\rho_{\gamma}(q)= \;:\rho_{\gamma}(q): -~_0\langle 0|\rho_{\gamma}(q)|0\rangle_0$. 
The left and right charge and spin density operators are defined as  
\begin{eqnarray}
\rho_{L/R}(q)&=&\rho_{L/R\;\uparrow}(q)+\rho_{L/R\; \downarrow}(q),\\
\sigma_{L/R}(q)&=&\rho_{L/R\;\uparrow}(q)-\rho_{L/R\; \downarrow}(q).
\end{eqnarray}
In terms of left and right movers, the kinetic energy  
part of the Hamiltonian reads
\begin{eqnarray}
\label{H-kin}
   H_0&=&\sum_{\sigma}\sum\limits_{k
   =-\infty}^{\infty}\left[\epsilon_F+kv_F\right]
   (\psi^+_{kR\sigma}\psi_{kR\sigma}+
   \psi^+_{kL\sigma}\psi_{kL\sigma})\nonumber\\
   &=&\sum_{\sigma}\int_0^L \textrm{d}x\left[ :\psi^+_{R\sigma}(x)
   i\partial_x\psi_{R\sigma}(x):+:\psi^+_{L\sigma}(x)
   i\partial_x\psi_{L\sigma}(x):\right].
\end{eqnarray}
As a consequence of  the linear dispersion, $H_0$ and the density
operators obey the following commutation relations:
\begin{eqnarray}
[H_0,\rho_{R\sigma} (q) ] & = & v_F q\rho_{R\sigma} (q), \qquad 
[H_0,\rho_{L\sigma} (q) ]  = -v_F q\rho_{L\sigma} (q),
\end{eqnarray}
which  allows expressing  the kinetic energy  part of the Hamiltonian
as a quadratic form of  charge and spin density operators.

Using the density operators, one defines the following operators: 
\begin{equation}
   b_{ qL\sigma }=\frac{i}{\sqrt n_q}\rho_{L\sigma}(q),\;\;\;\qquad
   b^+_{qL\sigma}=-\frac{i}{\sqrt n_q}\rho_{L\sigma}(-q),\qquad
\end{equation}
\begin{equation}
 b_{qR\sigma}=\frac{i}{\sqrt n_q}\rho_{R\sigma}(-q),\qquad
   b^+_{qR\sigma}=-\frac{i}{\sqrt n_q}\rho_{R\sigma}(q),\qquad   
\end{equation}  
where $n_q=qL/2\pi$, and $q>0$. These operators  fulfill the
bosonic commutation rules
\begin{equation}
        [b_{q\gamma},b_{q'\gamma'}^+]=\delta_{qq'}\delta_{\gamma\gamma'},
\end{equation}
where $\gamma,\gamma' \in \{R\uparrow, R\downarrow, L\uparrow, L\downarrow\}$.
\begin{figure}
\centerline{\psfig{figure=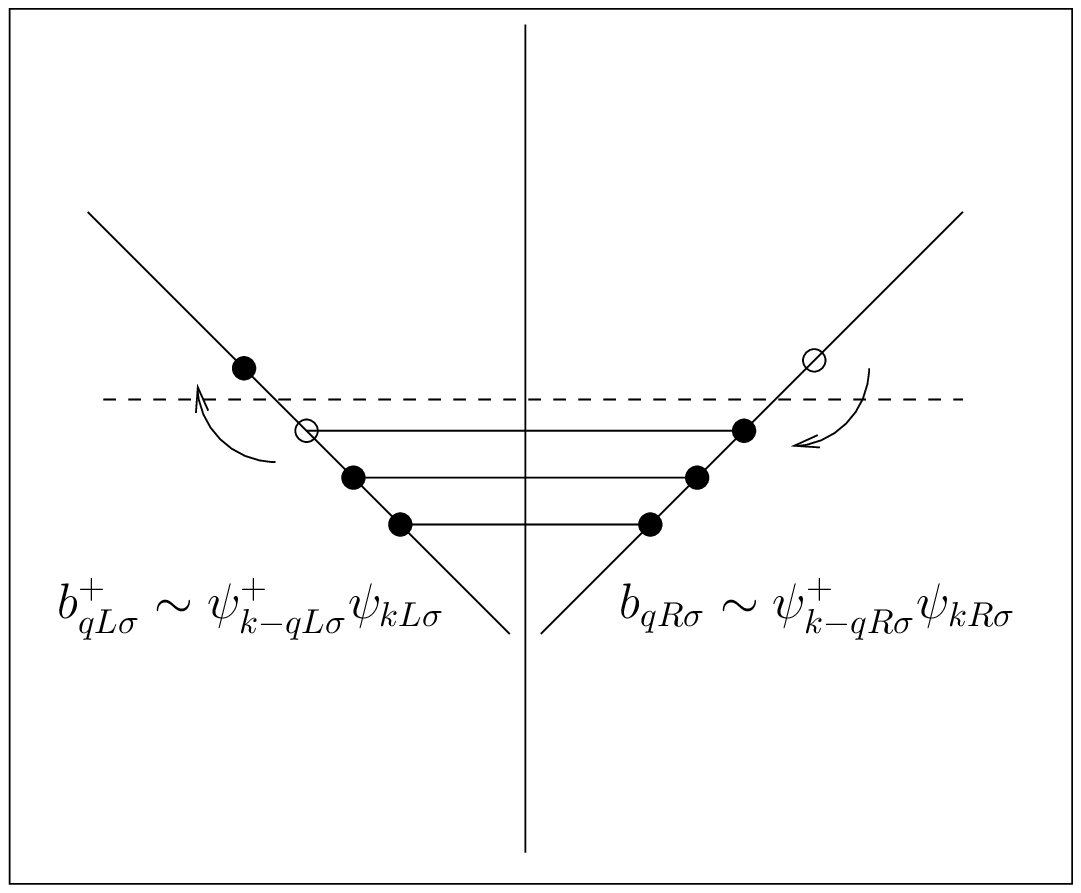,width=8.5cm,height=7.5cm}}
\caption{$b^+_{qL\sigma}$ creates a particle-hole excitation on the left side,
 while $b_{qR\sigma}$ annihilates a particle-hole
excitation on the right side.}
\end{figure}

Returning to real space, bosonic fields are defined as 
\begin{equation}
\label{varphiL}
      \varphi_{L\sigma}(x)=\sum_{q>0}\frac{1}{\sqrt n_q}
      \textrm{e}^{-iqx-aq/2}b_{qL\sigma},
\end{equation}
\begin{equation}
\label{varphiR}
     \varphi_{R\sigma}(x)=-\sum_{q>0}\frac{1}{\sqrt n_q}
      \textrm{e}^{iqx-aq/2}b_{qR\sigma},
\end{equation}
where $a\rightarrow 0$ is assumed. Formally, $a^{-1}$  is a cutoff for the $q$-summation;
physically, $a$ corresponds to the lattice constant.
It has been shown that  the fermionic field operators are related to the bosonic fields 
via the following bosonization
identity \cite{Haldane81, Delft98}:
\begin{eqnarray}
\label{bosident}
\Psi_{L \sigma} (x) & = & \frac{1}{\sqrt{L}} F_{L\sigma}
{\rm e}^{-i\varphi_{L\sigma }^+(x)} {\rm e}^{-i\varphi_{L \sigma}(x)}
{\rm e}^{-2\pi i N_{L\sigma} x/L} \\
\label{bosident2}
 & = &
\frac{1}{\sqrt{2 \pi a}} F_{L \sigma}
{\rm e}^{-i(\varphi_{L\sigma}^+(x) + \varphi_{L \sigma}(x))}
{\rm e}^{-2\pi i N_{L \sigma}x/L}, \\
\Psi_{R \sigma} (x) & = &
\frac{1}{\sqrt{2 \pi a}} F_{R \sigma}
{\rm e}^{i(\varphi_{R\sigma}^+(x) + \varphi_{R \sigma}(x))}
{\rm e}^{2\pi i N_{R \sigma}x/L}.
\end{eqnarray}
The factor $\sqrt{L/2\pi a}$ in Eq.\ (\ref{bosident2}) comes from the commutator 
$$[\varphi_{\gamma}^+(x),\varphi_{\gamma}(x)]/2=-\ln \sqrt{L/2\pi a}.$$
The  $N_\gamma$ operators introduced in the bosonization identity count
  the number of
 $\gamma$-electrons relative to the reference state $|0\rangle_0$:
\begin{eqnarray}
N_{\gamma} \equiv\sum_{k=-\infty}^{\infty}:\psi_{k\gamma}^+\psi_{k\gamma}:\;\;=
     \sum_{k=-\infty}^{\infty}\left[\psi_{k\gamma}^+\psi_{k\gamma}-
        ~_0\langle 0|\psi_{k\gamma}^+\psi_{k\gamma}0\rangle_0\right].
\end{eqnarray}
The $F^+_\gamma, F_\gamma$ operators, known as Klein factors or ladder operators, 
raise or lower the fermion numbers and assure that fermions of different species
anti-commute. A detailed discussion of their properties  
is given in  section \ref{kleinfactors}.

Later it will be convenient to introduce new charge and spin fields, 
$\phi_{c,s}$ and $\theta_{c,s}$, which are 
linear combinations of left and right movers
\begin{eqnarray}
\label{dual1}
\phi_{c,s} &= & \frac{1}{2\sqrt{2}}(\varphi_{L \uparrow} \pm \varphi_{L \downarrow }
                              + \varphi_{R \uparrow} \pm \varphi_{R \downarrow}+
			      {\rm h.c.} ), \\
\label{dual2}			      
\theta_{c,s} &= & \frac{1}{2\sqrt{2}}(\varphi_{L \uparrow} \pm \varphi_{L \downarrow }
                              - \varphi_{R \uparrow} \mp \varphi_{R \downarrow}+
			      {\rm h.c.} ),
\end{eqnarray}
where the index $c$ stands for charge and the index $s$ stands for spin. 
They satisfy the commutation relations
\begin{eqnarray}
[ \phi_{\beta}(x),\phi_{\beta'}(y)] &=&
      [\theta_{\beta}(x),\theta_{\beta'}(y)]=0, 
\end{eqnarray}
\begin{eqnarray}
 [\phi_{\beta}(x),\theta_{\beta'}(y)]&=& 
    -i\pi \Theta (x-y)\delta_{\beta\beta'}
    +\frac{i\pi}{2}\delta_{\beta\beta'}+\frac{i\pi}{L}(x-y) \delta_{\beta\beta'}, \;
 a\rightarrow 0,
\end{eqnarray} 
\begin{eqnarray}
[\phi_{\beta}(x),\partial_y\theta_{\beta'}(y)]&=&
 i\pi \delta(x-y)\delta_{\beta\beta'}-\frac{i\pi}{L} \delta_{\beta\beta'},\;
 a\rightarrow 0,
\end{eqnarray}
where $\beta,\beta'\in\{c,s\}$. 
In the thermodynamic limit, i.e.\ when the term
 $\frac{1}{L}\delta_{\beta\beta' }$ can be neglected, 
 the  fields $\phi_\beta(x)$ and $\partial_x \theta_\beta(x)$ are 
conjugate variables.

We express now the electron density using the bosonization formalism.
In real space, the density operator  reads
\begin{equation}
\rho_{\sigma}(x)=\sum\limits_{q}\textrm{e}^{iqx}\rho_{\sigma}(q)\approx
  \sum\limits_{|q|\ll 2k_F}\textrm{e}^{iqx}\rho_{\sigma}(q)+
  \sum\limits_{\pm}\sum\limits_{|q|\ll 2k_F}
  \textrm{e}^{i(\pm 2k_F+q)x}\rho_{\sigma}(\pm 2k_F+q),
\end{equation}
which in terms of bosonic fields is 
\begin{eqnarray}
\rho_{\sigma}(x)&\approx&
   \frac{N_{R \sigma}+N_{L \sigma}}{L}
    +\frac{1}{2\pi} 
    \partial _x[\phi_{L\sigma}(x) +\phi_{R\sigma}(x)]\nonumber\\      
   &+&\frac{1}{2\pi a}F^+_{L\sigma}F_{R\sigma} \textrm{e}^{-2ik_F x} 
	      \textrm{e}^{2\pi i(N_{R \sigma}+N_{L \sigma}+1)x/L}
	      \textrm{e}^{i(\phi_{L\sigma} +\phi_{R\sigma})}
	      \textrm{e}^{-ix(G-4k_F) }\nonumber\\  
     &+& \frac{1}{2\pi a}F^+_{R\sigma}F_{L\sigma}\textrm{e}^{-2ik_F x }
	      \textrm{e}^{-2\pi i(N_{R \sigma}+N_{L \sigma}+1)x/L}
        \textrm{e}^{-i(\phi_{L\sigma} +\phi_{R\sigma})}
         +\textrm{h.c.}.
\end{eqnarray}
Here, the term in the second line is important for commensurate band filling, when
the reciprocal lattice vector is $G\approx 4k_F$.
In the literature \cite{Fukuyama}, ignoring both the Klein factors
and the term in the second line, the density operator is simply given as
\begin{equation}
\rho_{\sigma}(x)=
   \frac{1}{2\pi} \partial _x[\phi_{L\sigma}(x) +\phi_{R\sigma}(x)]
    +\frac{1}{\pi a}\cos [-2k_F x+\phi_{L\sigma}(x) +\phi_{R\sigma}(x)]. 
\end{equation}

\section{Klein factors}
\label{kleinfactors}	
The Klein operators $F_{\gamma}$ and $F^+_{\gamma}$ lower
or raise the total fermion number by one, 
which no combination of bosonic operators can ever do, and
they assure also that fermions of different species anti-commute. In  the thermodynamic
limit and for  gapless systems,  for most  practical purpose, e.g.\ the computation
of correlation functions, there is  no difference between states containing 
$N_{\gamma}$ and $N_{\gamma}\pm 1$   particles.
However, when gaps open up, giving rise to finite correlation lengths, 
one must be careful  
when dealing with these operators \cite{Schonhammer01,Schonhammer02,Varma02}.  

\begin{figure}
 \centerline{\psfig{figure=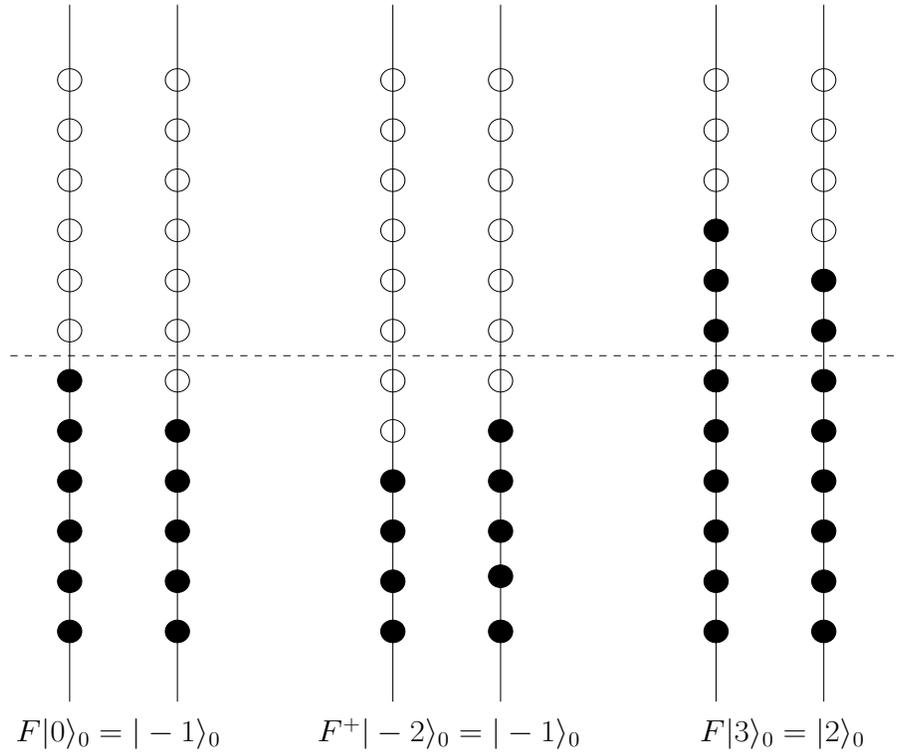,width=12cm,height=10cm}}
  \caption{The action of $F$ on the  states $|0\rangle_0$ and  $|3\rangle_0$, and the action of 
  $F^+$ on the state $|-2\rangle_0$.}
 \label{figklein1} 
\end{figure}

For systems with only one type of fermions the Klein factors 
$F_{\gamma}, F_{\gamma}^+$ simply  
modify the number of particles corresponding to the 
ground state on which the boson excitations are created, as depicted in 
Fig.\ \ref{figklein1}.
The Klein factors  commute with the bosonic operators $b_{q\gamma}$, 
$b_{q\gamma}^+$, but not
with the  particle number operator $N_{\gamma}$:
\begin{equation}
[F_{\gamma},N_{\gamma}]=F_{\gamma},\qquad [F_{\gamma}^+,N_{\gamma}]=
                -F_{\gamma}^+.
\end{equation}

If  more than one  species of fermions is considered, e.g.\ two
 in the case of  spinless fermions, 
 the corresponding 
($N_L,N_R$)-particle ground states are tensor products of left
and right movers:
\begin{equation}
|N_L,N_R\rangle_0 =|N_L\rangle_0  \otimes |N_R\rangle_0.
\end{equation}
In order to preserve the fermionic character  of 
the annihilation/creation operators, the $F_R/F^+_R$ 
operators are defined such that
they pick up a minus  sign   when they pass the left  fermionic operators:
\begin{eqnarray}
F_R|N_L,N_R\rangle_0 &\equiv & (-1)^{N_L}|N_L\rangle_0 
              \otimes |N_R-1\rangle_0,\\
F_R^+|N_L,N_R\rangle_0 &\equiv &(-1)^{N_L}|N_L\rangle_0  
         \otimes |N_R+1\rangle_0.
\end{eqnarray}	 
In general, for fermionic  systems  consisting of  $M$ species, 
 one has to  define
 a particular order for the 
species ($N_1, ..., N_M$), such that
\begin{equation} 
 |N_1,N_2,..., N_M\rangle_0 \equiv |N_1\rangle_0  \otimes |N_2\rangle_0
 \otimes .... \otimes|N_M\rangle_0
 \end{equation}
and the Klein factors pick up the total sign change 
from the preceding fermionic  operators:
\begin{equation}
F_\gamma|N_1,N_2,...N_M\rangle_0 
   \equiv  (-1)^{\sum\limits_{\alpha=1}^{\gamma-1}N_\alpha}
   |N_1,N_2,...,N_\gamma-1,..., N_M\rangle_0,
\end{equation} 
\begin{equation}
F^+_\gamma|N_1,N_2,...N_M\rangle_0 
  \equiv  (-1)^{\sum\limits_{\alpha=1}^{\gamma-1}N_\alpha}
  |N_1,N_2,...,N_\gamma+1,..., N_M\rangle_0.
\end{equation}
For fermions with spin $1/2$ (i.e.\ $M=4$), the following order is defined:
\begin{equation} 
 |N_{L\uparrow},N_{L\downarrow},N_{R\uparrow},N_{R\downarrow}\rangle_0. 
 \end{equation} 
According to  their definition, the Klein factors of different kind
anti-commute. The 
commutation relations fulfilled by the Klein factors
are
\begin{equation}
[b_{q\gamma},F_{\gamma'}]=
 [b^+_{q\gamma},F_{\gamma'}]=[b_{q\gamma},F^+_{\gamma'}]=
             [b^+_{q\gamma},F^+_{\gamma'}]=0,
\end{equation}
\begin{equation}
\{F^+_{\gamma},F_{\gamma'}\}=2\delta_{\gamma\gamma'},
\end{equation}
\begin{equation}
\{F_{\gamma},F_{\gamma'}\}= \{F^+_{\gamma},F^+_{\gamma'}\}=0 
         \quad ( \textrm{if} \quad\gamma \ne \gamma'),
\end{equation}	 
\begin{equation}
	 [F_{\gamma},N_{\gamma'}] = \delta_{\gamma\gamma'}F_{\gamma},
\end{equation}	 
\begin{equation}
[F^+_{\gamma},N_{\gamma'}] = - \delta_{\gamma\gamma'}F^+_{\gamma}.
\end{equation}
\section{Luttinger model}
\label{luttinger}

\begin{figure}
 \centerline{\psfig{figure=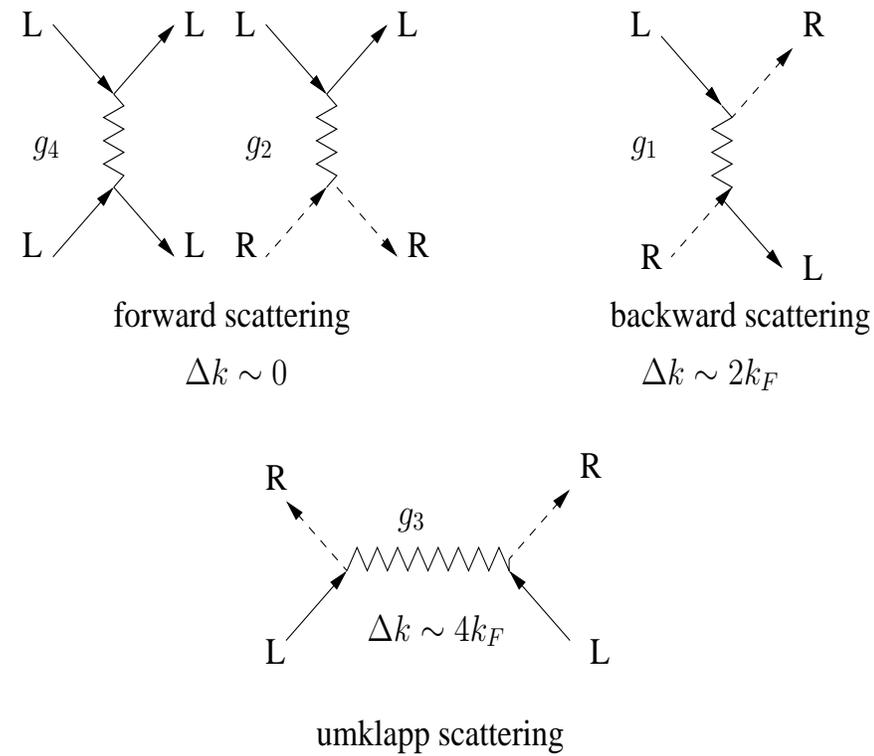,width=11.5cm,height=10cm}}
    \caption{Graphic representation of the scattering processes. The solid and the 
    dashed lines correspond to
    electrons belonging to the branches containing $-k_F$ and $+k_F$,
 respectively. }
\label{gology}
\end{figure}

As a result of the approximations made,
namely   the spectrum linearization 
 around the Fermi points and the introduction of a filled Fermi sea, some one-dimensional models  become exactly solvable in  certain cases, within
the bosonization method.
The Hamiltonian under consideration is 
\begin{equation}
H=H_0+H(g_1)+H(g_2)+H(g_3)+H(g_4),
\end{equation}
where $H_0$ is the kinetic energy given in Eq.\ (\ref{H-kin}) and 
$H(g_1),...,H(g_4)$
are interaction terms,  given by
\begin{eqnarray}
\label{g-ology}
    H(g_1)&=&\frac{1}{L}\sum_{kpq\sigma\sigma'} (g_{1\parallel}
      \delta_{\sigma\sigma'}+g_{1\perp}
      \delta_{-\sigma\sigma'})\psi_{kR\sigma}^+\psi_{pL\sigma'}^+
        \psi_{-p+qR\sigma'}\psi_{-k+qL\sigma},\nonumber\\
    H(g_2)&=&\frac{1}{L}\sum_{kpq\sigma\sigma'} 
       (g_{2\parallel}\delta_{\sigma\sigma'}+g_{2\perp}
      \delta_{-\sigma\sigma'})
       \psi_{kR\sigma}^+\psi_{pL\sigma'}^+\psi_{p-qL\sigma'}
       \psi_{k-qR\sigma},\nonumber\\
    H(g_3)&=&\frac{1}{2L}\sum_{kpq\sigma\sigma'} 
      (g_{3\parallel}\delta_{\sigma\sigma'}+g_{3\perp}
          \delta_{-\sigma\sigma'})\nonumber\\
          &\times&(\psi_{kL\sigma}^+\psi_{pL\sigma'}^+\psi_{q-pR\sigma'}
          \psi_{-k-q-(G-4k_F)R\sigma}\nonumber\\
        &&\quad\quad\quad\quad+\psi_{kR\sigma}^+
\psi_{pR\sigma'}^+\psi_{-p-q-(G-4k_F)L\sigma'}\psi_{-k+qL\sigma}),\nonumber\\
    H(g_4)&=&\frac{1}{2L}\sum_{kpq\sigma\sigma'} (g_{4\parallel}
             \delta_{\sigma\sigma'}+g_{4\perp}
          \delta_{-\sigma\sigma'}) \nonumber\\
             \quad &\times&(\psi_{kR\sigma}^+\psi_{pR\sigma'}^+\psi_{p+qR\sigma'}
             \psi_{k-qR\sigma}+
            \psi_{kL\sigma}^+\psi_{pL\sigma'}^
	    +\psi_{p-qL\sigma'}\psi_{k+qL\sigma}).\nonumber \\
\end{eqnarray}
For a graphical representation see Fig.\ \ref{gology}.

 The nomenclature in terms of $g_i$ is standard in the literature and is 
 called ``g-ology" \cite{Solyom}. 
 The $g_2$ and $g_4$
interactions are scattering processes  with small momentum transfer, i.e.\ $\ll 2k_F$,
while  the $g_1$ interaction
is a scattering process with a large  momentum transfer $\approx 2k_F$. 
 Umklapp scattering 
is relevant only for $G\approx4k_F, 6k_F,...$, where $G$ is a reciprocal lattice vector, i.e.\ for commensurate filling.
If the interactions are assumed to be instantaneous and local, the Pauli
principle forbids the processes corresponding to $g_{3\parallel}$ and 
$g_{4\parallel}$.
In the absence of the $g_{1\perp}$ and $g_{3\perp}$ processes  the bosonization 
of the terms  (\ref{H-kin}) and (\ref{g-ology}) leads to
 the  exactly solvable quadratic Hamiltonian 
\begin{eqnarray}
\label{densLutt}
H_{\textrm{Luttinger}}&=\;&\sum\limits_{q>0} 
    \left(\frac{\pi v_F}{L}+\frac{g_{4\parallel}+g_{4\perp}}{2L}\right)
        (\rho_L(q)\rho_L(-q)+\rho_R(q)\rho_R(-q))\nonumber\\
   & +&\sum\limits_{q>0}\frac{-g_{1\parallel}+g_{2\parallel}+g_{2\perp}}{2L}
       (\rho_L(q)\rho_R(-q)+\rho_R(q)\rho_L(-q))\nonumber\\
   &+ & \sum\limits_{q>0}\left(\frac{\pi v_F}{L}+\frac{g_{4\parallel}-g_{4\perp}}{2L}\right)
        (\sigma_L(q)\sigma_L(-q)+\sigma_R(q)\sigma_R(-q))\nonumber\\
   & +&\sum\limits_{q>0}\frac{-g_{1\parallel}+g_{2\parallel}-g_{2\perp}}{2L}
       (\sigma_L(q)\sigma_R(-q)+\sigma_R(q)\sigma_L(-q)) \nonumber\\ 
    & +& \textrm{``zero modes" }  (q=0), 
\end{eqnarray}
which in terms of dual charge and spin fields, Eqs.\ (\ref{dual1})-(\ref{dual2}), is known as Luttinger Hamiltonian
\begin{eqnarray}
\label{Lutt}
H_{\textrm{Luttinger}}&=&\sum_{\alpha=c,s}\int_0^L
    \frac{ dx }{2\pi}\left[\frac{v_\alpha}{g_\alpha}(\partial_x\phi_\alpha)^2 +
      v_\alpha g_\alpha(\partial_x\theta_\alpha)^2 \right]\nonumber\\
      &+&
      \frac{\pi}{4L}\sum_{\alpha=c,s}\left[\frac{v_\alpha}{g_\alpha}N^2_\alpha+
        v_\alpha g_\alpha J^2_\alpha\right].
\end {eqnarray} 
The coefficients $v_\alpha$ and  $g_\alpha$ are called  Luttinger parameters, and
they are related to the $g$'s  via \cite{Schulz}
\begin{eqnarray}
\label{lutt-param-vs}
v_s&=&v_F\sqrt{\left(1+\frac{g_{4\parallel}}{2\pi v_F}-\frac{g_{4\perp}}{2\pi v_F}\right)^2-
     \left(\frac{g_{2\parallel}-g_{1\parallel}-g_{2\perp}}{2\pi v_F}\right)^2},
 \\\label{lutt-param-vc}
 v_c&=&v_F\sqrt{\left(1+\frac{g_{4\parallel}}{2\pi v_F}+\frac{g_{4\perp}}{2\pi v_F}\right)^2
            -\left(\frac{g_{2\parallel}-g_{1\parallel}+g_{2\perp}}{2\pi v_F}
	         \right)^2 },
 \\\label{lutt-param-gs}
g_s&=&\sqrt{\frac
 {2\pi v_F+g_{4\parallel}-g_{4\perp}-g_{2\parallel}+g_{1\parallel}+g_{2\perp}}
 {2\pi v_F+g_{4\parallel}-g_{4\perp}+g_{2\parallel}-g_{1\parallel}-g_{2\perp}}},
\\\label{lutt-param-gc}
g_c&=&\sqrt{\frac
 {2\pi v_F+g_{4\parallel}+g_{4\perp}-g_{2\parallel}+g_{1\parallel}-g_{2\perp}}
 {2\pi v_F+g_{4\parallel}+g_{4\perp}+g_{2\parallel}-g_{1\parallel}+g_{2\perp}}}.
\end{eqnarray}
The Luttinger Hamiltonian has the same type of low energy
excitations as the harmonic chain, i.e.\ it corresponds to a free boson model 
(see appendix \ref{H-bosop}). It contains two decoupled sectors corresponding
to charge and spin excitations. The phenomenon of spin-charge separation is an
important feature of fermionic systems in one dimension. 
For example, a ``real"  electron which is injected into 
 an interacting system  will decay into its constituent
elementary excitations which are  charge and spin density modes that propagate at
different velocities ($v_c$ and $v_s$). 

Besides the bosonic part, the Hamiltonian contains  also a second part 
(given in the second line of (\ref{Lutt}))
which is a 
combination of $N_\eta$ operators, defined as follows:
\begin{equation}
\label{nc}
\textrm{total charge:}
 \quad  \quad  \;\; N_c=N_{L,\uparrow}+N_{L,\downarrow}+N_{R,\uparrow}+N_{R,\downarrow},
\end{equation}
\begin{equation}
 \label{jc}
\textrm{charge current:}\quad\;\; \;J_c=
 N_{L,\uparrow}+N_{L,\downarrow}-N_{R,\uparrow}-N_{R,\downarrow},
\end{equation} 
\begin{equation}
 \label{ns}
\textrm{total spin:}
\quad \quad \quad \;\; N_s=N_{L,\uparrow}-N_{L,\downarrow}+N_{R,\uparrow}-N_{R,\downarrow},
\end{equation} 
\begin{equation}
 \label{js}
\textrm{spin current:}
\quad \quad \;\;\;J_s=N_{L,\uparrow}-N_{L,\downarrow}-N_{R,\uparrow}+N_{R,\downarrow};
\end{equation} 
the fermionic ground state $|N_{L,\uparrow},N_{L,\downarrow},N_{R,\uparrow},N_{R,\downarrow}\rangle_0$
can be redefined in terms of these new operators as
\begin{equation} 
 |N_{L\uparrow},N_{L\downarrow},N_{R\uparrow},N_{R\downarrow}\rangle_0\equiv 
 |N_c,J_c,N_s,J_s\rangle_0.
\end{equation}  



\chapter{Spinless fermions }
\label{spinless}
In the  Luttinger Hamiltonian, 
 the Klein factors appear only in the combination  $F^+_{\eta}F_{\eta}=1$ and 
 hence they can be ignored.
 The 
Klein factors become important when nonlinear terms appear. 
 In  order to
 study their role in more detail,  we consider in the
 following a system  of interacting spinless fermions with nearest-neighbor
  interaction on a
 one-dimensional lattice with static dimerization. The static dimerization
 is  introduced as a
 periodic lattice distortion, which leads to a modulation of the hopping
  amplitude.
Both the  nearest-neighbor interaction and the static dimerization
 introduce nonlinearities
in the bosonized Hamiltonian, with non-trivial combinations of Klein factors  
$F^+_{L}F_{R}$,
which require a careful and rigorous treatment \cite{Mocanu1-04}. Using
 a variational ansatz in which the
bosonic fields  and the Klein factors  are treated on equal footing,
 we calculate the 
energy gap and the Drude weight. To verify the accuracy  of the  method 
 we compare 
our  results with numerical and  exact  results which are 
available for certain values of the model parameters.
\nop{The Hamiltonian of this model is bosonized and a trial
Hamiltonian is constructed which serve as as the basis for the SCHA.
It is  calculated the energy gap and the Drude weight and the results are compared 
with exact results that are available for certain values of the model parameters.}

\section{Model and formalism}
The system of interacting spinless fermions 
 with static
dimerization is described by the following Hamiltonian:
\begin{eqnarray}
 \label{eq:dim-term}
    H & = & -t\sum_{j}(1+(-1)^j u)(c^+_jc_{j+1} + c^+_{j+1}c_j )\nonumber\\
     & &{} + V \sum_{j}n_j n_{j+1},
\end{eqnarray}
where $u$ is the dimerization parameter that leads to a periodic
modulation of the hopping amplitude, and $V$ is the strength of the
nearest-neighbor interaction.
\begin{figure}[h]
\label{peierls}
\centerline{\psfig{figure=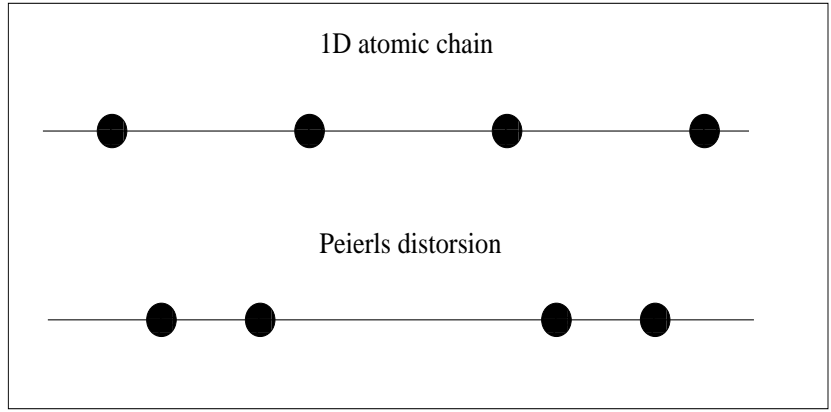,width=11cm,height=5.5cm}}
\caption{Modulation of the hopping amplitude}
\end{figure}  
In momentum space fermion operators are defined by 
\begin{equation}
c_j=\frac{1}{\sqrt{L/a}}\sum_{k}e^{ikR_j}c_k,\qquad \textrm{with }\; R_j=ja, 
\end{equation}
and correspondingly the Hamiltonian (\ref{eq:dim-term}) reads
\begin{eqnarray}
 \label{hamiltFourier}
      H & = &\sum_{k}\epsilon(k)c^+_kc_{k}
           +2itu\sum_{k}\sin(ka)c^+_kc_{k+\pi/a}\nonumber\\
         & &{}+\frac{a}{L}\sum_qV(q)\sum_{kp}c^+_kc^+_pc_{p-q}c_{k+q},
\end{eqnarray}
where $V(q)=Ve^{-iqa}$, $\epsilon (k)=-2t\cos(ka)$, and $a$ is the lattice spacing.

\begin{figure}[ht]
\label{peierls}
\centerline{\psfig{figure=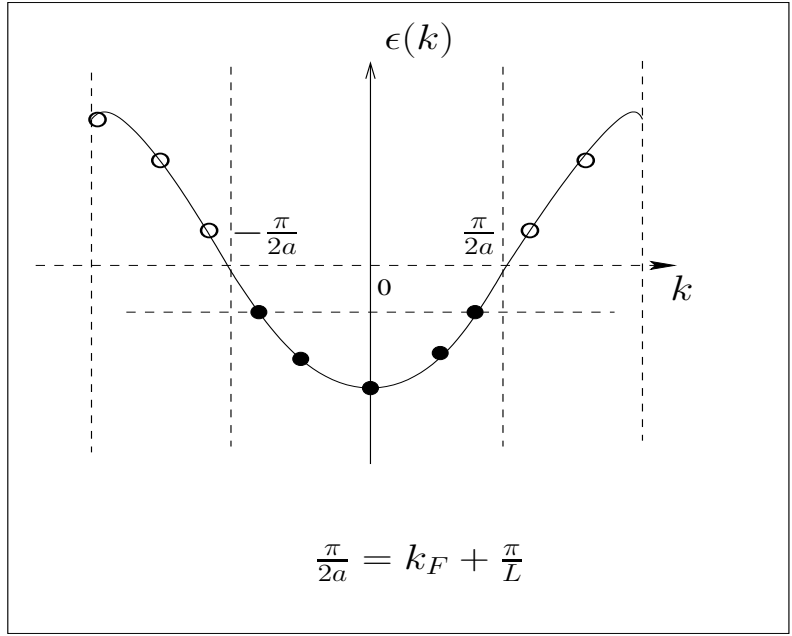,width=10.5cm,height=8.5cm}}
\caption{Reference state for a system with $10$ lattice sites,
 chosen as the filled Fermi
sea  for $5$ particles.}
\label{refstate}
\end{figure}

In order to have a non-degenerate ground state 
we consider a half-filled system with an odd  number of particles. Note that in this case,
$k_F=\pi/2a-\pi/L$. Before bosonizing 
the Hamiltonian we have to choose  a reference state with
 respect to which the boson
excitations are created; we choose the Fermi sea of the half-filled system 
as this 
  reference state, as illustrated
in Fig.\ \ref{refstate}. 
In order to bosonize the Hamiltonian (\ref{hamiltFourier}) we follow the 
notation introduced in chapter \ref{bosonization}.
 The left and
right spinless boson fields, and the bosonization identities are given by 
Eqs.\ (\ref{varphiL})-(\ref{bosident2}), where the spin index has to be dropped.
Correspondingly, the charge and spin fields introduced as linear combinations of the left and
right movers, given by Eqs.\ (\ref{dual1})-(\ref{dual2}), 
are replaced for spinless fermions by 
 \begin{eqnarray}
 \label{spinlessdual1}
 \phi(x)&=&\frac{1}{2}
    (\varphi_{L}+\varphi_{R}+\varphi^+_{L}+\varphi^+_{R}),
 \\
 \label{spinlessdual2}
 \theta(x)&=&\frac{1}{2}
    (\varphi_{L}-\varphi_{R}+\varphi^+_{L}-\varphi^+_{R}).
 \end{eqnarray}  
The bosonized Hamiltonian of
this model reads
\begin{equation}
\label{eq:totalhamilt}
H=H_0+H_{\rm{Peierls}}+H_{\rm{Umklapp}}.
\end{equation}
\begin{eqnarray}
 \label{spinless-lutt}
   H_0\qquad & = &\int_0^L \frac{{\rm d} x}{2\pi}\left\{\frac{v}{g}:(\partial_x\phi)^2:
     + \;v g :(\partial_x\theta)^2:\right\}+ \frac{\pi v}{2Lg}N^2
     +  \frac{\pi vg}{2L}J^2,
      \\
  \label{spinless-peierls}    
   H_{\rm{Peierls}} \;& =&
      - \tilde{u}\int_0^L {\rm d}x \ iF_L^+F_R  \ {\rm e}^{2i\phi(x)}
                               \ {\rm e}^{2i\pi xN/L}+{\rm h.c.},
      \\
   \label{spinless-umklapp}   
   H_{\rm{Umklapp}} & =&+ \tilde{V}\int_0^L {\rm d}x \ F_R^+F_R^+F_LF_L \ {\rm e}^{-4i\phi(x)} 
       \ {\rm e}^{-4i\pi xN/L}+{\rm h.c.},
\end{eqnarray}
where $\tilde{u}=tu/(\pi a)$, $\tilde{V}=V(2k_F)/(2\pi)^2a$ and
$V(2k_F)=-Ve^{ 2i\pi a/L}\approx-V$. 
 $\phi$ and $\theta$ are defined in Eqs.\ (\ref{spinlessdual1})-(\ref{spinlessdual2}) 
 and the cut-off  parameter $a$ 
  is  chosen equal to the  lattice spacing. $N$ and $J$ are the charge and
 current operators defined  as $N=N_L+N_R$ and $J=N_L-N_R=0,\pm 2,
 \pm 4,...$ . The particle number $N$ is a conserved quantity and is counted 
 relative to the reference state, i.e.\ $N=0$ in the case of half filling.
The  $g$ and $v$ are the Luttinger liquid parameters, given in the weak  coupling limit by 
\begin{equation}
v=v_F\sqrt{\left(1+\frac{g_4}{\pi v_F}\right)^2-\left(\frac{g_2}{\pi v_F}\right)^2},\quad
   g=\sqrt{\frac{1+\frac{g_4}{\pi v_F}-\frac{g_2}{\pi v_F}}
   {1+\frac{g_4}{\pi v_F}+\frac{g_2}{\pi v_F}}},
\end{equation}   
where we have  introduced the notation $g_4=V(0)$, $g_2=V(0)-V(2k_F)$.
At half filling, for the model of spinless fermions with only nearest neighbor interaction,
 the Luttinger parameters can
be exactly determined by comparison with the Bethe ansatz solution \cite{Lieb}. 
The renormalized Fermi velocity $v = \pi ta\sin(2\eta)/(\pi-2\eta)$ and the
Luttinger parameter $g = \pi/4\eta$ are related with the interaction
according to $V=-2t\cos(2\eta)$.

\begin{table}
\begin{center}
\begin{tabular}{|c||c|c|c|c|c|c|c|}
\hline
 & &  &  & & &  & \\
$V/t$ & $-2$ & $-\sqrt{2}$ & $-1$ & $0$& $1$& $\sqrt{2}$ & $2$\\
\hline
 & &  &  & & &  & \\
$\eta$ & $0$ & $\pi/8$ & $\pi/6$ & $\pi/4$& $\pi/3$& $3\pi /8$ & $\pi/2$\\
\hline
 & &  &  & & &  & \\
$g$ & $\infty$ & $2$ & $3/2$ & $1$& $3/4$& $2/3$ & $1/2$\\
\hline
 & &  &  & & &  & \\
$v/ta$ & $0$ & $0.94$ & $1.29$ & $2$& $2.59$& $2.82$ & $\pi$\\
\hline
\end{tabular}
\caption{Numerical examples for the relation between standard parameters of the
spinless-fermion model at half filling, on the basis of the Bethe ansatz results in the
regime $-2t\leq V\leq 2t:$ $V=-2t\cos (2\eta)$, $g=\pi/4\eta$, and 
$v= \pi ta\sin(2\eta)/(\pi-2\eta)$.}
\end{center}
\end{table}

The  Klein factors
and the term proportional to the current operator $\pi vgJ^2/(2L)$ do 
not commute. 
We define a new operator $A=F^+_RF_L$ and its hermitian conjugate
$A^+=F^+_LF_R $. The four-fermion terms arising from Umklapp scattering are 
$F_R^+F_R^+F_LF_L=-A^2$ and $F_L^+F_L^+F_RF_R=-(A^+)^2$. Since the Klein factors
are unitary operators, it is easy to show that also  the operator  $A$ is unitary and one
concludes that its eigenvalues are pure phase factors, i.e.
\begin{equation}
\label{evA}
    A|k\rangle=\textrm{e}^{ik}|k\rangle,\qquad
      A^+ |k\rangle=\textrm{e}^{-ik}|k\rangle,
\end{equation}
 with $0\le k<2\pi$. 

In the thermodynamic limit, $L\rightarrow\infty$,  the term $\sim J^2$ can be neglected if
we restrict the range of $J$ values to a finite interval.
We can choose a basis which diagonalizes the Klein factors and  replace
them 
 by their 
  eigenvalues. At the end one  obtains 
  \begin{eqnarray}
  \label{HP}
  H_{\rm{Peierls}} \;& =&
      - \tilde{u}\int_0^L {\rm d}x \ i{\rm e}^{ik} \ {\rm e}^{-2i\phi(x)}
                               \ {\rm e}^{-2i\pi xN/L}+{\rm h.c.},
    \\ 
    \label{HU} 
   H_{\rm{Umklapp}} & =&+ \tilde{V}\int_0^L {\rm d}x \ {\rm e}^{2ik}
      \ {\rm e}^{-4i\phi(x)} 
       \ {\rm e}^{-4i\pi xN/L}+{\rm h.c.}.
\end{eqnarray} 	  
As a result  the Hamiltonian  separates into different sectors of purely bosonic
Hamiltonians, labeled by $k$. 

As discussed e.g.\ by Schulz \cite{Schulz}, in the thermodynamic limit the Klein factors 
may be  replaced by  the Majorana (``real") 
fermion operators
$\eta_{\alpha}$, which assure  the proper anti-commutation between the different species of
fermions, but do not change the  particle numbers of  the ground state. The Majorana fermions satisfy
the anti-commutation relation 
\begin{equation}
\label{Majorana}
\{\eta_\alpha,\eta_\beta\}=2\delta_{\alpha,\beta},
\end{equation}
where $\alpha,\beta$ could be $R,L$ and $\sigma$ for fermions with spin. 
A typical fermion interaction term, involving  Klein factors, looks like
\begin{equation}
\psi^+_\alpha\psi^+_\beta\psi_\gamma\psi_\delta =
 F^+_\alpha F^+_\beta F_\gamma F_\delta \times 
(\textrm{boson operators}).
\end{equation}
Using Majorana operators a state with $N$ particles is equivalent to a  state with 
  $N \pm 1$ particles and a  fermion interaction term looks like 
\begin{equation}
\psi^+_\alpha\psi^+_\beta\psi_\gamma\psi_\delta =
   \eta_\alpha \eta_\beta \eta_\gamma \eta_\delta \times
     (\textrm{boson operators}),
\end{equation}     
with 
\begin{equation}
\label{h1h2h3h4}
( \eta_\alpha \eta_\beta \eta_\gamma \eta_\delta )^2 = 1\quad 
\Rightarrow \quad\eta_\alpha \eta_\beta \eta_\gamma \eta_\delta=\pm 1.
\end{equation}   
Using  Eqs.\ (\ref{Majorana}) and (\ref{h1h2h3h4}),
 one obtains only the eigenvalues $\eta_R\eta_L=-\eta_L\eta_R=\pm i$,
 instead of $\textrm{e}^{\pm ik}$,
i.e. the continuity is lost. Within the approximate approach using 
Majorana fermions the Hamiltonian 
(\ref{eq:totalhamilt}) can be written as
\begin{equation}
\label{maj-hamiltonian}
H=H_0 +2\tilde{u}\int_0^L {\rm d}x \cos 2\phi(x)
  -2\tilde{V}\int_0^L {\rm d}x \cos 4\phi(x).
\end{equation}   

\section{Phase diagram}
\begin{figure}
\centerline{\psfig{figure=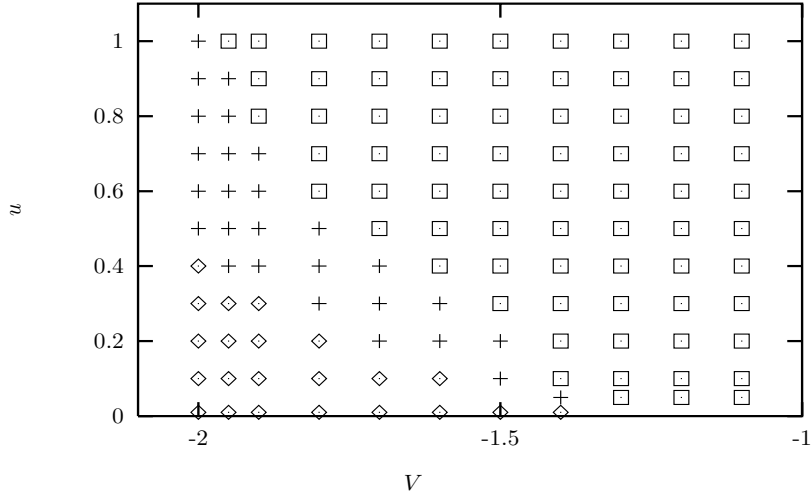,width=11cm,height=7cm}}
\caption{Phase diagram for dimerized spinless fermions as obtained using the DMRG \cite{Cosima98}.
The metallic phase of the system is denoted by $\diamond$. As demonstrated in \cite{Cosima98}
the metallic region increases when including not only a periodic modulation of the 
hopping parameter, but 
also in the interaction (+). The region with a gap  in the excitation spectrum is marked by 
$\boxdot$. }
\label{pd}
\end{figure}

The zero-temperature  phase  diagram 
of spinless fermions with static dimerization
has been investigated both numerically and analytically;
numerically it is often more convenient to study a dimerized spin $1/2$
chain, which however can be mapped onto the  model of interacting
one-dimensional spinless fermions by the  Jordan-Wigner transformation \cite{Jordan}. 
In the $V-u$ diagram one obtains  different phases: a metallic one
where the system behaves as a Luttinger liquid, and an insulating one where a
charge gap exists. 
For $u\rightarrow 0$
the critical point $V/t=-\sqrt{2}$ that separates 
the two phases has been obtained in \cite{Kohmoto81}. For $V/t\in (-\sqrt{2},2) $ the Umklapp scattering 
process is irrelevant and it
is renormalized to zero, while for $V/t>2$  this process becomes relevant and 
 results in the 
opening of a correlation gap. For $u\ne 0$, 
according to perturbative   renormalization group calculations \cite{Zang95} and confirmed
numerically using the  density matrix renormalization group (DMRG) 
\cite {Cosima98}, the phase boundary
in the $V-u$ plane shifts to more negative  values of $V$ with
increasing $u$, as can be seen in Fig.\ \ref{pd}.  

\section{Self-consistent harmonic approximation}
In the following we consider the bosonized form of the model given  in 
Eq.\ (\ref{eq:totalhamilt}). 
To discuss the gap formation, we use the self-consistent
harmonic approximation (SCHA) which has originally been introduced for
the sine-Gordon model  \cite{Coleman75}. 
The idea is to construct a trial Hamiltonian $H_{\textrm{tr}}$ where the nonlinear
terms $\sim {\rm e}^{\pm2i\phi}$ in Eq.\ (\ref{spinless-peierls}) 
are replaced by a quadratic form   
with a parameter $\Delta$ to be determined self-consistently according to
a variational principle for the  energy.

 As we discussed in the preceding section, in the 
thermodynamic limit we can replace the Klein factors by their eigenvalues Eq.\ (\ref{evA}), 
where we
choose the phase factor  $k=\pi/2$. The resulting Hamiltonian is equivalent to 
the Hamiltonian obtained
within the Majorana fermion approach and is given in Eq.\ (\ref{maj-hamiltonian}).

We introduce the trial Hamiltonian
\begin{eqnarray}
  \label{HtrialA}
      H_{\rm{tr}}^\Delta&=&\int_0^L \frac{\rm{d}x}{2\pi}
      \left\{\frac{v}{g}(\partial_x\phi)^2 + vg(\partial_x\theta)^2
      + \frac{\Delta^2}{v g} \phi^2(x)\right\}.
\end{eqnarray}
Since we are only interested in properties at zero temperature we use the
normalized ground state $|\psi_0\rangle$ of $H_{\rm{tr}}$ as a variational
wave function for the Hamiltonian (\ref{eq:totalhamilt}). This provides us
with an upper bound $\tilde E$ for the ground state energy $E$ of (\ref{eq:totalhamilt})
due to the inequality
\begin{equation}
  \label{eq:variational}
    E\le\tilde{E}=\langle \psi_0 |H  |\psi_0\rangle.
\end{equation}
The variational energy is given by
\begin{eqnarray}
  \label{eq:variational}
  \tilde{E}&=&\langle H_0\rangle_{\textrm{tr}}+
              \langle H_{\textrm{Peierls}} \rangle_{\textrm{tr}},\\
  \label{var-en}	      
  \frac{\tilde{E}}{L}&=&\frac{E_{\textrm{tr}}}{L}-\frac{\Delta^2}{2\pi v g}
                         \langle \phi^2\rangle_{\textrm{tr}} 
			 -2\tilde{u}{\rm e}^{-2\langle\phi^2\rangle_{\rm{tr}}},
\end{eqnarray}
where $E_{\textrm{tr}}$ is the ground state energy of $H^{\Delta}_{\rm{tr}}$. 
Minimizing $\tilde E$ with respect to the variational parameter $\Delta$ 
yields the gap equation
\begin{equation}
 \label{an-varpar}
  \Delta^2=  8\pi vg\tilde{u}\ {\rm e}^{-2\langle\phi^2\rangle_{\rm{tr}}}.
\end{equation}
Since $H_{\rm{tr}}^\Delta$ is bilinear in the field operators
it is straightforward to calculate the equal coordinate  correlation
function of the phase field $\phi(x)$ entering Eq.\ (\ref{var-en})
\begin{equation}
   \label{eq:phi2}
      \langle\phi^2\rangle_{\rm{tr}} = \frac{\pi v g}{L}
      \sum_{k>0} \frac{{\rm e}^{-k a}}{\sqrt{v^2 k^2 + \Delta^2}},
\end{equation}
where the $k$-values in the sum are multiples of $2\pi/L$.
Replacing the sum in Eq.\ (\ref{eq:phi2}) by an integral
we obtain (for $\Delta \ll \Delta_0$)
\begin{equation}
 \label{eq:phi2a}
     \langle\phi^2\rangle_{\rm{tr}} = \frac{g}{2} \ln{\frac{\Delta_0}{\Delta}},
\end{equation}
where $\Delta_0 = 2 v {\rm e}^{-\gamma} / a$ and $\gamma=0.5772$ is Euler's constant.
Inserting Eq.\ (\ref{eq:phi2a}) into Eq.\ (\ref{an-varpar}) yields
\begin{equation}
 \label{eq:delta}
    \frac{\Delta}{\Delta_0} = \left(\frac{u}{u_0}\right)^{1/(2-g)},
\end{equation}
with $u_0 = {\rm e}^{-\gamma} \Delta_0 / 4gt$. For $g > 2$ the right hand side
of Eq.\ (\ref{eq:delta}) diverges as $u \rightarrow 0$, and the equation  has only
the trivial solution $\Delta = 0$, i.e.\ the line $g = 2$
marks the transition from a gapless to a gapped phase
in the $g-u$ plane.
In the spinless fermion model
this corresponds to the line $V/t = - \sqrt{2}$ in the $V-u$ phase diagram.
The value $g_c = 2$ obtained within the variational approach is in accordance
with the exact result \cite{Kohmoto81}.
However, according to renormalization group \cite{Zang95} and DMRG calculations 
\cite {Cosima98}, the phase boundary
in the $g-u$ plane shifts to larger values of $g$ with
increasing $u$, while the SCHA gives a vertical line. 
On the other hand,
the exponent $1/(2 - g)$ characterizing the opening of the gap
is  exact \cite{Nakano81}.

For non-interacting fermions $V = 0$,
 the Luttinger  parameter is $g = 1$ and correspondingly  $\Delta
\propto u$, in agreement with the exact energy gap. The prefactor however 
depends on the cut-off procedure used. From Eq.\ (\ref{eq:delta}) we find 
$\Delta=4tu  {\rm e}^{\gamma}$, where the factor ${\rm e}^{\gamma}$ has its origins
in the soft cut-off used in Eq.\ (\ref{eq:phi2}). If we replace in Eq.\ (\ref{eq:phi2})
the soft cut-off by a hard cut-off the result is $\Delta=4tu/\pi$.

At $V/t = 2$, g=1/2,  where Umklapp scattering becomes relevant and leads to the
opening of a correlation gap, Eq.\ (\ref{eq:delta})
yields $\Delta \propto u^{2/3}$ which agrees up to a
logarithmic correction with the exact result
$\Delta_{ex}\propto u^{2/3}/ \sqrt{|\ln u|}$ \cite{Uhrig96}. 
\section{Finite systems}
In the following we consider a  finite system of length $L$. In this case it is not possible to
replace  the Klein factors by their eigenvalues, since the term proportional to the current
operator $J^2$ in the Luttinger Hamiltonian (\ref{spinless-lutt}) 
does not commute with the $F$'s. 
The trial Hamiltonian is now chosen as  a sum of two commuting parts
$ H_{\rm{tr}} =H_{\rm{tr}}^\Delta + H_{\rm{tr}}^B$,
 with  $H_{\rm{tr}}^\Delta$ given in Eq.\ (\ref{HtrialA})
 and the ``Klein Hamiltonian"
\begin{equation}
  \label{eq:HtrialB1}
     H_{\rm{tr}}^B = -iB(F^+_LF_R-F^+_RF_L) + \frac{\pi vg}{2L}J^2,
\end{equation}
where $\Delta$ and $B$ are  variational parameters to be determined
 self-consistently. $H_{\rm{tr}}^B$
is of the form of a tight-binding Hamiltonian  for a particle moving on 
a one-dimensional  lattice in a harmonic potential. A class of  similar
Hamiltonians has been studied in \cite{Schonhammer02}.
In this case we  decouple the Klein factors from the
bosonic fields and 
the minimum condition for the variational energy is equivalent to
the following substitution in the Peierls dimerization term
\begin{equation}
F^+_LF_R\textrm{e}^{2i\phi(x)}\rightarrow \langle F^+_LF_R\rangle \textrm{e}^{2i\phi(x)}+
                            F^+_LF_R \langle  \textrm{e}^{2i\phi(x)} \rangle .
\end{equation}
As a result, instead  of replacing the products of Klein factors by their {\it eigenvalues}
as in the thermodynamic limit we now have to
replace them by their {\it expectation values} with respect to the
ground state of $H_{\rm tr}^{B}$.
According to the Feynman-Hellmann theorem 
\cite{Feynmann,Hellmann}, the expectation value $\langle F^+_LF_R\rangle$  can be expressed in
terms of the derivative of the ground state energy $E_0$ of the Klein Hamiltonian 
$H_{\rm{tr}}^B $
\begin{equation}
\langle F^+_LF_R\rangle =i\frac{\partial E_0}{\partial B}.
\end{equation}
The variational energy obtained from $H_{\rm{tr}} $ has to be
minimized with respect to the parameters $\Delta$ and $B$ and we obtain the
gap equations
\begin{equation}
  \label{eq:varpar}
     \begin{array}{lll}
        \Delta^2 & = & -4\pi vg \tilde{u}E_0'(B) {\rm e}^{-2\langle\phi^2\rangle_{\rm{tr}}},
	    \\[2mm]
        B      & = & \tilde{u}L {\rm e}^{-2\langle\phi^2\rangle_{\rm{tr}}},
     \end{array}
\end{equation}
where $E_0(B)$ is the ground state energy of $H_{\rm{tr}}^B$ and 
$E_0'(B)=\partial E_0(B)/\partial B$.
In order to calculate the ground state energy of
$H_{\rm{tr}}^B$ it is convenient to
switch to the momentum representation, $J \rightarrow
-i {\rm d}/{\rm d}p_J $, where the  Hamiltonian reads
\begin{equation}
  \label{eq:HB}
    H_{\rm{tr}}^B=-\frac{\pi vg}{2L}\frac{d^2}{dp_J^2} +2B \cos(2p_J),
\end{equation}
with periodic boundary conditions for the wave function,
$\Psi(p_J+\pi)=\Psi(p_J)$. At this point, we note a close analogy with the quantum theory of Josephson
junctions, see \cite{tinkham} and appendix \ref{josephson}.
The corresponding  Schr\"odinger equation is the  Mathieu  equation.
Accordingly, the ground state energy of $H_{tr}^B$ has the following
asymptotic behavior \cite{Abramowitz}:
\begin{equation}
  \label{eq:E0}
     E_0(B) \approx
     \left\{
     \begin{array}{cc}
      -\frac{B^2L}{\pi vg}  & \textrm{for } LB \ll v,\\
       -2B+\sqrt{\frac{2B\pi vg}{L}}
       & \textrm{for } LB \gg v.
     \end{array}
   \right.
\end{equation}
Combining the two equations (\ref{eq:varpar}) one obtains
\begin{equation}
  \label{eq:delta2}
    \Delta^2 L^2 = - 4\pi v g E_0'(B) B L,
\end{equation}
which can be used to relate the two limiting cases of Eq.\ (\ref{eq:E0}) with
two different physical situations:
The condition of large system size $L \gg v/B$ is equivalent with 
$\Delta \gg v/L$ and therefore
the results for $\langle\phi^2\rangle_{\rm{tr}}$
and $\Delta(u)$ are the same as for the infinite system.
On the other hand, the condition of small systems, $L \ll v/B$ implies that $\Delta \ll v/L$.
In this case the size dependent energy gap $\Delta (L)$ 
is purely due to the finite system size, i.e.\  it is  of
order $ v/L$. Thus the crossover from a finite-size gap to a true
dimerization gap coincides with the crossover between the regions where Klein
factors are relevant or can be ignored.

\begin{figure}
   \label{fig1}
      \centerline{\includegraphics[width=10.5cm,height=6cm]{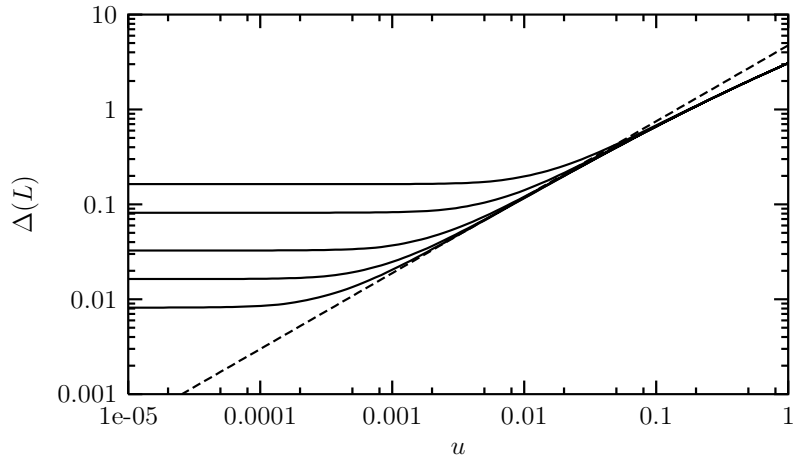}}
      \caption{ Energy gap $\Delta(L)$  (in units of $t$) as function of the dimerization parameter $u$
             for $L=100, 200, 500, 1000, 2000$ (from top to bottom). The dashed line is the analytic
          result of Eq.\ (\ref{eq:delta}) valid for $L \rightarrow \infty$
	   and $\Delta \ll \Delta_0$.}
\end{figure}

In general, the set of equations (\ref{eq:varpar}) can only be solved
numerically. Fig.\ \ref{fig1}
shows the size-dependent energy gap  $\Delta(L) = \sqrt{(2\pi v/L)^2 
+ \Delta^2}$  as 
 function of $u$ for several values of the system size $L$; notice that 
periodic boundary conditions are imposed, where the smallest
wave number is $2\pi /L$.
The Luttinger parameter is $g=3/4$  which corresponds to $V/t=1$.
\newpage
\section{Drude weight}
The Drude weight or charge stiffness is a physical quantity used to
characterize charge transport at zero frequency. At $T = 0$ the real part of the electrical conductivity at
frequency $\omega$ is of the form $\sigma(\omega) = 2\pi D \delta(\omega)
+ \sigma_{reg}(\omega)$ with $\sigma_{reg}(\omega)\rightarrow 0$ for
$\omega\rightarrow 0$ in a system without impurities.
Therefore $D = 0$ characterizes an
insulator while $D > 0$ describes an (ideal) conductor \cite{Kohn64}.
 The Drude weight is not a true order  parameter related to  a
 symmetry breaking, however, it can be used to distinguish  the insulating 
from  the metallic
 phase. 
The simplest way to calculate $D$ is due to
Kohn \cite{Kohn64} who has shown that the Drude weight can be
expressed as
\begin{equation}
 \label{eq:Kohn}
         D = \frac{L}{2} \left.\frac{{\partial}^2 E(\varphi)}
	 {\partial \varphi^2}\right|_{\varphi=0}=
	 \frac{L}{2}
	   \left(\frac{\Phi_0}{2\pi}\right)^2
	   \left.\frac{\partial^2 E(\Phi)}{\partial \Phi^2}\right|_{\Phi=0}, 
\end{equation}
where $E(\varphi)$ is the ground state energy of
a ring of circumference $L$ which is threaded by the magnetic flux 
$\Phi$, where $\varphi=2\pi \Phi/\Phi_0$ and $\Phi_0=h/e$.
In the fermionic model (\ref{eq:dim-term})
the hopping parameter $t$ picks up a phase factor $\exp(\pm i\varphi/L)$
when a flux $\Phi$ is applied.
Alternatively, within a proper gauge transformation,  
the parameter $\varphi$ can also be associated with a change of
boundary conditions, i.e.\ $\varphi = 0$ corresponds to periodic and
$\varphi = \pi$ to anti-periodic boundary conditions. 
The Drude weight is  directly related to the persistent current 
$I(\Phi)$ which flows
in a ring that is penetrated by a magnetic flux  \cite {Eckern96,Eckern02}
$$I(\Phi)=-\frac{\partial E(\Phi)}{\partial \Phi}. $$

\subsection{Drude weight within bosonization}
We now turn to the calculation of the Drude weight $D$ within the bosonization
 formalism.
In the presence of the magnetic flux the only modification
in the  bosonized Hamiltonian of the Luttinger model 
  appears in the current operator 
  $J$ which has to be replaced by
$J + \frac{\varphi}{\pi}$ \cite{Loss92} and the Drude weight is given by the
product of Luttinger parameters $D_0=vg/2\pi$.
Correspondingly, for the system with dimerization,  we modify the $B$-dependent part of the
the trial Hamiltonian and write
\begin{equation}
 \label{eq:HtrialB,phi}
    H_{\rm{tr}}^B(\varphi)=-iB(F^+_L F_R - F^+_R F_L) +
   \frac{\pi v g}{2L} \left({J}+\frac{\varphi}{\pi}\right)^2.
\end{equation}
Applying the same procedure as in the case $\varphi = 0$
yields a variational estimate $\tilde E$ for the ground state energy
which now depends on the flux $\varphi$, i.e.\ $\tilde E =
\tilde E(\Delta,B,\varphi)$ where $\Delta$ and $B$, obtained
from the gap equations (\ref{eq:varpar}), are also
functions of $\varphi$.
On first sight it might seem impossible to calculate the Drude weight using
Eq.\ (\ref{eq:Kohn}) since there is no analytical solution of the
gap equations (\ref{eq:varpar}) even for $\varphi = 0$. However, a closer
look reveals a great deal of simplification. 
From the second derivative
\begin{eqnarray}
   \nonumber
    \frac{{\rm d}^2 \tilde E}{{\rm d}\varphi^2}  & = &
     \frac{\partial^2 \tilde E}{\partial \Delta^2}
     \left(\frac{\partial \Delta}{\partial \varphi}\right)^2
     + \frac{\partial^2 \tilde E}{\partial B^2}
     \left(\frac{\partial B}{\partial \varphi}\right)^2
      + \frac{\partial \tilde E}{\partial \Delta}
       \frac{\partial^2 \Delta}{\partial \varphi^2}\\
    & &
    + \frac{\partial \tilde E}{\partial B}
    \frac{\partial^2 B}{\partial \varphi^2}
     + \frac{\partial^2 \tilde E}{\partial \varphi^2}
    + \makebox{mixed terms}
\end{eqnarray}
one retains only
\begin{equation}
   \left.\frac{{\rm d}^2 \tilde E}{{\rm d}\varphi^2}\right|_{\varphi=0} =
   \left.\frac{\partial^2 \tilde E(\Delta,B,\varphi)}{\partial \varphi^2}\right|_{\varphi=0}
\end{equation}
since {\it i)} $\frac{\partial \tilde E}{\partial \Delta} =
\frac{\partial \tilde E}{\partial B} = 0$ due to the minimum
condition of the energy and {\it ii)}
$\frac{\partial \Delta}{\partial \varphi} = \frac{\partial B}{\partial \varphi} = 0$
at $\varphi=0$ due to symmetry
($\Delta$ and $B$ are even functions of $\varphi)$.
Since the trial Hamiltonian $H_{\rm{tr}} = H_{\rm{tr}}^{\Delta} + H_{\rm{tr}}^B(\varphi)$
consists of two commuting parts we may write
$\tilde E(\Delta,B,\varphi) = \tilde E(\Delta) + E_0(B,\varphi)$
where $\tilde E(\Delta)$ depends only on $\Delta$
and $E_0(B,\varphi)$ is the ground state energy of $H_{\rm{tr}}^B(\varphi)$,
i.e. we obtain the simple result
\begin{equation}
   D = \frac{L}{2}
   \left.\frac{\partial^2 E_0(B,\varphi)}{\partial\varphi^2}\right|_{\varphi=0},
\end{equation}
where $B$ is given by Eq.\ (\ref{eq:varpar}).

To proceed we again represent $H_{\rm{tr}}^B(\varphi)$
in terms of the Mathieu equation (\ref{eq:HB}) where now
the boundary conditions are $\Psi(p_J+\pi)=e^{i\varphi}\Psi(p_J)$.
It is now straightforward to calculate the Drude weight
for a finite system of size $L$ in the gapped phase $g < 2$.
In the ``finite size gap" region ($L \Delta \ll v$), the confining potential of
 the trial Hamiltonian  (\ref{eq:HtrialB,phi}) is the dominant term and the
hopping can be regarded as a small perturbation. In second order perturbation
theory  we obtain
\begin{equation}
 \label{Drude1}
    D = D_0 (1 - \frac{q^2}{2} + \ldots),
\end{equation}
where $ q = 2LB / \pi v g \propto u L^{2-g}$ and $D_0 = vg / 2\pi$ is
the Drude weight of an unperturbed Luttinger liquid.  In the limit
$q\rightarrow 0$ which is equivalent to $u L^{2-g}\rightarrow 0$  the Drude weight 
is equal to the value $D_0$. 
\\
In the opposite limit ($L \Delta \gg v$) which corresponds to
a Mathieu equation with a large cosine potential the variation of the
ground state energy with change of boundary conditions is exponentially small
(as expected from the WKB approximation). A more careful treatment 
(see appendix \ref{mathieu}) yields the exponential dependence
\begin{equation}
  \label{Drude2}
        D \approx D_0 \frac{4 }{\pi} \left(\frac{L \Delta}{vg}\right)^{3/2}
        {\rm e}^{-{2 L \Delta}/{\pi vg}},
\end{equation}
where $\Delta$ and $u$ are related via Eq.\ (\ref{eq:delta}).

For $g>2$ the system remains in the Luttinger liquid phase 
for small dimerization $u$, however with renormalized Luttinger parameters.
These renormalizations  are not obtained within the SCHA and consequently the
Drude weight is not reduced in this region.

\subsection{Drude weight for free spinless fermions }
\label{drudefree}

\begin{figure}[ht]
   \centerline{\psfig{figure=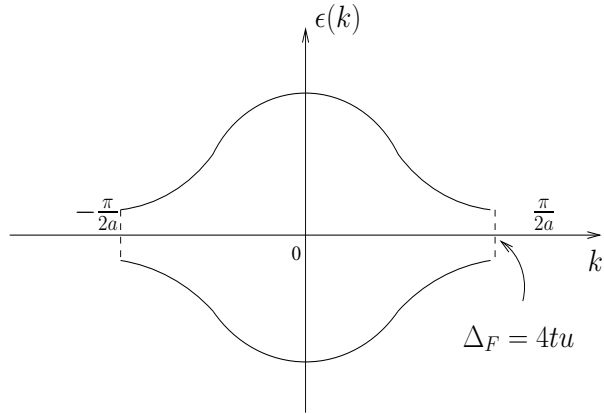,width=8cm,height=5.5cm}}
   \caption{Dispersion of  free fermions with dimerization $u$.}
 \label{band-spliting}
\end{figure}

As a test  case that can be  solved exactly, we consider the Hamiltonian 
(\ref{eq:dim-term}) in the limit $V=0$.
In this case we can calculate the  ground state
energy analytically. Our aim is to compare the exact result
for the Drude weight 
 with the result obtain within the
SCHA  in order to assess the accuracy of  the method. \\
When a flux $\Phi$ is applied, the system is described by the 
Hamiltonian
\begin{equation}
H=-t\sum_{j}(1+u(-1)^j)(c^+_jc_{j+1}\textrm{e}^{i\varphi/L}+{\rm h.c.}),  
\end{equation}
which in momentum representation reads
\begin{equation}
H=-2t\sum\limits_{k} {}^{'}
  (c^+_{k}\;,\quad c^+_{k+\pi})
\left(\begin{array}{cc}
  \cos k & -iu\sin k\\
 iu\sin k &-\cos k
 \end{array}
 \right)
\left(\begin{array}{c}
 c_{k}\\
 c_{k+\pi}
 \end{array}
 \right). 
 \end{equation}
 The prime on the sum  indicates the reduction of the Brillouin
 zone from $[-\frac{\pi}{a},\frac{\pi}{a}]$ to 
 $[-\frac{\pi}{2a},\frac{\pi}{2a}]$.
The corresponding dispersion 
  $\epsilon_{1,\;2} (k)=\pm 2t \sqrt{\cos^2 k+u^2\sin^2 k}$ is shown in
  Fig.\ \ref{band-spliting}.
The ground state energy is 
\begin{equation}
 \label{aenergy}
   E(\varphi) = - 2t\sum_{k} {}^{'} \sqrt{\cos^2 k + u^2 \sin^2 k},
\end{equation}
where for a system of length $L$, $k= (2 \pi n + \varphi)/L$,  and 
the lattice constant is set to one.
In the following  
we assume a 
 dependence of  the ground state energy on $\varphi$  of the form \cite{Nathanson92}
\begin{equation}
E(\varphi)=E_0+E_1 \cos(\varphi),
\end{equation}
The second term, in general, is  an infinite sum of harmonic terms, 
but for a ring with a Peierls distorsion,  threated by a magnetix flux, 
at $T=0$ the main contribution is given by the first
harmonic term \cite{Nathanson92} .
Under this assumption  the Drude weight is given by
\begin{equation}
\label{difD}
D=-\frac{LE_1}{2}=\frac{L}{4}[E(\pi)-E(0)].
\end{equation}

We  calculate now the difference $\Delta E=E(\pi)-E(0)$ which is needed  in the
Drude weight evaluation  (\ref{difD}). We introduce $z_m = {\rm e}^{ik_m}$
and the function $f(z) = 1/(z^L - {\rm e}^{i\varphi})$
which has single poles at $z_m$ with residua ${\rm Res} f(z)|_{z=z_m} 
 = z_m {\rm e}^{-i\varphi} /L$.
Expressing Eq.\ (\ref{aenergy}) in terms of $z_m$ we may replace the sum by
a contour integral and obtain
\begin{equation}
    E(\varphi) = - \frac{tL {\rm e}^{i\varphi}\sqrt{1 - u^2}}{2} \oint_{\cal C} \frac{{\rm d}z}{2\pi i} \;
    \frac{\sqrt{z^2 + z^{-2} + 2\gamma}}{z(z^L - {\rm e}^{-i\varphi})},
\end{equation}
where $\gamma = (1 + u^2)/(1 - u^2)$ and ${\cal C}$ is a contour that encloses
the singularities of $f(z)$.
\begin{figure}
   \centerline{\psfig{figure=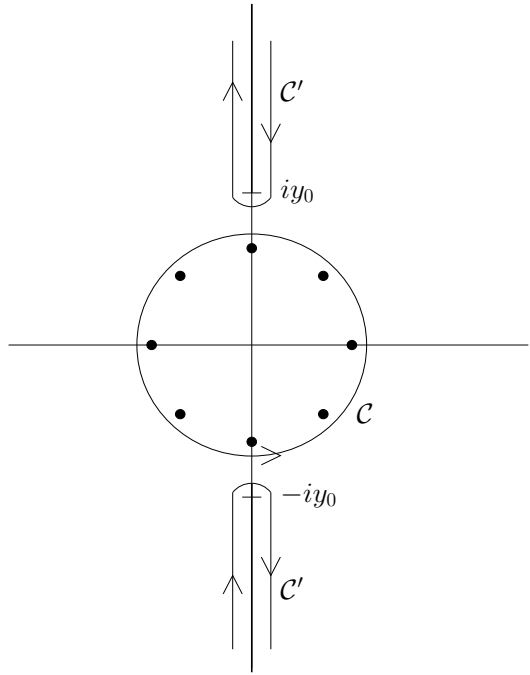,width=7cm,height=9cm}}
   \caption{The contour ${\cal C}$ is deformed into the contour ${\cal C'}$ along
   the branch cuts between  $\pm iy_0$  and $\pm i\infty$.}
 \label{contour}
\end{figure}
 We choose ${\cal C}$ to be composed of two circles
with radii slightly larger or smaller than one, respectively. Substituting
$z \rightarrow 1/z$ the integral along the inner circle can be mapped
onto the integral along the outer circle.
The square root in the numerator has
branch cuts along the imaginary axis in the interval
$[-y_1,y_1]$ and for $|y| > y_0$ where $\pm i y_0$ and $\pm i y_1$ are
the zeroes of the function under the square root.
Now we deform the integration contour along the branch cut 
from $\pm iy_0$ to $\pm i \infty$, as can be seen in Fig.\ \ref{contour}
and obtain for the energy difference $\Delta E = E(\pi) - E(0)$:
\begin{equation}
  \Delta E  = \frac{4tL\sqrt{1 - u^2}}{\pi}
  \int_{y_0}^\infty \; \frac{{\rm d}y}{y}
  \frac{\sqrt{y^2 + y^{-2} - 2\gamma}}{y^{L} - y^{-L}}.
\end{equation}
Substituting $y = {\rm e}^x$ yields
\begin{equation}
  \Delta E = \frac{4tL\sqrt{1 - u^2}}{\pi}  \int_{x_0}^\infty {\rm{d}}x \
   \frac{\sqrt{2\cosh (2x) - 2\gamma}}{{\rm e}^{Lx} -{\rm e}^{-Lx}},
\end{equation}
with $2x_0 = {\rm Arcosh} \gamma = \ln((1+u)/(1-u))$. For $Lx_0 \gg 1$ the integral is rapidly
cut off by the exponential and we may expand the square root around $x = x_0$.
The Drude weight is then
\begin{equation}
   D = \frac{L \Delta E}{4} \approx t \sqrt{\frac{2uL}{\pi}} {\rm e}^{-\frac{L}{2} \ln\frac{1+u}{1-u}}
     \left(1 +  \frac{3(1-u^2)}{8Lu} + \ldots\right).
\end{equation}
Expanding the logarithm for $u \ll 1$,
 and inserting $\Delta_{\textrm{ex}} = 4t u$
and $v_F=2t$ yields 
\begin{equation}
\label{Drudeex}
   D_{\textrm{ex}}\approx D_0 \left(\frac{\pi L\Delta_{\textrm{ex}}}{v_F}\right)^{1/2}
   {\rm e}^{-{L \Delta_{\textrm{ex}}}/{2 v_F}}.
\end{equation}
In the special case of free fermions, the Luttinger  parameters are 
 $g=1$ and $v=v_F=2t$. If we insert these values  in Eq.\ (\ref{Drude2})
 we can  compare the exact result, Eq.\ (\ref{Drudeex}), with the  Drude weight obtained
within  
the SCHA, Eq.\ (\ref{Drude2}).
We see that the exponential behavior $D\sim \exp (-\textrm{const}\cdot L\Delta/v)$ is
reproduced, however,  with a different constant $2/\pi$ instead of $1/2$. This slight
difference is not surprising since the energy scale  $\Delta_0$, see Eq.\ (\ref{eq:delta}),
is known to depend on the cutoff procedure within SCHA.

\chapter{Hubbard model}
\label{hub}
In order to analyze further the importance of Klein factors in bosonized Hamiltonians with
several  nonlinear perturbations,
we consider in the following  a  system of interacting 
spin $1/2$ fermions on a one-dimensional lattice. As in the case of spinless fermions,
we demonstrate how to handle the Klein factors in a systematic way, both in the thermodynamic limit and for finite
systems. 
Within the self-consistent harmonic approximation the decoupling of the Klein factors
from the  bosonic fields 
 results  in a two-dimensional 
tight binding Hamiltonian with a confining potential. While for spinless fermions, the
 Klein Hamiltonian can be mapped onto 
 a  Mathieu equation which can be solved analytically in certain limits, 
 for fermions with spin 
 the problem
becomes more complicated. 
As an application  we consider two different models, 
the Peierls-Hubbard model and the ionic Hubbard model. The first model 
contains  a static alternating Peierls distorsion  $u$
of the lattice,
i.e.\ a modulation of the hopping amplitudes in the kinetic term. 
In the second model,  an alternating on-site energy modulation is
introduced, i.e.\ a staggered potential.
\section{Hubbard model}
The Hubbard model is described by the Hamiltonian 
\begin{equation}
H=-t\sum_{i,\sigma}(c^+_{i\sigma}c_{i+1\sigma}+\textrm{h.c})+
   U\sum_in_{i\uparrow}n_{i\downarrow},
\end{equation} 
where $c^+_{i\sigma}$ creates an electron with spin  direction
$\sigma=\uparrow ,\downarrow$ and  $t$ is the hopping para-meter. $U$ is the 
interaction energy  of 
two electrons at the same site, ($U>0$).\\
In
momentum space the Hamiltonian reads
\begin{equation}
H=\sum_{k\sigma}\epsilon_kc^+_{k\sigma}c_{k\sigma}+
   \frac{U}{N}\sum_{k,p,q}c^+_{k+q\uparrow}c^+_{p-q\downarrow}
       c_{p\downarrow}c_{k\uparrow}
\end{equation}
where $\epsilon_k =-2t\cos(ka)$ and $N$ is the number of lattice sites.
Applying the bosonization  procedure, described in chapter \ref{bosonization}
one obtains
\begin{equation}
\label{hub-hamiltonian}
  H=H_0+H_1+H_2,
\end{equation} 
where $H_0$ is the Luttinger Hamiltonian, 
\begin{equation}
\label{Luttinger}
H_{0}  = \sum_{\alpha=c,s}\int_0^L
    \frac{dx}{2\pi}\left\{\frac{v_\alpha}{g_\alpha}(\partial_x\phi_\alpha)^2 +
      v_\alpha g_\alpha(\partial_x\theta_\alpha)^2 \right\}
      + \frac{\pi}{4L}\sum_{\alpha=c,s}\left\{\frac{v_\alpha}{g_\alpha}N^2_\alpha +
        v_\alpha g_\alpha J^2_\alpha\right\},
\end{equation}
and $H_1$ and $H_2$ are the backscattering and Umklapp contribution, respectively.
Backward and  Umklapp scattering, which represent scatterings with  large
  momentum transfer,  introduce nonlinear terms in the spin and
  the charge sector, respectively. As a consequence   the spin and the charge
  currents are no longer conserved.

\begin{figure}
\centerline{\psfig{figure=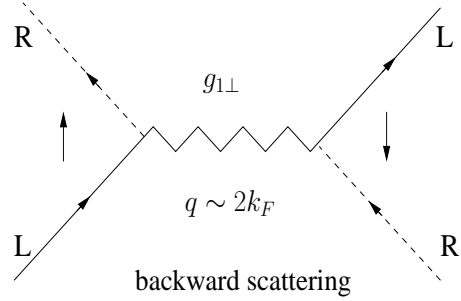,width=6cm,height=4cm}}
\caption{Schematic representation of backward scattering. Solid (dashed) 
lines refer to electrons close to the left (right) Fermi points. $g_{1\perp}$
couples electrons with opposite spin.}
\label{back}
\end{figure}

The Hamiltonian resulting from the backward scattering   
corresponds to an interaction  between left and right  Fermi points  with a
 momentum transfer $q\sim 2k_F$, see Fig.\ \ref{back}.   
In bosonized form the backward  scattering  term reads
\begin{eqnarray}
\label{backscat}
H_1&=&\tilde{U}\int_0^L dx
   \left[F^+_{R\uparrow}F^+_{L\downarrow}F_{R\downarrow}F_{L\uparrow}
   {\rm e}^{-i\frac{2 \pi}{L}N_sx}  {\rm e}^{-i2\sqrt{2}\phi_s}\right.\nonumber\\
    &+&\left. F^+_{L\uparrow}F^+_{R\downarrow}F_{L\downarrow}F_{R\uparrow}
     {\rm e}^{i\frac{2 \pi}{L}N_sx}  {\rm e}^{i2\sqrt{2}\phi_s}\right],
\end{eqnarray} 
with $ \tilde{U}=UL/[(2\pi a)^2N]$. 
   
\begin{figure}
\centerline{\psfig{figure=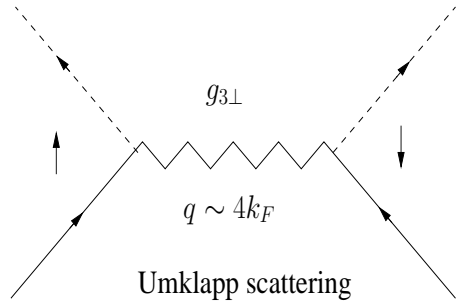,width=6cm,height=4cm}}
\caption{Schematic representation of Umklapp scattering. Solid (dashed) 
lines refer to electrons close to left (right) Fermi points. $g_{3\perp}$
couples electrons with opposite spin.}
\label{umklappsc}
\end{figure}

The Umklapp term $H_2$
 is only important for commensurate band fillings.
 This process comes about because the state at
$k_F$ is equivalent to the state $k_F+nG$, where $n$ is an integer and $G$ is a
reciprocal lattice vector, $G=2\pi/a$, with $a$ being the lattice spacing.
 For the half-filled band case, $k_F=\pi/2a$, and therefore the
 scatterings $k_F \rightarrow k_F +2k_F $ and $k_F \rightarrow -k_F$ are 
 equivalent.    
The Umklapp  process is characterized by  scattering of two particles in the same direction
across the Fermi surface with a momentum transfer $q\sim 4k_F$,
see Fig.\ \ref{umklappsc}. 
\\
The bosonized
form of the Umklapp term is 
\begin{eqnarray}
\label{Umklappscat}
H_2&=&\tilde{U}\int_0^L dx
   \left[F^+_{R\uparrow}F^+_{R\downarrow}F_{L\downarrow}F_{L\uparrow}
    {\rm e}^{-i\frac{2 \pi}{L}N_cx}  {\rm e}^{-i2\sqrt{2}\phi_c}\right.\nonumber\\
   & + &
    \left.F^+_{L\uparrow}F^+_{L\downarrow}F_{R\downarrow}F_{R\uparrow}
     {\rm e}^{i\frac{2 \pi}{L}N_cx}  {\rm e}^{i2\sqrt{2}\phi_c}\right].
\end{eqnarray}  
The Hubbard model is a particular case of the ``g-ology" model given by Eq.\ (\ref{gology}), 
for which the
coupling parameters are $g_{i\perp}=U$ and 
$g_{i\parallel}=0$. From Eqs.\ (\ref{lutt-param-vs})-(\ref{lutt-param-gc})
 the corresponding Luttinger parameters are obtained as
$$v_s=v_F\sqrt{1-\frac{U}{\pi v_F}}, \qquad
  v_c=v_F\sqrt{1+\frac{U}{\pi v_F}} $$ 
$$ g_s=\frac{1}{\sqrt{1-\frac{U}{\pi v_F}}},\qquad 
 g_c=\frac{1}{\sqrt{1+\frac{U}{\pi v_F}}},$$
where $v_F=2ta\sin(k_F a)$ is the Fermi velocity. 
The ground state phase diagram of the ``g-ology" model has been investigated 
in detail using renormalization group techniques \cite{Solyom}.
Schulz \cite{Schulz}  noticed  that in the 
small $U$ 
perturbative regime, 
for all filling factors  the backward scattering 
 renormalizes to the weak-coupling fixed point  $g_{1\perp}^*=0$ and the
$g_s$ parameter scales to $g_s^*=1$. In lowest order of the renormalization group equations the charge
parameter $g_c$ remains unrenormalized and its perturbative value is $g_c=1-U/2\pi v_F$ 
\footnote{There is a  misprint in  \cite{Schulz}, where the perturbative value of
the charge parameter is given by $g_c=1-U/\pi v_F$.}.
Schulz also pointed out that there is no phase transition in the Hubbard model 
between weak and strong coupling, and hence the small and large $U$ limits
belong to the same phase. As a consequence, the equality $g_s=1$ remains valid 
for all repulsive interactions.\\
In the  strong  coupling limit ($U\gg t$), there are no  
low-energy charge excitations involving doubly occupied sites. In a restricted Hilbert space
containing only singly  occupied sites,  in  second order in $t/U$
the Hubbard Hamiltonian can be mapped  onto the $t-J$  model, 
where the  spins are coupled through
an effective antiferromagnetic exchange integral $J=4t^2/U$ \cite{Klein74} 
\begin{equation}
  H_{t-J}=-t\sum\limits_{j} \left[(1-n_{j-\sigma})
  c^+_{j\sigma}c_{j+1\sigma}(1-n_{j+1-\sigma})+{\rm h.c.}\right] +
  J\sum\limits_{j}\left[{\bf  S_{j}}\cdot 
  {\bf S_{j+1}}-\frac{1}{4}n_jn_{j+1} \right].
\end{equation}
At half filling the $t-J$ model reduces to the 
antiferromagnetic Heisenberg model 
\begin{equation}
H=J\sum\limits_{j} {\bf S_{j}}\cdot {\bf S_{j+1}},
\end{equation} 
 which  can be transformed  
 to the spinless-fermion Hamiltonian with nearest neighbor interaction 
using the
Jordan-Wigner transformation \cite{Jordan}, discussed in the previous chapter, for the special 
case $V=2t$, $g=1/2$.

The exact solution of the one-dimensional Hubbard model via  Bethe ansatz 
has been found by Lieb and Wu \cite{Lieb}.
However, the  Bethe ansatz wave function is in many cases  too complicated for the evaluation
of expectation values or correlation functions. Still, the Bethe ansatz is a
convenient starting point to determine the Luttinger parameters for arbitrary
band filling and interaction strength \cite{Schulz}. 

\section{Peierls-Hubbard model}
\label{Peierls}
The coupling of a one-dimensional metal to an elastic lattice results in an instability towards a lattice distortion
known as  Peierls instability \cite{Peierls55}.
A lattice distorsion decreases the electronic energy which overweights
an increase in lattice energy. Due to an opening of a band gap at the Fermi energy the system becomes an insulator.
This phenomenon has been  observed in various
quasi one-dimensional materials, e.g.\ conjugated polymers \cite{Keiss92} like polyacetylene, charge-transfer salts 
\cite{Ishiguro90} and more recently in the one-dimensional spin systems  CuGeO$_3$ \cite{Hase93} and NaV$_2$O$_5$
\cite{Isobe96}.

From a theoretical point of view, a first step towards a quantitative description of the Peierls transition has been
made by Su, Schrieffer and Heeger \cite{Su79}. However, in their model no  electron-electron interactions  are taken into account.

The  effects of electronic correlations on the Peierls transition have been studied using a variety of methods including variational 
wave functions \cite{Horsch81, Baeriswyl87}, Hartree Fock  \cite{Kivelson82}, quantum Monte
Carlo \cite{Hirsch83}, numerical diagonalization of small systems \cite{Mazumdar83,Hayden88,Waas90}, bosonization 
\cite{Sugiura02, Mocanu04-2}, and incremental expansion \cite{Malek98,Malek03}.

In the following we neglect the lattice dynamics and consider a model with static 
lattice distortion  given by the Hamiltonian 
\begin{equation}
\quad \quad H = -t \sum_{j,\sigma} (1 + (-1)^j u)
     (c_{j\sigma}^+ c_{j+1\sigma}^{}
+ c_{j+1\sigma}^+ c_{j\sigma})
+ U \sum_{j} n_{j\uparrow} n_{j\downarrow},
\end{equation}
which is the usual Hubbard Hamiltonian with an additional periodic  
 modulation
of the hopping
described by the dimerization parameter $u$. For the case of a half-filled band,
which we consider here,
 the
lattice modulation is alternating.
In  momentum space, the Peierls contribution  reads 
\begin{equation}
H_{\textrm{Peierls}}=t u \sum_{k,\sigma} ({\rm e}^{ika} c_{k\sigma}^+ c_{k+\pi\sigma}^{} + {\rm h.c.})
\end{equation}
and in bosonized form
\begin{eqnarray}
H_{\textrm{Peierls}}&=&
    \frac{itu}{\pi a}\int_0^L dx F^+_{R\uparrow}F_{L\uparrow}
     {\rm e}^{-i\frac{\pi}{L}(N_c+N_s)x} {\rm e}^{-i\sqrt{2}(\phi_c+\phi_s)}\nonumber\\
&&+\frac{itu}{\pi a}\int_0^L dx F^+_{R\downarrow}F_{L\downarrow}
     {\rm e}^{-i\frac{\pi}{L}(N_c-N_s)x} {\rm e}^{-i\sqrt{2}(\phi_c-\phi_s)}+{\rm h.c.}.
\end{eqnarray}    
At half filling, $N_c=N_s=0$, and using  the notation $\tilde{u}=tu/\pi a$ we obtain  
\begin{equation}
\label{Peierls}
H_{\rm{Peierls}} =
   \tilde u \int_0^L dx \;  \{i F^+_{R\uparrow}F_{L\uparrow}
   {\rm e}^{-i\sqrt{2}(\phi_c+\phi_s)}
   +                                  i F^+_{R\downarrow}F_{L\downarrow}
   {\rm e}^{-i\sqrt{2}(\phi_c-\phi_s)}
   + {\rm h.c.} \}.
   \;\; 
\end{equation}

In the following we focus on the role of the Klein factors.
Due to the conservation of charge and spin all combinations of
Klein factors appearing in the Hamiltonian (\ref{Peierls}) can be expressed
in terms of the operators $A_{\uparrow} = F_{R\uparrow}^+ F_{L\uparrow}$ and
$A_{\downarrow} = F_{R\downarrow}^+ F_{L\downarrow}$ plus their hermitian conjugates.
In particular, the four-fermion terms arising from Umklapp and backscattering
read
\begin{eqnarray}
\label{fourterms}
F^+_{R\uparrow} F^+_{R\downarrow} F_{L\downarrow} F_{L\uparrow}
& = & F^+_{R\uparrow} F_{L\uparrow} F^+_{R\downarrow} F_{L\downarrow}
\;\; = \;\; A_{\uparrow} A_{\downarrow}, \;\;  \\
F^+_{R\uparrow} F^+_{L\downarrow} F_{R\downarrow} F_{L\uparrow}
& = & F^+_{R\uparrow}  F_{L\uparrow} F^+_{L\downarrow} F_{R\downarrow}
\;\; =\;\; A_{\uparrow} A_{\downarrow}^+. \;\; 
\end{eqnarray}
Since the Klein factors are unitary, $F_{\alpha}^+F_{\alpha} = F_{\alpha}F_{\alpha}^+ = 1$,
it is easy to show that
\begin{equation}
[A_{\uparrow},A_{\uparrow}^+] = [A_{\downarrow},A_{\downarrow}^+] = 0,\;\; 
\end{equation}
and we may choose a basis where $A_{\sigma}$ and $A_{\sigma}^+$ are both diagonal.
From $ A_{\uparrow}^+ A_{\uparrow} = A_{\downarrow}^+ A_{\downarrow} = 1$ one concludes that the eigenvalues
of $A_{\sigma}$ are pure phase factors, i.e.
\begin{eqnarray}
\label{eigenvalues}
A_{\uparrow}   |k_{\uparrow},k_{\downarrow}\rangle  
     & = & {\textrm e}^{ik_{\uparrow}}  |k_{\uparrow},k_{\downarrow}\rangle, \quad\quad
A_{\uparrow}^+ |k_{\uparrow},k_{\downarrow}\rangle \;\; = \;\;
    {\textrm e}^{-ik_{\uparrow}} |k_{\uparrow},k_{\downarrow}\rangle, \\
A_{\downarrow}   |k_{\uparrow},k_{\downarrow}\rangle 
     & = & {\textrm e}^{ik_{\downarrow}}      |k_{\uparrow},k_{\downarrow}\rangle, \quad\quad 
A_{\downarrow}^+ |k_{\uparrow},k_{\downarrow}\rangle \;\; = \;\; 
    {\textrm e}^{-ik_{\downarrow}}|k_{\uparrow},k_{\downarrow}\rangle, 
\end{eqnarray}
with $0 \le k_{\sigma} < 2 \pi$.
Although the terms $\sim J_{c,s}^2$ appearing in (\ref{Luttinger}) 
do not commute with the Klein factors
it seems reasonable to neglect them  in the thermodynamic limit $L \rightarrow \infty$. 
We may thus replace the Klein factors in $H_1, H_2$ and $H_{\rm{Peierls}}$ by their eigenvalues,
and obtain
\begin{equation}
\label{Backscattering1}
H_1 = \tilde U \int_0^L dx \;
   \{ {\rm e}^{i(k_{\uparrow}-k_{\downarrow})} {\rm e}^{-i 2\sqrt{2}\phi_s} + {\rm h.c.}\},
\end{equation}
\begin{equation}
\label{Umklapp1}
H_2  = \tilde U \int_0^L dx \;
   \{ {\rm e}^{i(k_{\uparrow}+k_{\downarrow})} {\rm e}^{-i 2\sqrt{2}\phi_c} + {\rm h.c.}\},
\end{equation}
\begin{equation}
\label{Peierls1}
H_{\textrm{Peierls}} =  \tilde u \int_0^L dx \;  \{
   i {\rm e}^{ik_{\uparrow}}{\rm e}^{-i\sqrt{2}(\phi_c+\phi_s)} +
   i {\rm e}^{ik_{\downarrow}}{\rm e}^{-i\sqrt{2}(\phi_c-\phi_s)} + {\rm h.c.}\}. \;\; 
\end{equation}
As a result the Hamiltonian of the dimerized Hubbard model separates into
different sectors of purely bosonic Hamiltonians which are
labeled by $k_{\uparrow}$ and $k_{\downarrow}$. In an 
 approach where the Klein factors are replaced by Majorana
fermions \cite{Schulz00} one obtains only the eigenvalues $\pm i$ for
the two-fermion terms, and $\pm 1$ for the four-fermion terms, 
i.e.\ the continuous  symmetry is lost.
\nop{Shifting the field operators according to $\phi_{c,s} \rightarrow \phi_{c,s}
+ (k_{\uparrow} \pm  k_{\downarrow})/2\sqrt{2}$,
the phase factors can be absorbed into the field operators and
one obtains
\begin{equation}
\label{sine_Gordon}
H = H_0 + 2 \tilde U \int_0^L dx \;
       (\cos 2\sqrt{2}\phi_c + \cos 2\sqrt{2}\phi_s)
       + 4 \tilde u \int_0^L dx \;
        \sin \sqrt{2}\phi_c \cos \sqrt{2}\phi_s \;\; 
\end{equation}
In this sine-Gordon-like Hamiltonian the operator constraint
that $\phi_{c,s}$ has no $q=0$ component (see Eq.\ (\ref{varphiL}-\ref{varphiR}))
has to be replaced by
\begin{equation}
\int_0^L dx \;  \phi_{c,s} = \frac{L}{2\sqrt{2}}(k_{\uparrow} \pm  k_{\downarrow}) \;\; .
\end{equation}}

\subsection{Self-consistent harmonic approximation}

In order to study the opening of charge and spin gaps in the dimerized Hubbard model we use the
self-consistent harmonic approximation (SCHA) in which the exponentials of field operators
appearing in (\ref{Umklapp1})-(\ref{Peierls1}) are replaced by quadratic forms. In the thermodynamic limit 
we replace the Klein factors 
by their eigenvalues 
and introduce the trial Hamiltonian
\begin{equation}
\label{Htrial}
H_{\rm tr}  = \sum_{\alpha=c,s}\int_0^L
    \frac{dx}{2\pi}\left\{\frac{v_\alpha}{g_\alpha}(\partial_x\phi_\alpha)^2 +
      v_\alpha g_\alpha(\partial_x\theta_\alpha)^2 +
      \frac{\Delta_\alpha^2}{v_\alpha g_\alpha} \phi_\alpha^2 \right\},
\end{equation}
which provides us with a variational estimate for the ground state energy
\begin{eqnarray}
\label{varenergy}
\tilde E & = &
      \langle H_0 \rangle_{\rm tr} + \langle H_1 \rangle_{\rm tr}
     +\langle H_2 \rangle_{\rm tr} + \langle H_{\rm{Peierls}} \rangle_{\rm tr}, \\
\frac{\tilde E}{L} & = & \frac{E_{\rm tr}}{L} - \sum_{\alpha=c,s}
\frac{\Delta_\alpha^2}{2\pi v_\alpha g_\alpha} \langle\phi_\alpha^2 \rangle_{\rm tr}
+ e(k_\uparrow, k_\downarrow),
\end{eqnarray}
where $E_{\rm tr}$ is the ground state energy of $H_{\rm tr}$, and
\begin{equation}
\label{energy}
e(k_\uparrow, k_\downarrow) = 2 B_c \cos(k_\uparrow + k_\downarrow) + 
    2 B_s \cos(k_\uparrow - k_\downarrow) - 2 B_{cs} (\sin k_\uparrow + \sin k_\downarrow ),
\end{equation}
with
\begin{eqnarray}
\label{Bdef}
B_c & = & \tilde U \; {\rm e}^{-4 \langle \phi^2_c \rangle_{\rm tr}}, \\
B_s & = & \tilde U \; {\rm e}^{-4 \langle \phi^2_s \rangle_{\rm tr}}, \\
B_{cs}& = & \tilde u \; {\rm e}^{- \langle \phi^2_c \rangle_{\rm tr}}
{\rm e}^{- \langle \phi^2_s \rangle_{\rm tr}}. \;\; 
\end{eqnarray}
Minimizing the variational ground state energy with respect to $\Delta_c$ and $\Delta_s$
yields the gap equations
\begin{eqnarray}
\label{gap-equation1}
  \frac{\Delta_c^2}{2\pi v_c g_c } & = & - 4 B_c \frac{\partial e_0}{\partial B_c}
 - B_{cs} \frac{\partial e_0}{\partial B_{cs}}, \\
\label{gap-equation2}
  \frac{\Delta_s^2}{2\pi v_s g_s } & = & - 4 B_s \frac{\partial e_0}{\partial B_s}
 - B_{cs} \frac{\partial e_0}{\partial B_{cs}},
\end{eqnarray}
where $e_0$ is the minimum of $e(k_\uparrow , k_\downarrow)$ 
with respect to $k_{\uparrow}$ and $k_{\downarrow}$
(see Table\ \ref{Peierls-tabel}).
\begin{table}
\begin{center}
\begin{tabular}{|c|c|c|c|}
\hline
range & $k_{\uparrow}$ & $k_{\downarrow}$ & $e_0(B_c,B_s,B_{cs})$ \\
\hline
 & & & \\
$0 < B_{cs} < 2B_s$ & $\arcsin\frac{B_{cs}}{2B_s}$ & $\pi - \arcsin\frac{B_{cs}}{2B_s}$ &
$- 2 B_c - 2 B_s - \frac{B_{cs}^2}{B_s}$ \\
 & & & \\
\hline
& & & \\
$2B_s < B_{cs}$ & $\frac{\pi}{2}$ & $\frac{\pi}{2}$ & $-4B_{cs} - 2 B_c + 2 B_s$ \\
 & & & \\
\hline
\end{tabular}
\caption{Minimum $e_0$ of $e(k_\uparrow, k_\downarrow) = 2 B_c \cos(k_\uparrow + k_\downarrow) +
2 B_s \cos(k_\uparrow - k_\downarrow) - 2 B_{cs} (\sin k_\uparrow + \sin k_\downarrow )$
which is used in the gap equations (\ref{gap-equation1}) and (\ref{gap-equation2}).
The first line is only relevant for $u=0$ (Hubbard model without dimerization)
whereas the second line is relevant for $u > 0$.}
\label{Peierls-tabel}
\end{center}
\end{table}

The first minimum is only relevant for $u=0$ where  $ B_{cs}=0$, i.e.\ the first line of \\
Table\ \ref{Peierls-tabel} with $\partial e_0/\partial B_c = -2$,
$\partial e_0/\partial B_s = -2$  corresponds to the Hubbard model without dimerization.
In this case the equations for the charge and spin parameters are
\begin{eqnarray}
\label{hc1} 
\frac{\Delta_c^2}{2\pi v_cg_c }&=&8B_c,\\
\label{hs1}   
\frac{\Delta_s^2}{2\pi v_sg_s }&=&8B_s.
\end{eqnarray}
The numerical solution of these equations is displayed  in Fig.\ \ref{Hub1},
where here and in all the following figures both  $\Delta_c$ and $\Delta_s$
are given in units of $t$.

Since $H_{\rm tr}$ is quadratic in the bosonic fields it is straightforward to calculate 
$\langle \phi^2_c \rangle_{\rm tr} $ and $\langle \phi^2_s \rangle_{\rm tr} $ 
analytically. In the thermodynamic limit, where the sums are
replaced by integrals  one obtains
\begin{eqnarray}
\label{phi2c}
\langle \phi^2_c \rangle_{\rm tr} & = & \frac{g_c}{2} \ln\frac{\Delta_{0c}}{\Delta_c}, \\
\label{phi2s}
\langle \phi^2_s \rangle_{\rm tr} & = & \frac{g_s}{2} \ln\frac{\Delta_{0s}}{\Delta_s},
\end{eqnarray}
where $\Delta_{0c,0s}$ are cutoff-dependent energy scales of the order of the bandwidth,
and $\Delta_{c,s}<\Delta_{0c,0s}$ is assumed.
The  solution of the gap equations (\ref{hc1}) and (\ref{hs1}) is  
\begin{eqnarray}
\Delta_c&=&\Delta_{c0}\left(\frac{16\pi v_cg_c\tilde{U}}{\Delta_{c0}^2}\right)^{1/(2-2g_c)},\\
\label{hubbardspin}
\Delta_s & =&0.
\end{eqnarray}
Using the perturbative result $g_c=1-U/2\pi v_F$ and $v_F=2t$, we obtain 
\begin{equation}
\Delta_c \propto \textrm{e}^{-2\pi t\ln (t/U)/U}
\end{equation}
in the small $U$ limit.
This should be compared with the  exact solution from Bethe ansatz where
  for small $U$ the charge gap is given by
\begin{equation}
\Delta_c^{\textrm{B.a.}}=\frac{8U}{\pi t}\textrm{e}^{-2\pi t/U}.
\end{equation}
The SCHA result deviates from the exact solution by the factor
$\ln (t/U)$ in the exponent.
We   show the charge gap obtained within the SCHA in comparison with 
 the  exact solution from the Bethe ansatz 
in Fig.\ \ref{Hub2}.
\begin{figure}
\centerline{\includegraphics[width=10.0cm,height=7.0cm]{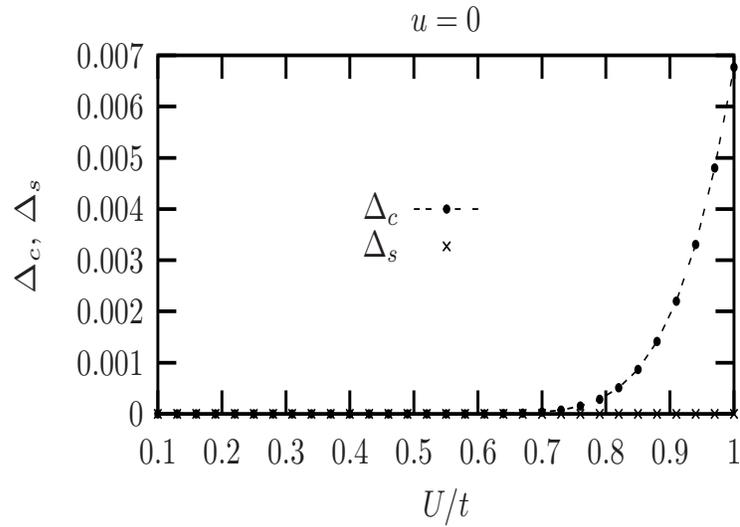}}
\caption{\label{figU}The charge gap  $\Delta_c$ (in units of $t$), 
    obtained within SCHA as function of the Hubbard $U$ for $u=0$.  
    In this case $\Delta_s=0$.  }
\label{Hub1}   
\end{figure}
\begin{figure}
\centerline{\includegraphics[width=10.0cm,height=7.0cm]{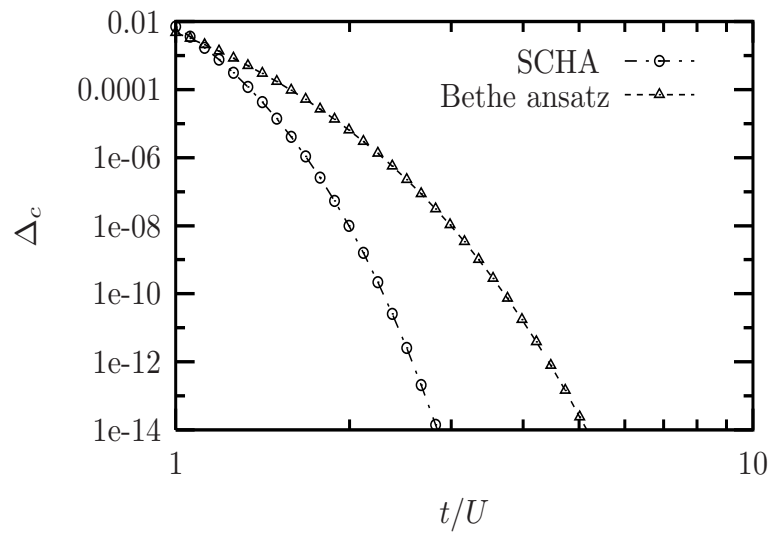}}
\caption{\label{figU}The gap parameter $\Delta_c$  (in units of $t$)
   as function of $t/U$: the  SCHA  solution 
    $\Delta_c^{\textrm{SCHA}}\sim \textrm{e}^{-2\pi t\ln (t/U)/U}$, and the 
   Bethe ansatz solution $\Delta_c^{\textrm{B.a.}}\sim \textrm{e}^{-2\pi t/U}$, both
   valid in the limit $U/t\ll 1$ .}
\label{Hub2}   
\end{figure}

\begin{figure}
\centerline{\includegraphics[width=10.0cm,height=7cm]{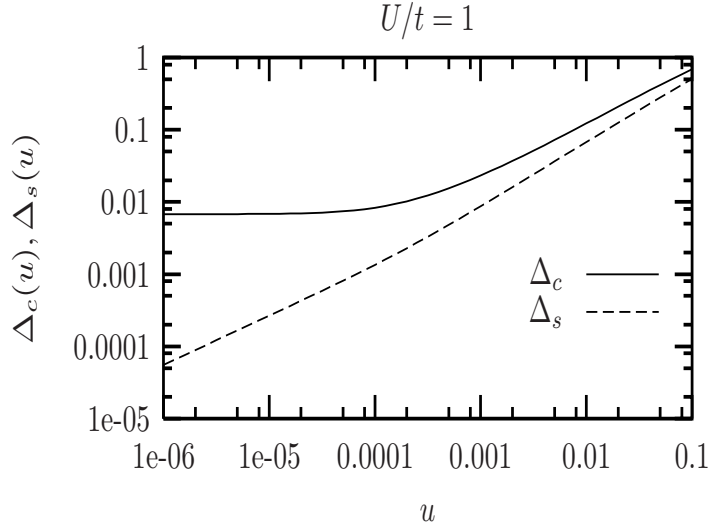}}
\caption{Charge gap $\Delta_c$ and spin gap $\Delta_s$ (in units of $t$) of the
dimerized Hubbard model for $U/t=1$ obtained within SCHA. We have used the value $g_s = 1$
and the expressions $g_c = 1/\sqrt{1 + U/(\pi v_F)}$,
$v_{c,s} = v_F \sqrt{1 \pm U/(\pi v_F)}$.}
\label{figU=1}   
\end{figure}

For non-zero dimerization $u > 0$
a solution of the gap equations exists only for $B_{cs} > 2B_s$, i.e.\ the
second line of Table \ref{Peierls-tabel} has to be used,  with $\partial e_0/\partial B_c = -2$,
 $\partial e_0/\partial B_s = 2$
and $\partial e_0/\partial B_{cs} = -4$. The corresponding gap equations are
\begin{equation}
\label{scha-eq1}
\frac{\Delta_c^2}{2\pi v_cg_c }=4B_{cs}+8B_c=
          4\tilde u \; {\rm e}^{- \langle \phi^2_c \rangle_{\rm tr}- \langle \phi^2_s \rangle_{\rm tr}}+
	  8\tilde U \; {\rm e}^{-4 \langle \phi^2_c \rangle_{\rm tr}},
\end{equation}
\begin{equation}
\label{scha-eq2}	 	 
\frac{\Delta_s^2}{2\pi v_sg_s }=4B_{cs}-8B_s=
          4\tilde u \; {\rm e}^{- \langle \phi^2_c \rangle_{\rm tr}- \langle \phi^2_s \rangle_{\rm tr}}
	  -8\tilde U \; {\rm e}^{-4 \langle \phi^2_s \rangle_{\rm tr}}.
\end{equation}
The  numerical solution of these  equations is  represented  in Fig.\ \ref{figU=1} 
for $U/t=1$. For very small dimerization $u$ the charge gap $\Delta_c$ approaches a
constant while the spin  gap goes to zero, as expected for the Hubbard model without
dimerization. On the other hand, for strong dimerization $\Delta_c$ and $\Delta_s$
approach each other as expected for a band insulator without correlations. 
The transition
between the two regimes is smooth and continuous.

\begin{figure}
\centerline{\includegraphics[width=10cm,height=6.5cm]{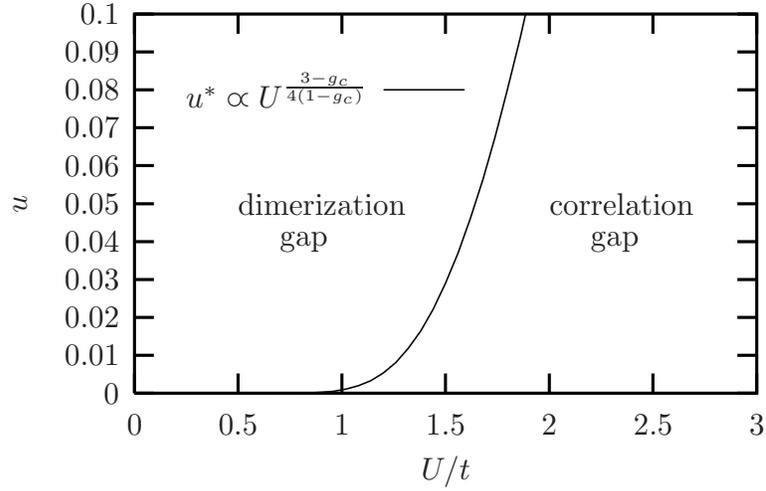}}
\caption{Phase diagram of the dimerized Hubbard model. $u^*$ marks the 
crossover
 from a dimerization gap to a
correlation gap.}
\label{crossover} 
\end{figure}

\begin{figure}
\centerline{\includegraphics[width=11cm,height=6.5cm]{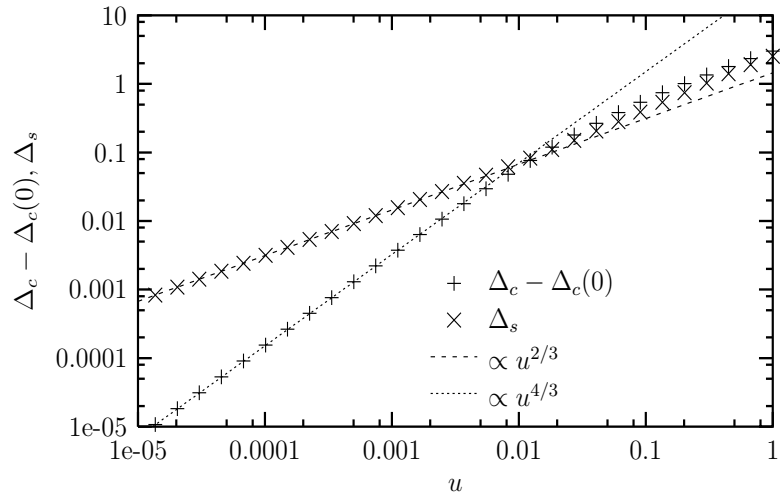}}
\caption{ Charge gap $\Delta_c(u)-\Delta_c(0)$ and spin gap $\Delta_s(u)$ 
(in units of $t$)
of the dimerized Hubbard model for $U/t = 1$ obtained within the SCHA.
The straight lines are the analytic results of Eqs.\ (\ref{chargegap}) and (\ref{spingap})
valid for $u < u^*$. Note that $\Delta_c(0)$ is subtracted from
the charge gap in  order to highlight the power law behavior.}
\label{anPeierls}
\end{figure}

For small $U$ these results can be verified analytically. 
In order to solve the gap equations analytically we consider the case $U > 0$
where $g_s = 1$ and $g_c < 1$, and restrict ourselves to the limit of small dimerization $u$.
Using the analytical results for $\langle \phi^2_c \rangle_{\rm tr} $ and
 $\langle \phi^2_s \rangle_{\rm tr}$,
given by Eqs.\ (\ref{phi2c})-(\ref{phi2s}), one obtains 
 that for  $u \rightarrow 0$ the spin gap vanishes while the charge gap approaches
a constant according to
\begin{eqnarray}
\label{chargegap}
\Delta_c(u) - \Delta_c(0) & \propto & u^{4/3}, \\
\label{spingap}
\Delta_s(u) &\propto & u^{2/3},
\end{eqnarray}
with cutoff-dependent prefactors (for details see appendix \ref{an-solution}).
 The exponent $2/3$ that characterizes the
opening of the spin gap is in accordance with
the corresponding exponent of the dimerized antiferromagnetic
Heisenberg chain up to a logarithmic correction in the prefactor \cite{Uhrig96}.
Since the Heisenberg model corresponds to the $U \rightarrow \infty$ limit of
the half-filled Hubbard model, this indicates that 
the SCHA result (\ref{spingap})
is exact and persists even in the strong-coupling
regime $U/t \gg 1$.
For $u > u^*$ the behavior of the gaps is changed to 
\cite{Schuster99}
\begin{eqnarray}
\label{gap1}
\Delta_c(u) \approx \Delta_s(u) & \propto & u^{2/(3-g_c)},
\end{eqnarray}
where the crossover value
$u^*$ is defined by $\Delta_s(u^*) = \Delta_c(0)$. The $u-U$ phase diagram
is shown in Fig.\ \ref{crossover}.
In Fig.\ \ref{anPeierls} we show  $\Delta_c(u) - \Delta_c(0)$ and $\Delta_s(u)$
as function of $u$ for $U/t = 1$ as obtained from the
numerical solution of the gap equations.
For comparison, the analytical results
of Eqs.\ (\ref{chargegap}) and (\ref{spingap}) are also shown.

\clearpage

\subsection{Finite systems}
\label{FinPeierls}
As in the case of spinless fermions,
for a finite system  it is not  possible to simply
replace the Klein factors by their eigenvalues
since the terms proportional to the spin and charge current,
 $J_c$ and $J_s$, in the Luttinger Hamiltonian (\ref{Luttinger})
do not commute with the $F$'s.
One may, however, decouple the Klein factors from the bosonic fields using a variational
ansatz. We introduce  the Klein Hamiltonian
\begin{eqnarray}
\label{HKleinP}
H_{\rm tr}^{B_{cs}B_cB_s}  & = & i B_{cs}L(F_{R\uparrow}^+ F_{L\uparrow} + F_{R\downarrow}^+
 F_{L\downarrow})
+ B_c LF_{R\uparrow}^+F_{R\downarrow}^+F_{L\downarrow}F_{L\uparrow} \nonumber \\
 &   &  + B_sL F_{R\uparrow}^+F_{L\downarrow}^+F_{R\downarrow}F_{L\uparrow}  +{\rm h.c.}
 \nonumber \\
&   & +\frac{\pi}{4L}(v_c g_c J^2_c + v_s g_s J^2_s),
\end{eqnarray}
where $B_c, B_s$ and $B_{cs}$  are now variational parameters
to be determined self-consistently.
$H_{\rm tr}^{B_{cs}B_cB_s}$ is of the form of a tight-binding Hamiltonian
for a particle moving on a $2d$ lattice in a harmonic potential.
The explicit representation of this Hamiltonian in the $|J_c,J_s\rangle$
is given in appendix \ref{KH-tbm}.
The minimum condition for the variational energy is equivalent to
the following substitution in the backward and Umklapp terms 

\begin{equation}
\label{meanfield}
F^+_{\alpha}F^+_{\beta}F_{\gamma}F_{\rho} 
  {\rm e}^{\pm i k_a\phi_a}
\rightarrow
\langle F^+_{\alpha}F^+_{\beta}F_{\gamma}F_{\rho} 
    \rangle {\rm e}^{\pm i 2\sqrt{2}\phi_a}
+ F^+_{\alpha}F^+_{\beta}F_{\gamma}F_{\rho}  
 \langle {\rm e}^{\pm i 2\sqrt{2}\phi_a}\rangle
\end{equation}
and to the substitution 
\begin{equation}
\label{meanfield}
F^+_{\alpha}F_{\beta}
   {\rm e}^{\pm i \sqrt{2}\phi_a \mp i \sqrt{2}\phi_b}
\rightarrow
\langle F^+_{\alpha}F_{\beta}
    \rangle {\rm e}^{\pm i \sqrt{2}\phi_a \mp i \sqrt{2}\phi_b}
+ F^+_{\alpha}F_{\beta}
 \langle {\rm e}^{\pm i \sqrt{2}\phi_a \mp i \sqrt{2}\phi_b}\rangle
\end{equation}
in the Peierls term, where 
$\alpha,\beta,\gamma,\rho \in\{R\uparrow,L\uparrow,R\downarrow,L\downarrow \}$ and 
 $a,b\in\{c,s\}$.
As a result, instead of
replacing the products of Klein factors by their eigenvalues
as in the thermodynamic limit we now have to
replace them by their  expectation values with respect to the
ground state of $H_{\rm tr}^{B_{cs}B_cB_s}$ divided by the system size.  
The details of the calculations are given
in appendix \ref{finSCHA}. 
The Klein Hamiltonian (\ref{HKleinP}) 
can only be diagonalized numerically.
In the bosonic sector one ends up with
a sine-Gordon type model like in the thermodynamic limit.
In order to evaluate the mean field
parameter, 
in the framework of the SCHA,
the quantity
$e_0$ which enters the gap equations (and which is
explicitly given in Table\ \ref{Peierls-tabel}) is replaced  by  $e^0_{\textrm{tr}}$, the ground state energy of 
$H_{\rm tr}^{B_{cs}B_cB_s}$ divided by $L$ (see appendix \ref{finSCHA}).
Apart from this modification Eqs.\ (\ref{gap-equation1})-(\ref{gap-equation2}) 
remain unchanged.
 
\begin{figure}
\centerline{\includegraphics[width=12.0cm,height=9.0cm]{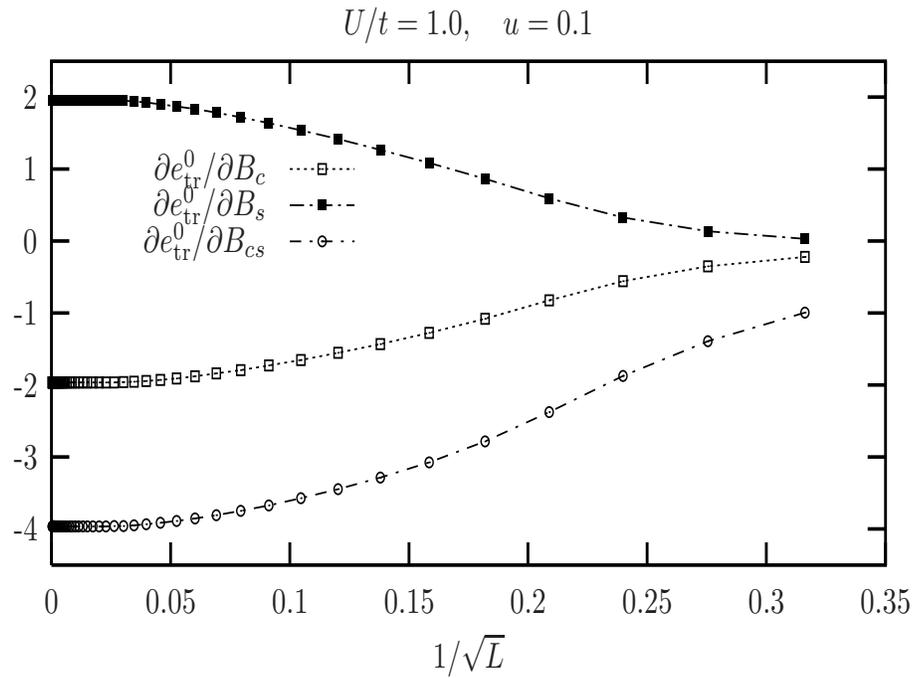}}
\caption{The derivative of $e^0_{\textrm{tr}}$ with respect to the variational parameters 
         $B_c, B_s, B_{cs}$ needed to find the solution of equation
	  (\ref{gap-equation1}), (\ref{gap-equation2}), for $u\ne 0$.
	  With  increasing
	  $L$ they converge to the values corresponding to the thermodynamic 
	  limit given in 
	 Table\ \ref{Peierls-tabel}: $ \partial e^0_{\textrm{tr}}/\partial B_c=2,\; 
	     \partial e^0_{\textrm{tr}}/\partial B_s=-2, 
	  \; \partial e^0_{\textrm{tr}}/\partial B_{cs}=-4$. }
\label{fpklein}
\end{figure}

We consider a system with both spin and charge gap, i.e.\ $u\ne 0$. For
large systems the parameters $B_{cs}, B_c$ and $B_s$ become size-independent,
with the consequence that the kinetic energy in $H_{\rm tr}^{B_{cs}B_cB_s}$
dominates the  confining potential. While approaching  the thermodynamic limit, the
 $e^0_{\textrm{tr}}$, is given by the minimum of
$e(k_{\uparrow},k_{\downarrow})$, see Eq.\ (\ref{energy}),  and
the expectation values of the Klein factors 
converge to the corresponding eigenvalues as can
be seen in Fig.\ \ref{fpklein}.

\begin{figure}
\centerline{\includegraphics[width=13cm,height=10cm]{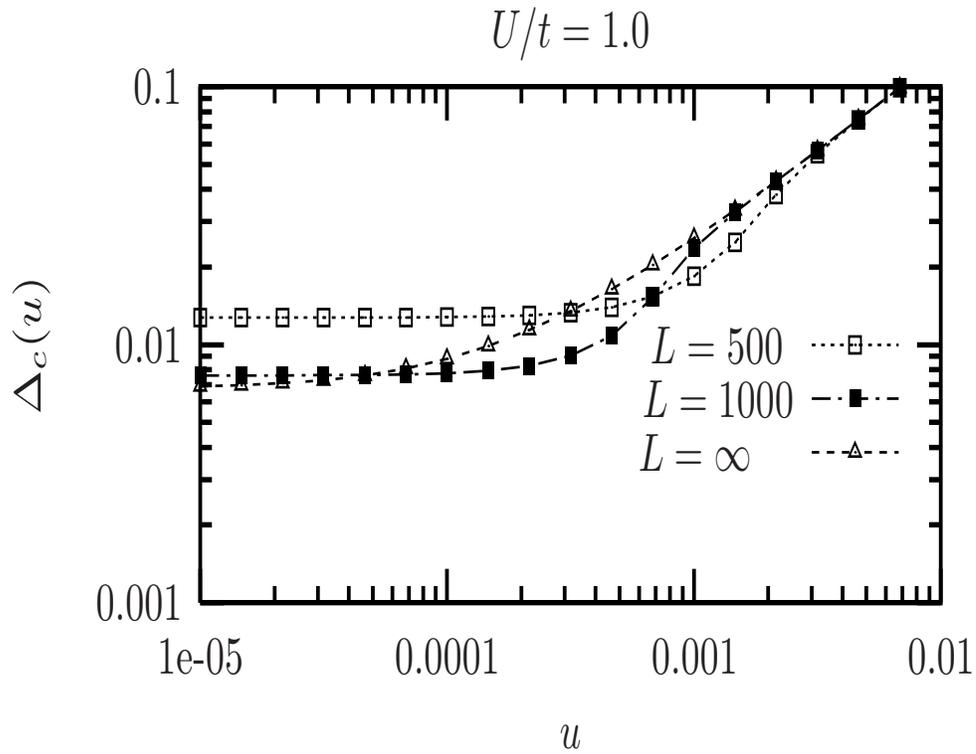}}
\caption{The  gap  parameter $\Delta_c$ (in units of $t$)  as a function of the  dimerization $u$  
for different 
 system sizes compared with the thermodynamic limit.}
\label{fpcharge}
\end{figure}

\begin{figure}
\centerline{\includegraphics[width=13cm,height=10cm]{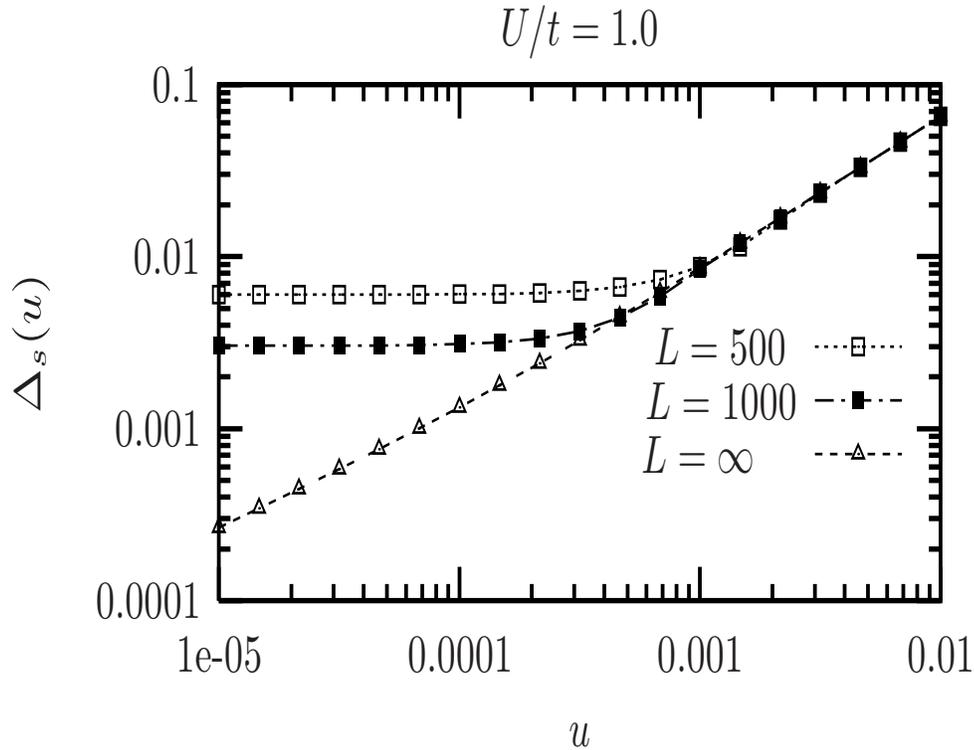}}
\caption{The gap parameter $\Delta_s$  (in units of $t$) as a function of the dimerization $u$ 
for different 
 systems sizes compared with the thermodynamic limit; $\Delta_s$  has to be compared to the finite 
 size energy $v_F/L$, which is e.g.\ $v_F/L=2\cdot 10^{-3}t$ for $L=1000$.}
\label{fpspin}
\end{figure}

In Figs.\ \ref{fpcharge} and  \ref{fpspin} we show the dependence
of the gap  parameters $\Delta_c$ and $\Delta_s$ on the system size.
Finite size effects
are most pronounced in the small $u$ region that corresponds to the Mott insulating
phase. In particular the spin gap parameter $\Delta_s$ approaches a finite value for all
finite systems for $u\rightarrow 0$, while it goes to zero in the thermodynamic limit.
\section{Ionic Hubbard model}
\label{Ionic}
The Hubbard model with an additional  alternating on-site
energy modulation  has been used to  study
 the organic mixed-stack charge transfer crystals with alternating donor  and acceptor molecules
 \cite{Hubbard81,Nagaosa86,LeCointe95,Caprara00}
or to understand the ferroelectric transition in perovskite materials such as BaTiO$_3$ \cite{Egami93}
or KNbO$_3$\cite{Neumann92}. 
The model is described  by the
 Hamiltonian
\begin{equation}
\label{ionicHam}
H=-t\sum_{j,\sigma}(c^+_{j\sigma}c_{j+1\sigma}+\textrm{h.c.})+
   U\sum_j n_{j\uparrow}n_{j\downarrow}
     +\Delta\sum_{j,\sigma}(-1)^jc^+_{j\sigma}c_{j\sigma},
\end{equation} 
where $\Delta$ is the amplitude of the staggered potential.
At $U=0$ and for $\Delta >0$, 
the model corresponds to  a conventional band insulator with a band gap $2\Delta$. The one-dimensional
half-filled Hubbard model without an alternating potential ($\Delta =0$) and with $U>0$ describes a
correlated insulator with vanishing spin gap.
 
In the atomic limit, $t=0$, and for  $U < 2\Delta$, every second site of the lattice with on-site energy $-\Delta$ is
occupied by two electrons while the sites with energy $\Delta$ are empty. The energy difference between the
ground state and the highly degenerate first excited state is $2\Delta-U$. For $U > 2\Delta$, each site is
occupied by one electron and the energy gap is $U-2\Delta$. Within  this 
simplified picture one obtains a critical point
$U_c(\Delta)=2\Delta$, for which the excitation gap disappears.
These  considerations suggest that the system will be in two
quantitatively different phases in the limits $U\ll\Delta$ and $U\gg\Delta$, with a quantum phase transition
from a band insulator to a correlated insulator in between.

The early numerical \cite{Soos78,Resta95} and analytical \cite{Strebel70,Ortiz96} 
results reported that at $T=0$ and for fixed $\Delta$ a single phase transition  
occurs if $U$ is varied.
Using bosonization Fabrizio et al.\ \cite{Fabrizio99} predicted  a $two-transition \; scenario$ 
for the ground state phase diagram. For $U<U_{c1}$ the system is a band insulator with 
finite charge 
and spin  gaps. The first critical point $U_{c1}$ is an Ising critical point, where the charge gap
vanishes.  The intermediate phase, for $U_{c1}<U<U_{c2}$, is a spontaneously  dimerized insulator phase, in which the
bosonic spin and charge gaps are finite. This phase is characterized by a non-zero expectation value of the dimerization operator 
\begin{equation}
D=\sum_{i,\alpha}(-1)^{i}(c^+_{i\sigma}c_{i+1\sigma}+\rm{h.c.}).
\end{equation}
The second transition, at  $U=U_{c2}$,  is of the Kosterlitz-Thouless type,
and the system goes over into a  correlated insulator phase with a finite charge gap and a
vanishing spin gap. Various attempts based on numerical tools have been made  
to verify this scenario, without leading to a definite conclusion. In particular, exact diagonalization \cite{Gidopoulos00,Torio01}, valence bond
techniques \cite{Anusooya01}, quantum Monte Carlo \cite{Wilkens01} and the DMRG 
\cite{Takanada01, Manmana03, Kampf03} were used with different results in favor of one or two critical
points.

In momentum space the  Hamiltonian  (\ref{ionicHam}) reads 
\begin{equation}
H=\sum_{k,\sigma}\epsilon_kc^+_{k\sigma}c_{k\sigma}+
   \frac{U}{N}\sum_{k,p,q}c^+_{k+q\uparrow}c^+_{p-q\downarrow}
       c_{p\downarrow}c_{k\uparrow}+H_{\textrm{ionic}},
\end{equation}
where $H_{\textrm{ionic}}$ is   the contribution due to  the alternating on-site
energy modulation 
\begin{equation}
H_{\textrm{ionic}}=\Delta \sum_{k\sigma} ( c_{k\sigma}^+ c_{k+\pi\sigma} + {\rm h.c.}),
\end{equation}
which is referred to as  the  ``ionic" term.
In bosonized form it reads
$$H_{\textrm{ionic}}=
    \frac{\Delta}{2\pi a}\int_0^L dx F^+_{R\uparrow}F_{L\uparrow}
    \textrm{e}^{-i\frac{\pi}{L}(N_c+N_s)x}\textrm{e}^{-i\sqrt{2}(\phi_c+\phi_s)}$$ 
$$+\frac{\Delta}{2\pi a}\int_0^L dx F^+_{R\downarrow}F_{L\downarrow}
    \textrm{e}^{-i\frac{\pi}{L}(N_c-N_s)x}
    \textrm{e}^{-i\sqrt{2}(\phi_c-\phi_s)}+\rm{h.c.}.$$
At half filling, $N_c=N_s=0$,  the ionic contribution is
\begin{equation}
\label{Ionic}
H_{\rm{ionic}} =
   \tilde \Delta \int_0^L dx \;  \{ F^+_{R\uparrow}F_{L\uparrow}
   {\rm e}^{-i\sqrt{2}(\phi_c+\phi_s)}
   +                                   F^+_{R\downarrow}F_{L\downarrow}
   {\rm e}^{-i\sqrt{2}(\phi_c-\phi_s)}
   + {\rm  \rm {h.c.} } \},
   \;\; 
\end{equation}
where $\tilde{\Delta}=\Delta/2\pi a$.
Similar to the case of the Peierls-Hubbard model, we express the Klein factors 
 in terms of the operators  
$A_{\uparrow} = F_{R\uparrow}^+ F_{L\uparrow}$ and
$A_{\downarrow} = F_{R\downarrow}^+ F_{L\downarrow}$. 
We choose a basis in which these operators are diagonal, 
and using
Eqs.\ (\ref{fourterms}) and (\ref{eigenvalues}),
 we replace the Klein factors in $H_1, H_2$ and  $H_{\rm{ionic}}$ by their eigenvalues
 to  obtain
\begin{equation}
\label{Backscattering2}
H_1 = \tilde U \int_0^L dx \;
   \{ {\rm e}^{i(k_{\uparrow}-k_{\downarrow})} {\rm e}^{-i 2\sqrt{2}\phi_s} + {\rm h.c.}\},
\end{equation}
\begin{equation}
\label{Umklapp2}
H_2  = \tilde U \int_0^L dx \;
   \{ {\rm e}^{i(k_{\uparrow}+k_{\downarrow})} {\rm e}^{-i 2\sqrt{2}\phi_c} + {\rm h.c.}\},
\end{equation}
\begin{equation}
\label{Ionic1}
H_{\textrm{ionic}} =  \tilde \Delta \int_0^L dx \;  \{
    {\rm e}^{ik_{\uparrow}}{\rm e}^{-i\sqrt{2}(\phi_c+\phi_s)} +
    {\rm e}^{ik_{\downarrow}}{\rm e}^{-i\sqrt{2}(\phi_c-\phi_s)} + {\rm h.c.}\}. \;\; 
\end{equation}
As in the case of the dimerized Hubbard model, the Hamiltonian of the ionic 
Hubbard model separates  into different
sectors of purely bosonic Hamiltonians, labeled by $k_{\uparrow}$ and $k_{\downarrow}$.
Using  Majorana
fermions \cite{Schulz00} to represent the Klein factors, only
one sector can be obtained and again the continuity is lost.
\nop{ Shifting the field operators according 
 to $\phi_{c,s} \rightarrow \phi_{c,s}
+ (k_{\uparrow} \pm  k_{\downarrow})/2\sqrt{2}$,
the phase factors can be absorbed into the field operators and
one obtains
\begin{equation}
\label{sine-Gordon2}
H = H_0 + 2 \tilde U \int_0^L dx \;
       (\cos 2\sqrt{2}\phi_c + \cos 2\sqrt{2}\phi_s)
       + 4 \tilde \Delta \int_0^L dx \;
        \cos \sqrt{2}\phi_c \cos \sqrt{2}\phi_s \;\; .
\end{equation}}
\subsection{Self-consistent harmonic approximation}

\begin{figure}
\centerline{\includegraphics[width=11.5cm,height=8.5cm]{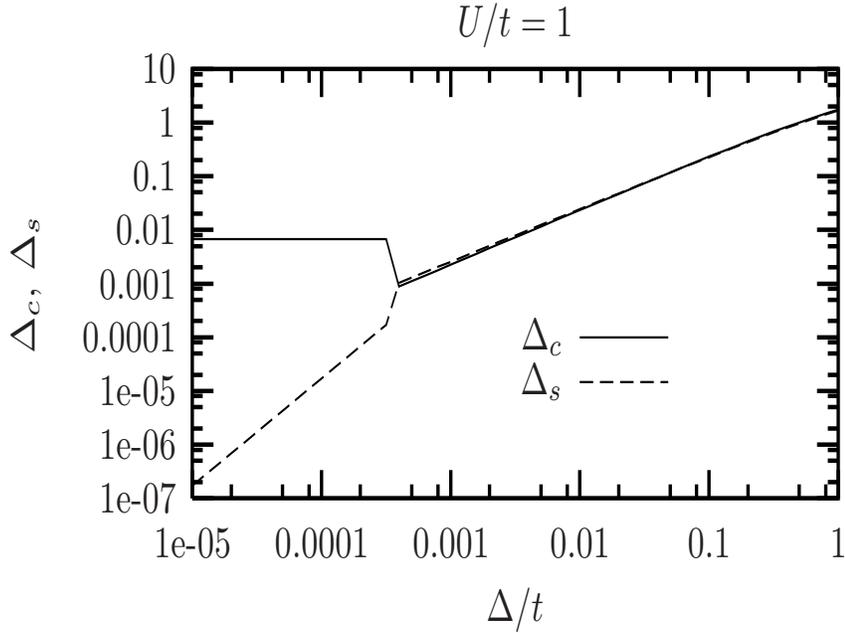}}
\caption{The mean field parameters $\Delta_c$ and $\Delta_s$ (in units of $t$) 
   as function of $\Delta$, the staggered potential, in the  thermodynamic limit. At $\Delta\sim 5\cdot 10^{-4}t$ we find a discontinous
   jump of the parameters.}
\label{an-ionU1}
\end{figure}

\begin{figure}
\centerline{\includegraphics[width=11.5cm,height=8.5cm]{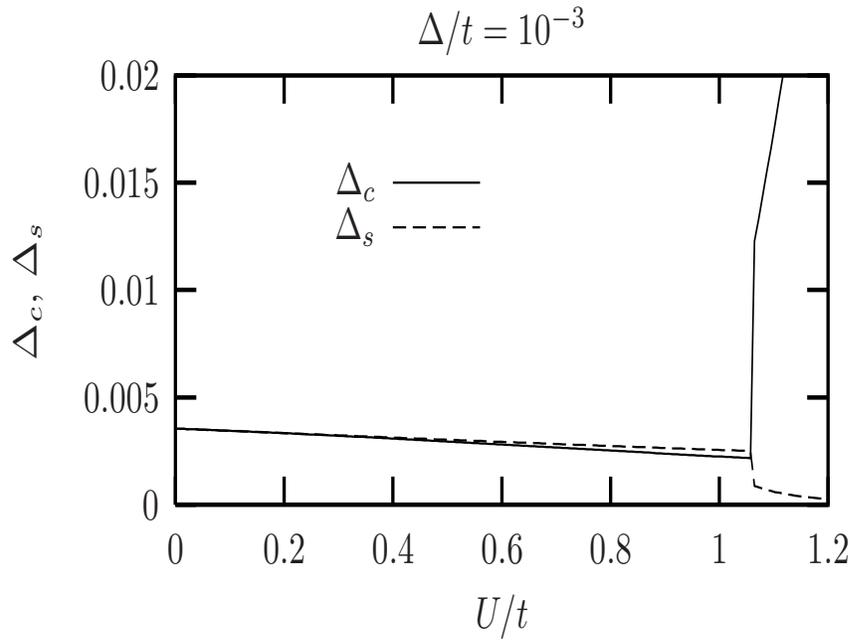}}
\caption{The mean field parameters $\Delta_c$ and $\Delta_s$ (in units of $t$), 
   as function of $U$,   in the thermodynamic limit. }
\label{an-iond1}
\end{figure}

\begin{figure}
\centerline{\includegraphics[width=11.5cm,height=8.5cm]{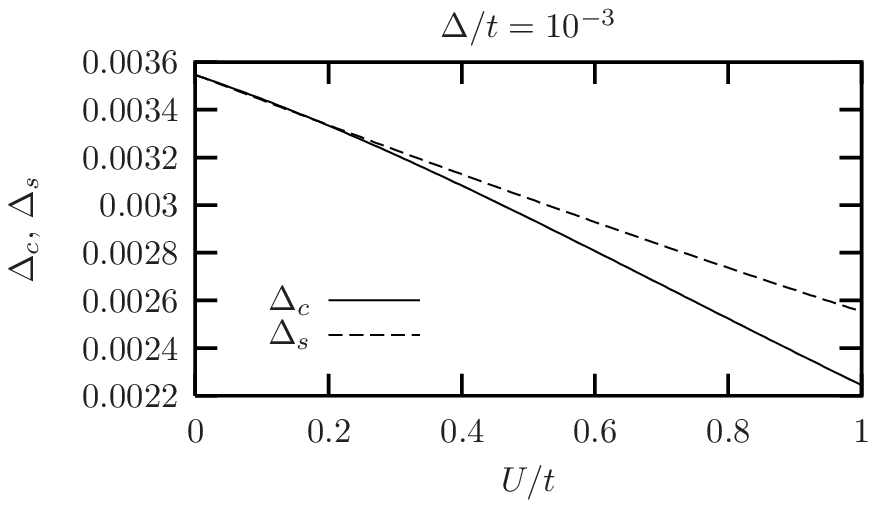}}
\caption{The mean field parameters $\Delta_c$ and $\Delta_s$ (in units of $t$),
   as function of $U$, in the thermodynamic limit. Note the different scales compared to the previous 
   figure. }
\label{an-iond2}
\end{figure}

\begin{figure}
\centerline{\includegraphics[width=11.5cm,height=8.5cm]{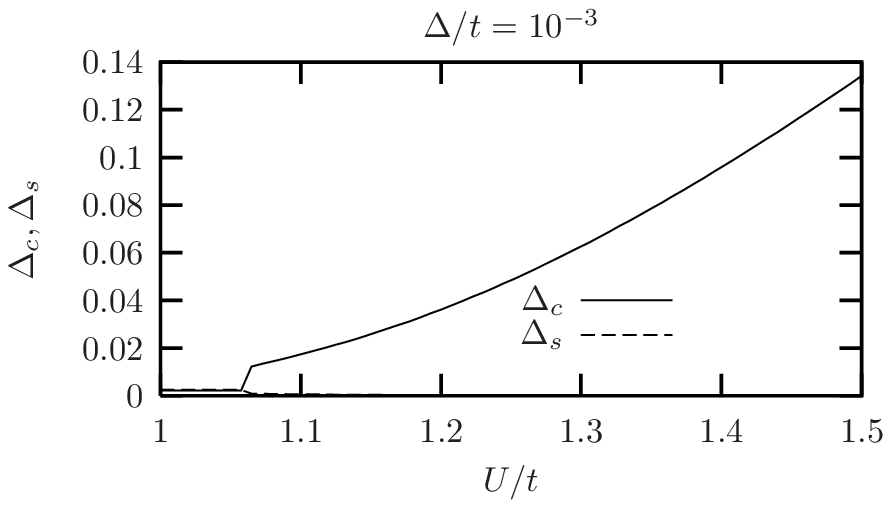}}
\caption{The mean field parameters $\Delta_c$ and $\Delta_s$ (in units of $t$), 
   as function of $U$ in the thermodynamic limit. Note the different scales compared to the previous 
   figures. }
\label{an-iond3}
\end{figure}

Since the nonlinear terms of the ionic Hubbard model
cannot be treated exactly we again employ the SCHA as in the case of Peierls Hubbard
model.  
In the thermodynamic limit 
we replace the Klein factors 
by their eigenvalues 
and introduce the trial Hamiltonian
\begin{equation}
\label{Htrial2}
H_{\rm tr}  = \sum_{\alpha=c,s}\int_0^L
    \frac{dx}{2\pi}\left\{\frac{v_\alpha}{g_\alpha}(\partial_x\phi_\alpha)^2 +
      v_\alpha g_\alpha(\partial_x\theta_\alpha)^2 +
      \frac{\Delta_\alpha^2}{v_\alpha g_\alpha} \phi_\alpha^2 \right\},
\end{equation}
which provides us with a variational estimate for the ground state energy
\begin{eqnarray}
\label{varenergy2}
\tilde E & = &
      \langle H_0 \rangle_{\rm tr} + \langle H_1 \rangle_{\rm tr}
     +\langle H_2 \rangle_{\rm tr} + \langle H_{\rm{ionic}} \rangle_{\rm tr}, \\
\frac{\tilde E}{L} & = & \frac{E_{\rm tr}}{L} - \sum_{\alpha=c,s}
\frac{\Delta_\alpha^2}{2\pi v_\alpha g_\alpha} \langle\phi_\alpha^2 \rangle_{\rm tr}
+ e(k_\uparrow, k_\downarrow),
\end{eqnarray}
where $E_{\rm tr}$ is the ground state energy of $H_{\rm tr}$, and
\begin{equation}
e(k_\uparrow, k_\downarrow) = 2 B_c \cos(k_\uparrow + k_\downarrow) + 
    2 B_s \cos(k_\uparrow - k_\downarrow) + 2 B_{cs} (\cos k_\uparrow + \cos k_\downarrow ),
\end{equation}
with
\begin{eqnarray}
\label{Bdef2}
B_c & = & \tilde U \; {\rm e}^{-4 \langle \phi^2_c \rangle_{\rm tr}}, \\
B_s & = & \tilde U \; {\rm e}^{-4 \langle \phi^2_s \rangle_{\rm tr}}, \\
B_{cs}& = & \tilde \Delta \; {\rm e}^{- \langle \phi^2_c \rangle_{\rm tr}}
{\rm e}^{- \langle \phi^2_s \rangle_{\rm tr}} \;\; .
\end{eqnarray}
Minimizing the variational ground state energy with respect to $\Delta_c$ and $\Delta_s$
yields the gap equations
\begin{eqnarray}
\label{gapequation12}
  \frac{\Delta_c^2}{2\pi v_c g_c } & = & - 4 B_c \frac{\partial e_0}{\partial B_c}
 - B_{cs} \frac{\partial e_0}{\partial B_{cs}}, \\
\label{gapequation22}
  \frac{\Delta_s^2}{2\pi v_s g_s } & = & - 4 B_s \frac{\partial e_0}{\partial B_s}
 - B_{cs} \frac{\partial e_0}{\partial B_{cs}},
\end{eqnarray}
where $e_0$ is the minimum of $e(k_\uparrow , k_\downarrow)$ 
with respect to $k_{\uparrow}$ and $k_{\downarrow}$
(see Table \ref{Ionic-tabel}).
Depending on the values of $B_c$, $B_s$ and $B_{cs}$, 
different physical situations arise:
\begin{table}
\begin{center}
\begin{tabular}{|c|c|c|c|}
\hline
$B_c>B_s$, range & $k_{\uparrow}$ & $k_{\downarrow}$ & $e_0(B_c,B_c,B_{cs})$ \\
\hline
 & & & \\
$0 < B_{cs} < 2\sqrt{B_cB_s}$ & $0$ & $\pi $ &
$- 2 B_c - 2 B_s$ \\
 & & & \\
\hline
& & & \\
$2\sqrt{B_cB_s} < B_{cs}<2B_c$ & $\arccos\frac{B_{cs}}{2B_c}+\pi$ & $\arccos\frac{B_{cs}}{2B_c}-\pi$ & 
                  $- 2 B_c + 2 B_s - \frac{B_{cs}^2}{B_c}$ \\
& & & \\
\hline
& & & \\
$ 2B_c<B_{cs}$ & $\pi$ & $\pi$ & 
                  $ 2 B_c + 2 B_s - 4B_{cs}$ \\		  
& & & \\
\hline
\end{tabular}
\caption{Minimum $e_0$ of $e(k_\uparrow, k_\downarrow) = 
2 B_c \cos(k_\uparrow + k_\downarrow) +
2 B_s \cos(k_\uparrow - k_\downarrow) + 2 B_{cs} (\cos
 k_\uparrow + \cos k_\downarrow )$
which is used in the gap equations (\ref{gapequation12})
 and (\ref{gapequation22}).
The first line is only relevant for $\Delta=0$ (Hubbard model 
without on-site energy modulation),
whereas the second and the third lines are  relevant for 
$\Delta > 0$.}
\label{Ionic-tabel}
\end{center}
\end{table}
\begin{itemize}
\item[1)] The solution $e_0=-2B_c-2B_s$ corresponds  to the Hubbard model without on-site
energy modulation, $\Delta =0$.  The system is a Mott insulator with a vanishing
spin gap, as in the Peierls-Hubbard model with $u=0$,
see  Fig.\ \ref{Hub1} and Eqs.\ (\ref{hc1})-(\ref{hubbardspin}).
\item[2)] The solution $e_0=2B_c+2B_s-4B_{cs}$ corresponds to a band insulator,
where 
the  charge and spin gap decrease with increasing $U$. In this case 
the gap equations (\ref{gapequation12})-(\ref{gapequation22}) read 
\begin{eqnarray}
\frac{\Delta_c^2}{2\pi v_cg_c}&=&4B_{cs}-8B_c,\\
\frac{\Delta_s^2}{2\pi v_sg_s}&=&4B_{cs}-8B_s.
\end{eqnarray}
As shown in appendix \ref{an-solution}, this case corresponds to the limit $\Delta/U\gg 1$,
and one obtains the following dependence of the gap parameters $\Delta_c$ and $\Delta_s$
on the ionic potential $\Delta$:
\begin{equation}
\label{AIc2}
\Delta_c \propto \Delta^{2/(3-g_c)},
\end{equation}
\begin{equation}
\label{AIs2}
\Delta_s \propto \Delta^{2/(3-g_c)}.
\end{equation}
This is the same result as obtained for the Peierls-Hubbard model with $u$
replaced by $\Delta$. For $U=0$, i.e.\ $g_c=1$, Eqs.\ (\ref{AIc2})-(\ref{AIs2})
yield $\Delta_{c,\; s}\propto \Delta$ in accordance with the known solution in the non-interacting limit.
\item[3)] The solution $e_0= -2 B_c + 2 B_s - B_{cs}^2/B_c$ corresponds
 to the spontaneously dimerized insulator 
  phase where the dimerization operator has the
expectation value
\begin{equation}
\langle D \rangle \sim \sin (k_{\uparrow})+\sin (k_{\downarrow})
  =-2\sqrt{1-\left(\frac{B_{cs}}{2B_c}\right)^2}.
\end{equation}
In this case the gap equations are
\begin{eqnarray}
\frac{\Delta_c^2}{2\pi v_cg_c}&=&8Bc-2\frac{B_{cs}^2}{B_c},\\
\frac{\Delta_s^2}{2\pi v_sg_s}&=&-8Bs+2\frac{B_{cs}^2}{B_c},
\end{eqnarray}
\end{itemize}
which are solved in appendix \ref{an-solution}.
The result is
\begin{equation}
\label{AIc3}
\Delta_c (\Delta)- \Delta_c (0)\propto \Delta^4,
\end{equation}
\begin{equation}
\label{AIs3}
\Delta_s  (\Delta)\propto \Delta^2,
\end{equation}
which is similar to the behavior of the Peierls-Hubbard model,
Eqs.\ (\ref{chargegap})-(\ref{spingap}), however with different exponents, 
$ 4$ instead of $4/3$ for the charge gap, and $2$ instead of $2/3$ for the spin gap. 
This different behavior 
can be understood in the strong coupling limit $U\rightarrow \infty$ where
in second order perturbation theory the Peierls term yields an alternating 
contribution to the Heisenberg exchange
coupling $J=4t^2/U$, while for the ionic Hubbard model 
the strong coupling limit is the antiferromagnetic Heisenberg model with uniform coupling. In
this case the opening of the spin gap is associated with higher than second order
processes
in the strong coupling limit.
Fig.\ \ref{an-ionU1} shows the dependence of $\Delta_c$ and $\Delta_s$ on
$\Delta$ for fixed Hubbard interaction $U/t=1$. Contrary to the case of the 
Peierls-Hubbard model there is no smooth crossover between the small and large 
$\Delta$ region but a discontinuous transition at some intermediate value $\Delta^*$.
%
%
In Fig.\ \ref{an-iond1} the ionic parameter $\Delta=10^{-3}$ is kept fixed and $\Delta_c$
and $\Delta_s$ are plotted as functions of $U$. For small values of $U$
spin and charge gaps are almost equal and decrease with increasing $U$, as highlighted in 
Fig.\ \ref{an-iond2}. At $U/t \approx 1.05 $ there is a discontinuous transition to a phase 
where $\Delta_s$ goes to zero while $\Delta_c$ increases  strongly, as can be seen in Fig.\ \ref{an-iond3}.

\begin{figure}
\centerline{\includegraphics[width=12.0cm,height=7.50cm]{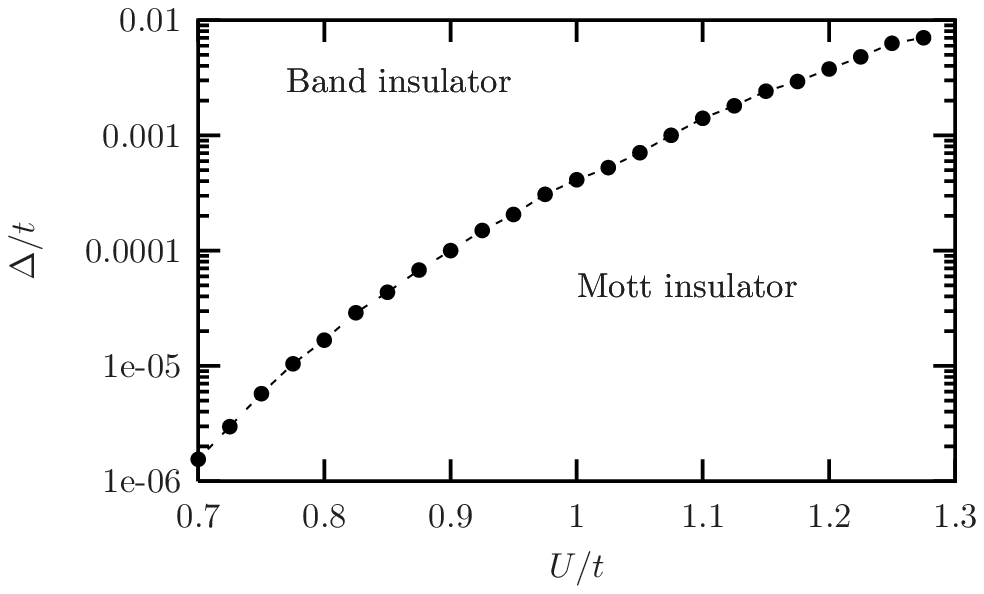}}
\caption{Phase diagram of the ionic Hubbard model. The dots correspond 
to the discontinuity of $\Delta_c$ and $\Delta_s$, as  for example 
 shown in Fig.\ \ref{an-ionU1} 
for $U/t=1$.}
\label{phase-dig-ion}
\end{figure}
Connecting the values of $\Delta$ where the discontinuous transition occurs for
different values of $U$ one obtains the phase boundary between a Mott insulator 
($\Delta_s \rightarrow 0$) and a band insulator ($\Delta_c \approx \Delta_s$)
as shown in Fig.\ \ref{phase-dig-ion}. 
We find only one phase transition, and not two, in contrast to the  phase diagram proposed by 
Fabrizion et al.\ \cite{Fabrizio99}. However,  in our bosonization approach, 
using SCHA to estimate the charge and the spin gap, we are limited  to small values of $U$ and $\Delta$.
\clearpage
\subsection{Finite systems}
To calculate the gaps of the ionic Hubbard model for a finite system we apply
the same procedure as in the case  of the Peierls-Hubbard model. We will find,
unexpectedly, strong
finite size effects near the phase transition. At present it is still not clear 
to us whether this behavior is an artefact of the SCHA or if it has physical reality. 

We decouple the Klein factors 
from the bosonic fields introducing the  Klein Hamiltonian 
\begin{eqnarray}
\label{HKlein}
H_{\rm tr}^{B_{cs}B_cB_s}  & = & B_{cs}L(F_{R\uparrow}^+ F_{L\uparrow} + F_{R\downarrow}^+
 F_{L\downarrow})
+ B_c LF_{R\uparrow}^+F_{R\downarrow}^+F_{L\downarrow}F_{L\uparrow} \nonumber \\
 &   &  + B_sL F_{R\uparrow}^+F_{L\downarrow}^+F_{R\downarrow}F_{L\uparrow}+{\rm h.c.}\nonumber \\ 
 & &+ 
 \frac{\pi}{4L}(v_c g_c J^2_c + v_s g_s J^2_s),
\end{eqnarray}
where $B_c, B_s$ and $B_{cs}$ are variational parameters.
$H_{\rm tr}^{B_{cs}B_cB_s}$ is again of the form of a tight-binding Hamiltonian
for a particle moving on a $2d$ lattice in a harmonic potential.
The explicit representation of this hopping Hamiltonian in the $|J_c,J_s\rangle$ basis is given in
appendix \ref{KH-tbm} for the Peierls-Hubbard model. For the ionic Hubbard model one has to
replace $\pm iB_{cs}$ by $B_{cs}$ to obtain the corresponding matrix elements.
As a result of the variational procedure, 
the Klein factors are replaced by their expectation values  with respect to
the  ground state of the Klein Hamiltonian.
Similar to the case of the Peierls-Hubbard model, in the framework of
the SCHA,   the quantity $e_0$ which enters the gap equations
 (and which is explicitly given in Table \ref{Ionic-tabel}) is replaced by  $e^0_{\textrm{tr}}$, the
 ground state energy of $H_{\rm tr}^{B_{cs}B_cB_s}$ divided by L, which can be calculated
 only numerically.
\begin{figure}
\centerline{\includegraphics[width=13.0cm,height=11.0cm]{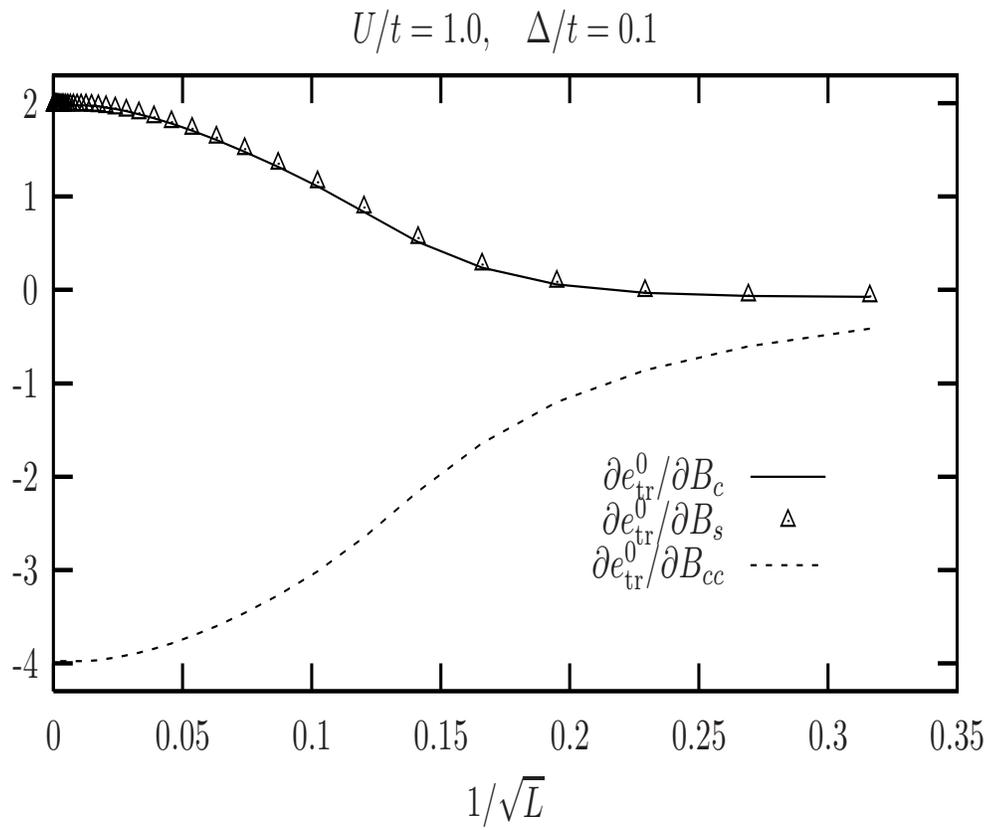}}
\caption{The derivative of $e^0_{\textrm{tr}}$ with respect to the 
     variational parameters 
         $B_c, B_s, B_{cs}$ needed in the  equations
	 (\ref{gapequation12}) and (\ref{gapequation22}).}
\label{fhiklein1}
\end{figure}
Fig.\ \ref{fhiklein1} shows the derivatives of $e^0_{\textrm{tr}}$
with respect to $B_c, B_s$ and $B_{cs}$ as function of the system size for fixed
model parameters $U/t=1$ and $\Delta/t=0.1$. For small $L$ the derivatives are
close to zero while with  increasing
 $L$ they converge to the values corresponding to 
the thermodynamic limit given in 
Table \ref{Ionic-tabel},  
$\partial e^0_{\textrm{tr}}/\partial B_c=2,\; \partial e^0_{\textrm{tr}}/\partial B_s=2, 
	  \; \partial e^0_{\textrm{tr}}/\partial B_{cs}=-4$. The scale  $1/\sqrt{L}$
is chosen in order to enlarge the crossover region.	  
	  
\begin{figure}
\centerline{\includegraphics[width=15.0cm,height=12cm]{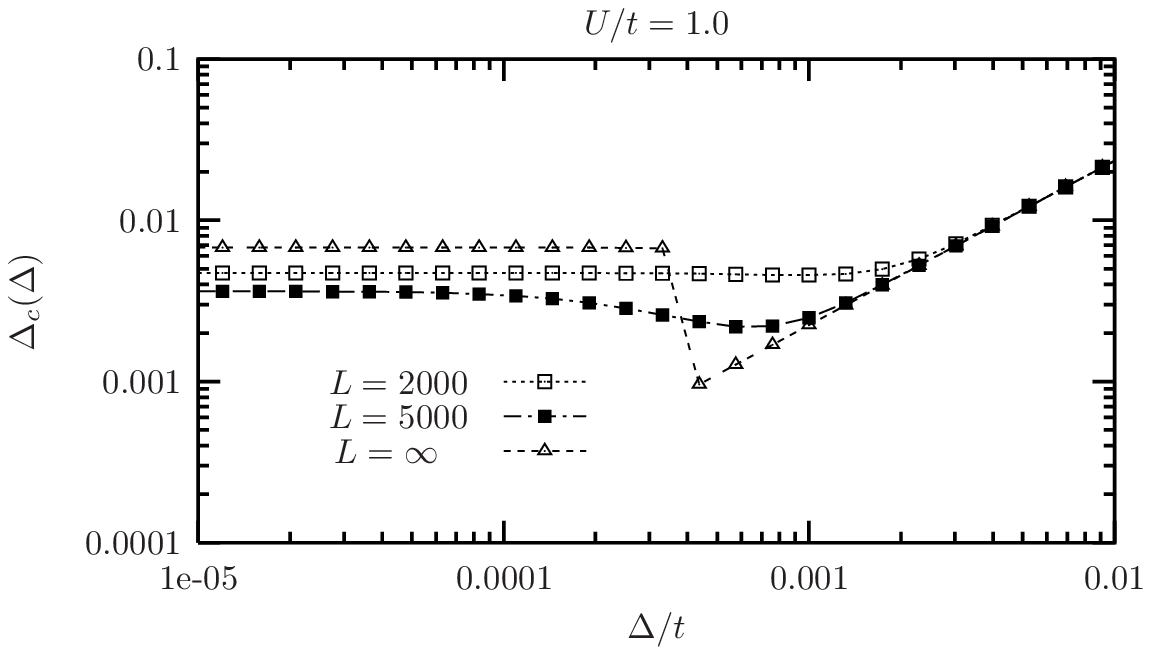}}
\caption{ The charge gap  parameter $\Delta_c$  (in units of $t$) as a function of $\Delta$ 
for different values of the system  size $L$  compared
with the thermodynamic limit.}
\label{ficharge1}
\end{figure}
\begin{figure}
\centerline{\includegraphics[width=15.0cm,height=12cm]{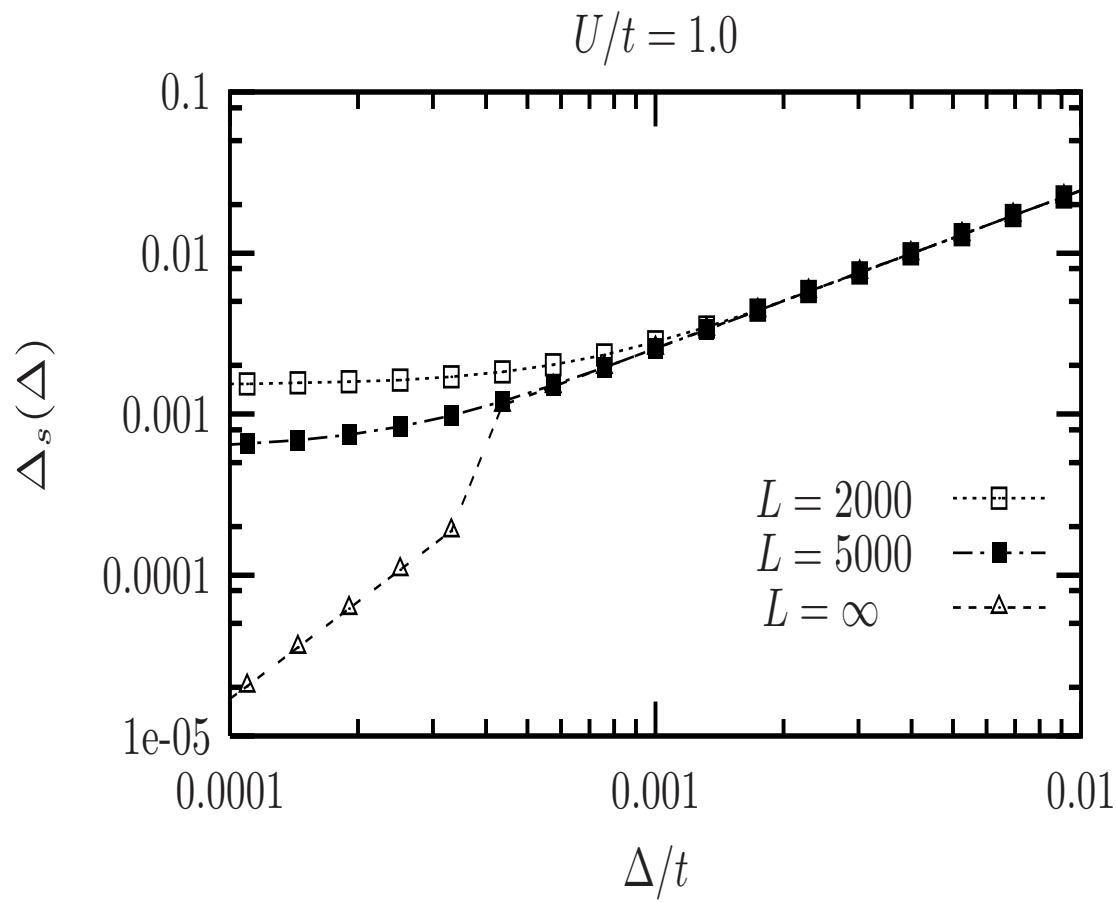}}
\caption{ The spin parameter $\Delta_s$ (in units of $t$) as a function of $\Delta$ for 
different values of the  system  size $L$ compared
with the thermodynamic limit.}
\label{fispin1}
\end{figure}

In Fig.\ \ref{ficharge1} we show the gap parameter $\Delta_c$ as function of
$\Delta$ and fixed Hubbard interaction $U/t=1$ for different system sizes. While
for large values of $\Delta$, corresponding to the band insulator phase, there are
nearly no finite size effects, for small $\Delta$, corresponding to a Mott
insulator phase, even for $L=5000$ the values
 have not yet converged to the thermodynamic
limit. The transition between the Mott insulator and the band insulator phase
appears continuous for small values of the system size $L$ but becomes sharper
with increasing $L$ and finally turns over into a discontinuous one. The same
observation holds for the spin gap parameter $\Delta_s$, which is displayed in
Fig.\ \ref{fispin1} for the same set of parameters as before. In particular for
$\Delta\rightarrow 0$ the gap parameter $\Delta_s$ approaches a constant for
finite systems while in the thermodynamic limit it goes to zero according to
$\Delta_s \propto \Delta^2$. 

\begin{figure}
\centerline{\includegraphics[width=15.0cm,height=12.0cm]{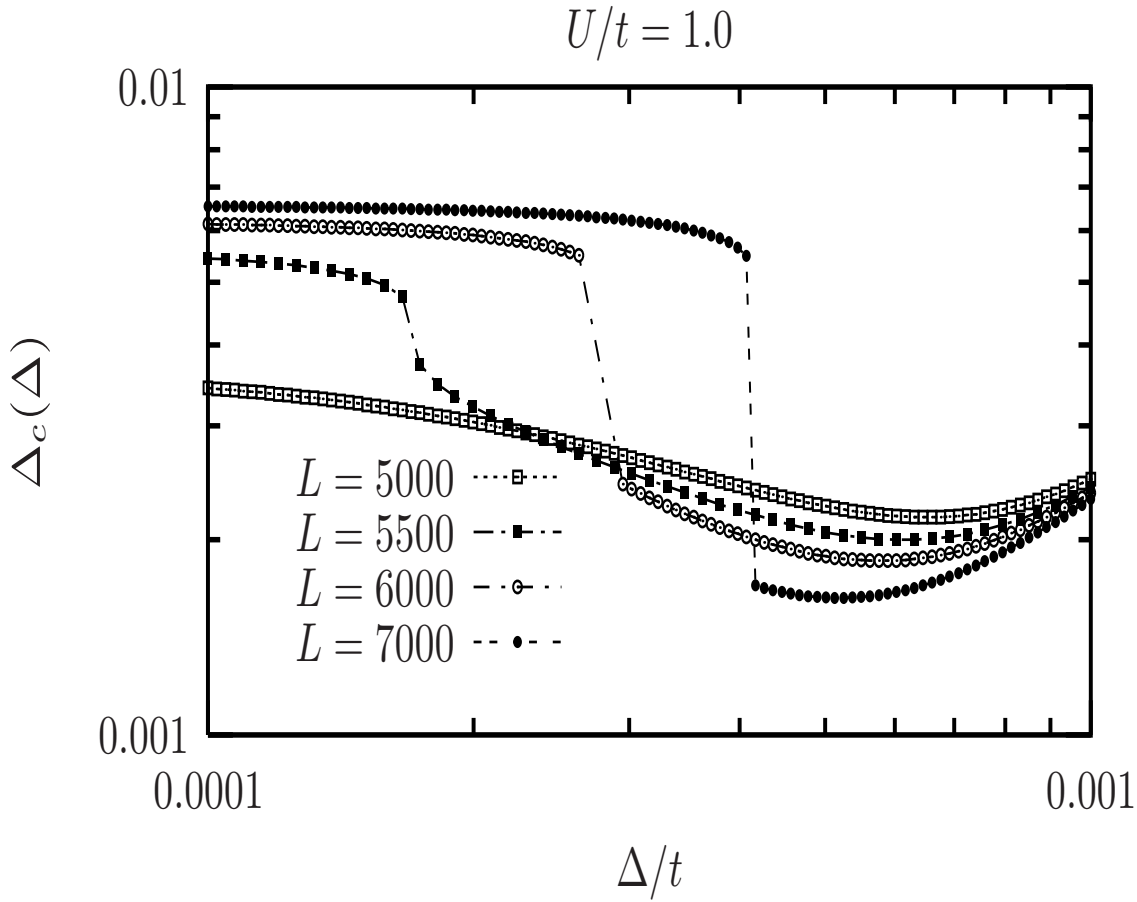}}
\caption{ The charge parameter $\Delta_c$ (in units of $t$) as a function of $\Delta$ 
for different values of the size system size  $L$ in the
region where the transition from a band insulator to a  Mott insulator appears.}
\label{ficharge2}
\end{figure}
\begin{figure}
\centerline{\includegraphics[width=15.0cm,height=12.0cm]{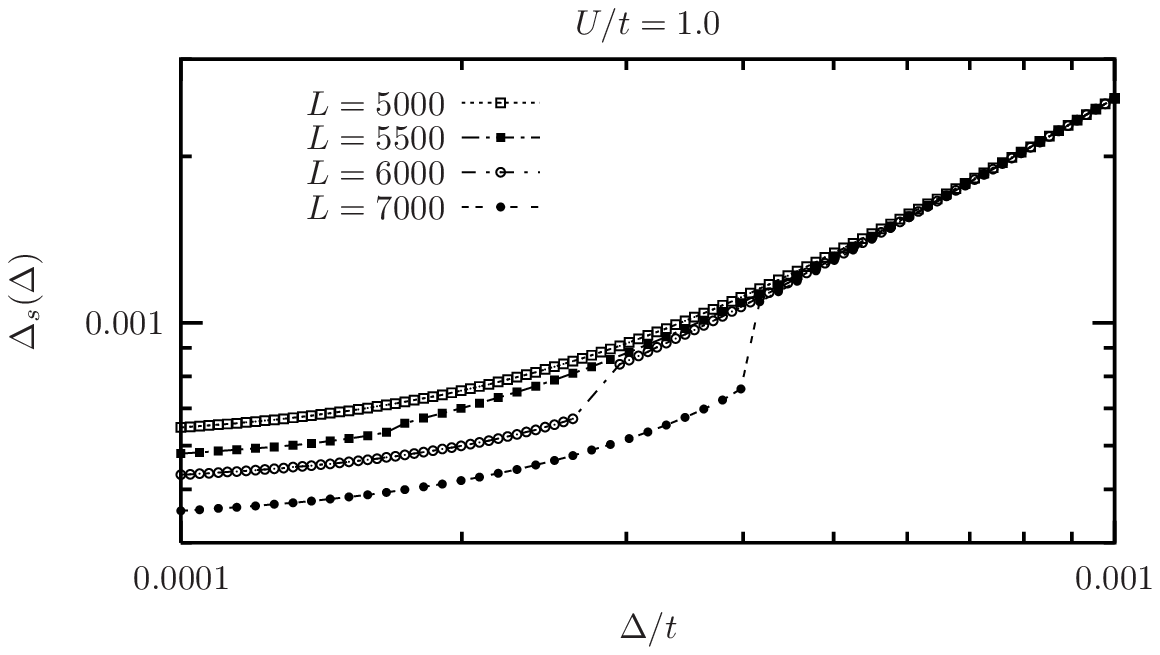}}
\caption{ The spin parameter $\Delta_s$ (in units of $t$) as a function of $\Delta$ 
for different  system size  $L$ in the
region where the transition from a band insulator to a  Mott insulator appears.}
\label{fispin2}
\end{figure}

In order to show in more detail how the discontinuous transition emerges from
the continuous one we plot the gap parameters $\Delta_c$ and $\Delta_s$ for
several values of $L$ in the region where the transition occurs in 
Figs.\ \ref{ficharge2} and \ref{fispin2}, respectively. For $U/t=1$, the 
discontinuity appears somewhere between $L=5000$ and $L=5500$ and becomes more
pronounced while $L$ is increased further.

\begin{figure}
\centerline{\includegraphics[width=15.0cm,height=12.0cm]{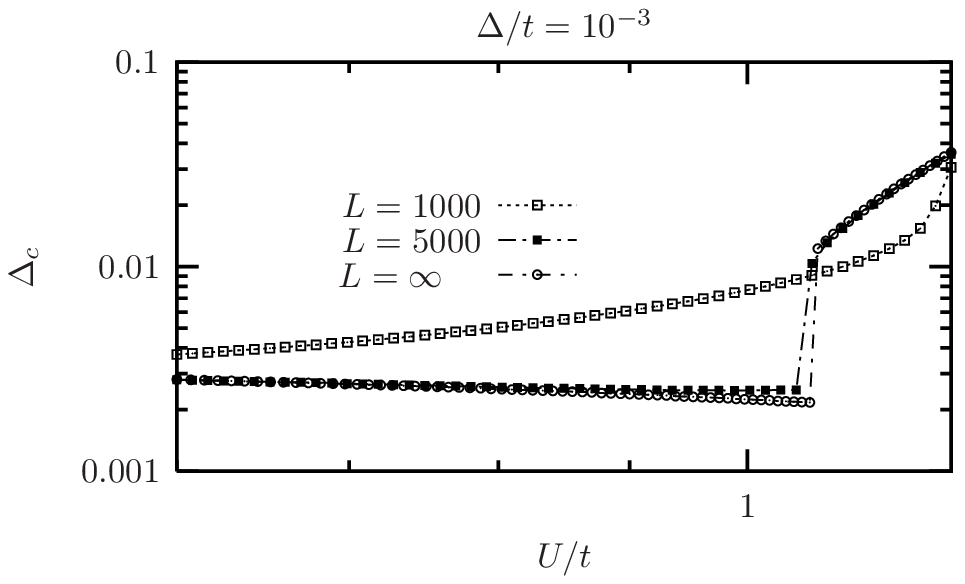}}
\caption{The charge parameter $\Delta_c$ (in units of $t$) as a function of the Hubbard interaction  $U/t$ for 
different system sizes compared
with the thermodynamic limit.}
\label{ficharge3}
\end{figure}

\begin{figure}
\centerline{\includegraphics[width=15cm,height=12.0cm]{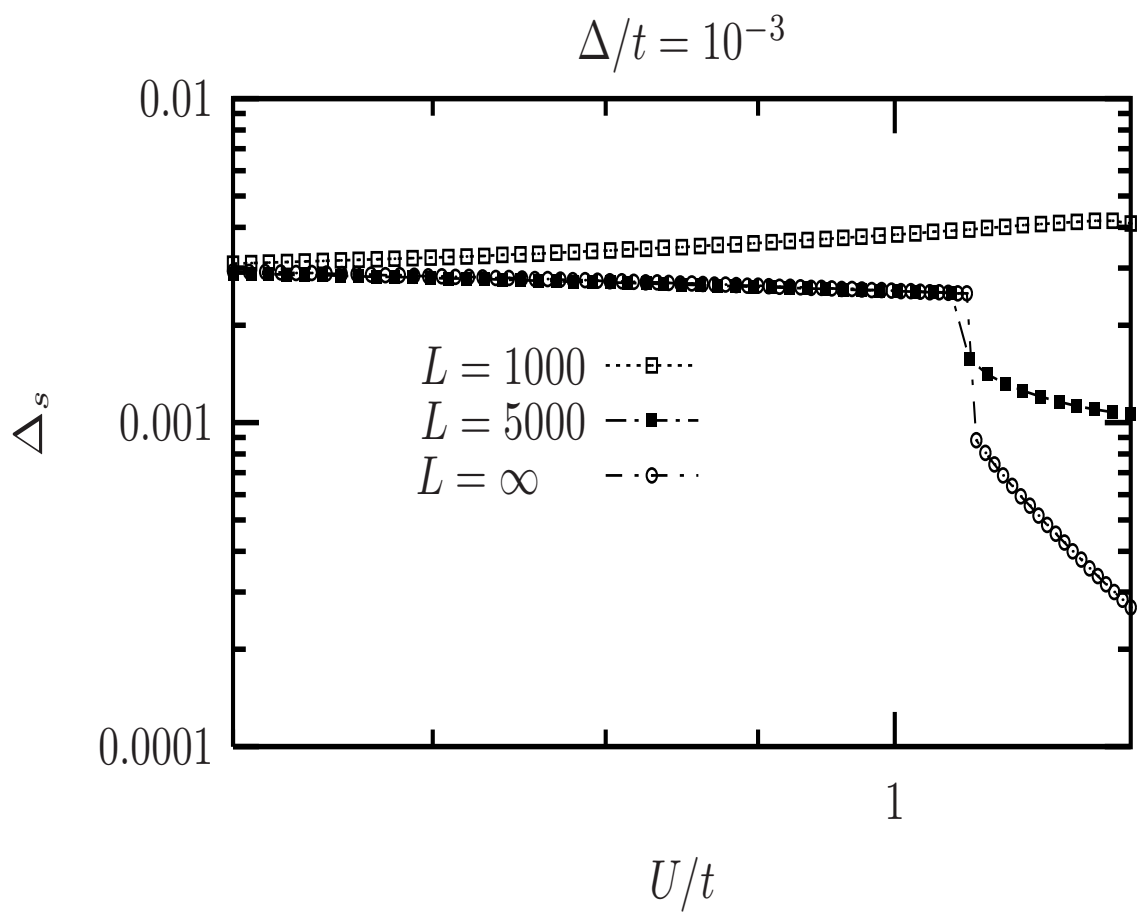}}
\caption{The spin parameter $\Delta_s$ (in units of $t$) as a function of the Hubbard interaction  $U/t$ for 
different system sizes  compared
with the thermodynamic limit}
\label{fispin3}
\end{figure}

%
%
%
%
%
%
In Figs.\ \ref{ficharge3} and \ref{fispin3} we show $\Delta_c$ and $\Delta_s$,
respectively, as function of the Hubbard interaction $U/t$ for fixed value of
the ionic parameter $\Delta/t=10^{-3}$ and different system sizes $L$. For this
small value of $\Delta/t$ there are still quite pronounced finite size effects
for $L=1000$. This indicates that numerical methods which are devised for finite
systems like exact diagonalization or DMRG would probably fail in the small $\Delta$
 regime.

\section{Conclusions}

In this chapter we considered  two different one-dimensional  Hubbard models, 
 the Peierls-Hubbard  and the  ionic Hubbard
model. In the bosonized version of these models two kinds of bosonic fields have
to be introduced which are associated with the spin and the charge degrees of
freedom, respectively.
The backward and the Umklapp scattering yield nonlinear terms in the 
spin and the charge sector, respectively, 
while the Peierls distorsion and the ionic potential 
introduce  a coupling between the spin and the charge sectors. This nonlinear terms
lead to the opening of spin or charge gap, depending on the model parameters. 

Our first goal was  to stress the importance of  Klein factors in bosonized
Hamiltonians with nonlinear perturbations.  In the study of such types 
of Hamiltonian, one has to keep in
mind that the Klein factors which accompany the nonlinear terms  
do not commute with the total spin and charge currents,
$J_{c,s}$, and therefore Klein factors and current operators 
cannot acquire a fixed value at the same time.
In the thermodynamic limit  the ground state of the gapped
system is a superposition of many states with different $J_{c}$ and $J_{s}$. 
In this situation it
is possible  to choose a fixed phase for the Klein operators. 
The bosonized
Hamiltonian is then the conventional sine-Gordon Hamiltonian. 
For finite size systems the 
Klein factors  can be  replaced by their  expectation values with
respect to the ground state of a  trial Hamiltonian, and the resulting bosonic
Hamiltonian is also of a sine-Gordon type, however with size-dependent
couplings.

Using a self consistent variational scheme, we decoupled the bosonic Hamiltonian
from the Klein Hamiltonian and  
 we calculated the  gap parameters 
$\Delta_c$ and $\Delta_s$ as a function of the strength of the perturbation 
(the dimerization $u$ in the case of the Peierls-Hubbard model, and the ionic
potential $\Delta$ in the case of the ionic Hubbard model). 
In both cases we
found a transition from  a Mott insulator phase
with vanishing or very small spin
gap to a band insulator phase 
where spin and charge gaps are of the same order of magnitude.
The main difference between the two models is that 
the transition between the Mott and the Peierls phase is continuous 
in the case of Peierls-Hubbard model and discontinuous for the ionic Hubbard
model.

\chapter{Screening in low-dimensional electron systems}
\label{vanadium}
In order to extend  the model calculations of the preceding sections towards a
more realistic description of quasi one-dimensional systems we 
consider in the following a one-dimensional interacting system, 
characterized by a Luttinger model, coupled to a three-dimensional environment, 
and calculate the spin and the charge susceptibility. 
The motivation is an old discussion on the role of electronic correlations 
for the metal-insulator transition in  VO$_2$.
\nop{In section \ref{dataVO2} we will discuss in same detail the electronic structure
of VO$_2$ near the the Fermi energy. In sections
\ref{modelVO2}-\ref{discussionVO2} we .....}

\section {Phase transition in VO$_2$}
\label{dataVO2}
A detailed analysis of electronic structure of vanadium dioxide (VO$_2$)
is given in \cite{Eyert98,Eyert02}. In the following we recall some of the
theoretical and experimental findings.
Vanadium dioxide 
in the paramagnetic metallic phase above $T_c=340$ K  has a rutile 
crystal structure,
each vanadium atom being located at the center of an oxygen octahedron.
The fivefold-degenerate $d$ levels of the V$^{4+}(3d^1)$ ion are  
split into doubly degenerate $e_g$ levels and triply degenerate $t_{2g}$ 
levels in the octahedral crystal field. The $e_g$ orbitals are strongly 
hybridized with the O $2p\sigma$ orbitals and have a large bandwidth,
and are pushed far away from the Fermi surface. The $t_{2g}$ 
levels are split into the $d_{\parallel}$ and $\pi^*$ levels by the 
orthorhombic component of the tetragonal crystal field. 
Thus the $d_{\parallel}$ and $\pi^*$ bands are situated at the lowest 
energies around the Fermi level. 
According to LDA calculations \cite{Eyert98,Eyert02} the $d_{\parallel}$ 
bands show a strongly one-dimensional dispersion, whereas the $\pi^*$ bands are
three-dimensional.
These  bands play the crucial role in electronic (transport and magnetic)
properties VO$_2$, and will be further studied in section \ref{modelVO2}.

\begin{figure}
\centerline{\psfig{figure=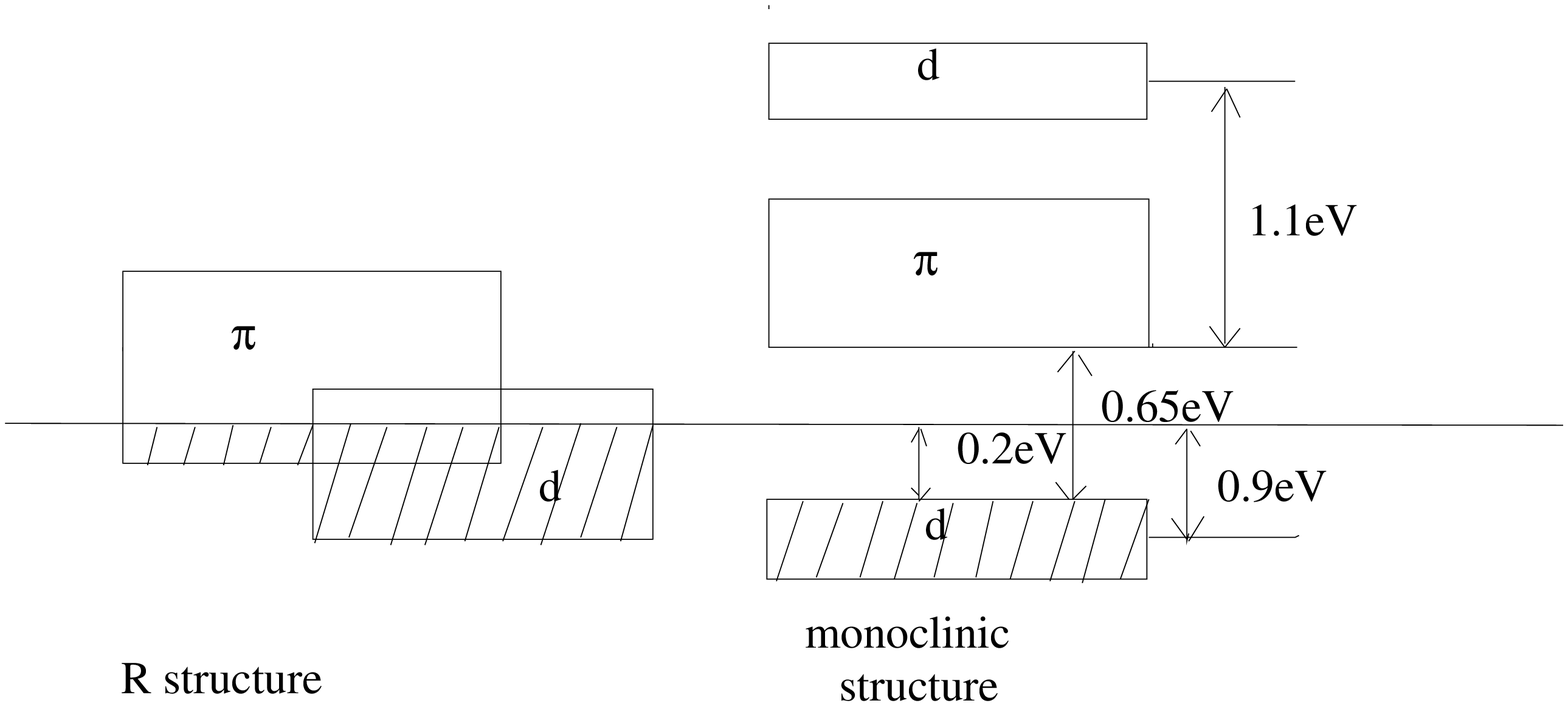,width=9cm,height=5.5cm}}
\caption{Schematic energy diagram of the $3d$ bands around the Fermi level for
VO$_2$ \cite{Shin90}.
The energy of $0.2$ eV is obtained by the UPS spectra. The energies of $0.5$ 
and $1.1$ eV are obtained by ultraviolet reflectance spectra, while  $0.65$ eV is a
value from the infrared absorption spectra \cite{Ladd69}}
\label{Shin-picture}
\end{figure}

In the insulating non-magnetic  phase below $T_c$, VO$_2$ is distorted to
a monoclinic crystal structure, involving a pairing of vanadium ions
V$^{4+}$ along the $c_r$ axis. Because of the change in the V-O hybridization,
the energy of the  hybridized $\pi^*$ band rises above the Fermi level and 
it becomes empty. Furthermore, the $d_{\parallel}$ band is split into two
 bands. A schematic energy diagram of the $3d$ bands around the Fermi level for
VO$_2$ is shown in Fig.\ \ref{Shin-picture}.

There is an old discussion concerning the driving mechanism for the
metal-insulator transition in this material and  the role of electronic 
correlations.
Some authors    
argued that 
electronic correlation
  effects are in VO$_2$  less pronounced than in  the other vanadium oxides, like V$_2$O$_3$. 
Goodenough pointed out that the metal-insulator transition of VO$_2$
may be explained by considering the change in crystal structure,
 which leads to the splitting of 
the $d_{\parallel}$ band and the rising of the $\pi^*$ bands \cite{Goodenough71}.  
Modern LDA \cite{Eyert98,Eyert02} and molecular dynamics 
\cite{Wentzcovitch} calculations support this point of view, although the
splitting of the $d_{\parallel}$ bands is underestimated, so that they predicted
a metallic instead of an insulating state.

 Further experimentals, however, demonstrated that electronic correlations play an important
 role also in VO$_2$. 
Pouget et al.\ \cite{Pouget74,Pouget75} performed nuclear magnetic 
    resonance (NMR)  measurements  under uniaxial stress in VO$_2$ and  
    V$_{1-x}$Cr$_x$O$_2$. They  found for  V$_{1-x}$Cr$_x$O$_2$  three different insulating phases (M$_1$, M$_2$
     and T). The M$_1$ phase is identical to the insulating phase of pure VO$_2$
     at normal pressure, while  in the  M$_2$ and T phases magnetic moments
     were observed. The appearance of the magnetic  moments was considered as
     a clear  sign for correlation effect in the insulating phase. 

Zylbersztejn and Mott \cite{Mott75} emphasized  the important role
    of the $\pi^*$ electrons in the metal-insulator transition. They argued 
    that, in the insulating phase, the hybridization between the the  $\pi^*$
    and O $2p\sigma$ orbitals increases due to the crystal distortion so that
    the $\pi^*$ band rises in  energy and becomes empty. Therefore the
    correlation energy of the $d_\parallel $ electrons becomes enhanced leading
    to a splitting into the lower and the upper Hubbard bands. They argued that
    the metallic phase is adequately described within the framework of 
    band theory, since the interaction of the $d_\parallel$ electrons is
    screened by the $\pi^*$ electrons. They estimated the effective 
    intra atomic Coulomb repulsion in the metallic phase from 
    the Stoner enhancement of
    the magnetic susceptibility as
\begin{equation}
      \frac{\chi_{\Vert}}{\chi_{\Vert}^P}
       =\frac{1}{1-U_{\rm eff}{\cal N}_{d_{\Vert}}}=3.2
\end{equation} 
   yielding  $ U_{\rm eff}=0.13$ eV. Here 
    $\chi_{\parallel}^P$ is the Pauli spin susceptibility and ${\cal N}_{d\parallel}=5.25$ 
   states/eV is
   the density of states per spin at the Fermi surface  in the $d_{||}$ band.

Shin et al.\ \cite{Shin90} measured the vacuum-ultraviolet reflectance and
    photoemission spectra of VO$_2$ in order to investigate the $3d$ bands
    and electron-correlation effects. The splitting energy between the
    two $d_\parallel$ bands in the insulating phase  was found to be about $2.5$ eV, bigger  than the value obtained 
     from the cluster band
    calculation \cite{Sommers75}, namely  $0.5$ eV.
      They argued that in the insulating
     phase the band splitting of the $\pi^*-d_{\parallel}$ and 
     $d_\parallel-d_{\parallel}$ bands cannot be explained by the crystal
     distortion alone, correlation effects  playing  an 
     important role also. They estimated the correlation energy in the 
     insulating phase of the 
     $d_\parallel$ band as about $U(d_\parallel,d_\parallel)\approx 2.1$ eV. 
 From the photoemission data they concluded that 
  the correlation energy of the $d_\parallel$ electrons must  not be
     neglected  even in the metallic phase, and $U(d_\parallel,d_\parallel)$ 
     was estimated to be not less than $1.3$ eV.
    
\section{The model}
\label{modelVO2}

Following the ideas of Mott and Zylbersztejn that in the metallic
    state of VO$_2$ the electron-correlation effects in the one-dimensional 
    $d_\parallel$ bands are screened by the three-dimensional $\pi^*$ electrons,
    we consider a one-dimensional, interacting  Fermi system,
    which is coupled
    to a three-dimensional environment through a local interaction. 
    The aim
    is to see how the presence of the three-dimensional system affects the
    one-dimensional system. We calculate the spin  and charge susceptibility,
    with the result  that the spin susceptibility
    is not changed due to the coupling to the environment, whereas the charge
    susceptibility shows screening. 
    
We consider  the
Hamiltonian 
\begin{equation}
\label{ham1+3}
H=H_{\textrm{Luttinger}}+H_{0\pi^*}+H_{d-\pi^*},
\end{equation}
where $H_{\textrm{Luttinger}}$ is the Hamiltonian corresponding to the Luttinger model, 
\begin{eqnarray}
H_{0\pi^*}&=&\sum_{k,\sigma}\epsilon _{\pi} ({\bf k})\pi^{*+}_{{\bf k}\sigma}\pi^*_{{\bf k}\sigma}
\end{eqnarray} 
corresponds to a band of  non-interacting $\pi^*$ electrons with a three-dimensional 
dispersion $\epsilon_{\pi}(k)$, and
\begin{eqnarray}
H_{d-\pi^*}&=&\frac{V}{N}\sum\limits_{{\bf kpq},\sigma \sigma'}d_{{\bf k}\sigma}^+d_{{\bf k}-{\bf q}\sigma}
     \pi^{*+}_{{\bf p}\sigma'}\pi^*_{{\bf p}+{\bf q}\sigma'}
\end{eqnarray}
describes 
coupling between $\pi^*$ and $d_{\parallel}$ electrons via an interaction of strength $V$; $N$
is the number of lattice points  (in the three dimensional lattice).
  In order to make a connection between our 
 calculation
 and the discussion on VO$_2$, we use the notation $d^+,d$ for the  
  creation  and annihilation 
 operators  of $d_{\parallel}$ (one-dimensional)  electrons,  and the notation $\pi^{*+},\pi^*$ 
 for the 
  creation and  annihilation 
 operators of $\pi^*$ (three-dimensional) electrons. A similar Hamiltonian has been studied by 
 Paquet and Leroux-Hugon \cite{Paquet80}. In this work, electronic correlations in the 
 one-dimensional $d_{\parallel}$ band were taken into account in terms of a spatially
 inhomogenous mean field theory. In contrast, we describe the $d_{\parallel}$ electrons as
 Luttinger liquid.

\section{Charge and spin  susceptibility}

The dynamic susceptibility $\chi$ is a measure of the system's response to an external
field. Generally, the dynamic susceptibility is given by a retarded commutator of the form 
\begin{equation}
\chi_{AA} ( {\bf q},t)=i\theta (t)
  \langle [A({\bf q},t),A(-{\bf q},0)]\rangle.
\end{equation}  
 For the spin susceptibility $A$ is the spin density  operator, 
$\sigma ({\bf q})=\rho_{\uparrow}({\bf q})-\rho_{\downarrow}({\bf q})$,
 while for the charge susceptibility $A$ is the charge density operator,
$\rho ({\bf q})=\rho_{\uparrow}({\bf q})+\rho_{\downarrow}({\bf q})$. The density operators  
$\rho_{\uparrow, \downarrow}$ are given by Eq.\ (\ref{rhosigma}) where for the general 
case the one dimensional momenta have to be replaced by the three dimensional vectors.

\subsection{Luttinger model}
For the calculation of the charge susceptibility in the   
Luttinger model we express  the charge
density in terms of left and right charge densities,
\begin{equation}
\rho (q)=\rho_R (q)+\rho_L (q).\nonumber\\
\end{equation}
Corresponding to this separation in left and right movers the  correlator  
$\chi_{\rho\rho}$ can be
expressed as 
\begin{equation}
 \chi_{\rho\rho}=i\theta (t) \langle [\rho(q,t),\rho(-q,0) ]\rangle =
  \chi_{LL} +\chi_{LR} +\chi_{RL} +\chi_{RR}.
\end{equation}
By taking the time derivative one finds for all four components $\chi_{\eta\gamma}$ 
an equation of motion of the type
\begin{equation}
\label{suscept}
-i\partial_t\chi_{\eta\gamma}=
  \delta(t)\langle[\rho_{\eta}(q,0),\rho_{\gamma}(-q,0) ]\rangle
-i\theta(t)
\langle \left[[\rho_{\eta}(q,t) ,H] , \rho_{\gamma}(-q,0) \right]\rangle.
\end{equation}
For the ``g-ology" Hamiltonian
 the commutators $[\rho_{\eta},H]$  are 
\begin{equation}
\label{comL}
[\rho_{L} ,H]=qv_F(1+\gamma_{4\parallel}+\gamma_{4\perp})\rho_L+
    qv_F(-\gamma_{1\parallel}+\gamma_{2\parallel}+\gamma_{2\perp})\rho_R,
\end{equation}
\begin{equation}
\label{comR}
[\rho_{R} ,H]=-qv_F(1+\gamma_{4\parallel}+\gamma_{4\perp})\rho_R-
    qv_F(-\gamma_{1\parallel}+\gamma_{2\parallel}+\gamma_{2\perp})\rho_L,
\end{equation}
where $\gamma_{i}=g_i/(2\pi v_F)$. With this  simple form of the commutators 
one  can  calculate
the susceptibility explicitly, and  the final result is
\begin{equation}
\label{chi-charge}
\chi_{\rho\rho}(q,\omega)=2Na{\cal N}_{1d}(\epsilon_F) \frac{g_c v_F}{v_c}
\frac{-(qv_c)^2}{\omega^2- (qv_c)^2}=\chi_{\rho\rho}^{1d},
\end{equation}
where ${\cal N}_{1d}(\epsilon_F)=(\pi v_F)^{-1}$ is the density of states per
spin at the Fermi surface, $a$ is the lattice constant, and  $v_c$ and $g_c$ are the Luttinger parameters in the charge sector as
given in Eqs.\ (\ref{lutt-param-vc}) and (\ref{lutt-param-gc}).

In a similar way,  the spin density can be separated into left and right spin 
densities
\begin{equation}
\sigma (q)=\sigma_R (q)+\sigma_L (q).
\end{equation}
Following the same steps as for the charge susceptibility one  obtains for the spin susceptibility
\begin{equation}
\label{chi-spin}
\chi_{\sigma\sigma}(q,\omega)=2Na{\cal N}_{1d}(\epsilon_F) \frac{g_s v_F}{v_s}
\frac{-(qv_s)^2}{\omega^2- (qv_s)^2},
\end{equation}
where $v_s$ and $g_s$ are the Luttinger parameters in the spin sector as
given in Eqs.\ (\ref{lutt-param-vs}) and (\ref{lutt-param-gs}).


\subsection{Luttinger model coupled to $3d$ electrons}
Now we  calculate  the charge susceptibility of a  one-dimensional electronic
system  which is coupled to a three-dimensional environment. 
The model is described  by  the
Hamiltonian (\ref{ham1+3})
\begin{equation}
H=H_{\textrm{Luttinger}}+H_{0\pi^*}+H_{d-\pi^*}.
\end{equation}
We consider in the following only interactions with small momentum transfer; 
in this case, in terms of density operators, the
Hamilton  $H_{d-\pi^*}$ reads
\begin{equation}
H_{d-\pi^*}=
\frac{V}{N}\sum\limits_{{\bf q}}\rho_d(-{\bf q})\rho_{\pi^*}({\bf q}),
\end{equation}
where ${\bf q}$ is the momentum transfer with 
${\bf q}=(q_{\parallel},\;{\bf q}_{\perp} )$.    
 Since the total charge  (and spin) density is a sum of three contibutions, 
 $\rho=\rho_L+\rho_R+\rho_{\pi^*}$, then    
 the correlator $\chi_{\rho\rho}$ is   
\begin{equation}
\chi_{\rho\rho}=\chi_{L L}+\chi_{R L}+\chi_{L R}+\chi_{R R}
        +\chi_{L \pi}+\chi_{R \pi}+\chi_{\pi \pi}+\chi_{\pi L}+\chi_{\pi R}.
\end{equation}    
To calculate the correlators $\chi_{\eta\gamma}$, we use again Eq.\ (\ref{suscept}), 
this time for $\eta,\gamma\in \{L,R,\pi^*\}$. 
In addition we need the commutators
\begin{equation}
\langle [\rho_{\pi^*}({\bf q},0), \rho_{\pi^*}(-{\bf q},0)]\rangle= 
  \sum\limits_{{\bf k}\sigma}( \langle n_{\pi^*\sigma}({\bf k}+{\bf q})\rangle 
            - \langle n_{\pi^*\sigma}({\bf k})\rangle)\approx
       -2{\cal N}_{3d}(\epsilon_F) {\cal V} {\bf v}_F^{3d}\cdot{\bf q},
\end{equation} 
\begin{equation}       
\langle [[\rho_{\pi^*}({\bf q},t),H_{0 \pi^*}],
                   \rho_{\eta}({-\bf q},0)]\rangle 
            \approx-{\bf v}_F^{3d}\cdot{\bf q}
	       \langle [\rho_{\pi^*}({\bf q},t),\rho_{\eta}({-\bf q},0)]\rangle ,
 \end{equation}
\begin{eqnarray}    
&&\langle [[\rho_{\pi^*}({\bf q},t),H_{d-\pi^*}],
                   \rho_{\eta}({-\bf q},0)]\rangle =\nonumber\\
  && \qquad \qquad  -2{\cal N}_{3d}(\epsilon_F)  {\bf v}_F^{3d}\cdot{\bf q} \frac{V}{N}{\cal V}
   \langle[\rho_L({\bf q},t)+\rho_R({\bf q},t),\rho_{\eta}({-\bf q},0)] \rangle ,
  \end{eqnarray}
\begin{equation}  
 [ \rho_L({\bf q},t),H_{ d-\pi^* } ]=
    -[\rho_R({\bf q},t),H_{ d-\pi^* } ]=
 \frac{V }{N} \frac{LN_{\perp}q_{\parallel}}{\pi}\rho_{\pi^* }({\bf q},t).
\end{equation}
where $N_{\perp}$ is the number of chains , 
${\cal N}_{3d}(\epsilon_F)$  is the density of states per spin and per volume at the Fermi surface, and 
${\cal V} $ is the volume.
After some calculations we obtain the following expressions for the charge and spin 
 susceptibility: 
\begin{equation}
\label{rpa1}
\chi_{\rho\rho}=\frac{\chi^{1d}_{\rho\rho}-2\chi^{1d}_{\rho\rho}V\chi_{\pi^*}/N
               +\chi_{\pi^*}}
               {1-\chi^{1d}_{\rho\rho}V\chi_{\pi^*}V/N^2},
\end{equation}
\begin{equation}
\label{rpa2}
\chi_{\sigma\sigma}=\chi_{\sigma\sigma}^{1d}+\chi_{\pi^*},
\end{equation}
where $\chi^{1d}_{\rho\rho}$ ($\chi^{1d}_{\sigma\sigma}$) is the charge 
(spin) susceptibility of the Luttinger model 
given by Eq.\ (\ref{chi-charge}) (Eq.\ (\ref{chi-spin})) with $Na=LN_{\perp}$,
and $\chi_{\pi^*}$ is the  susceptibility 
corresponding to the free  $\pi^*$ electrons. The results  (\ref{rpa1}) and (\ref{rpa2}) are
  equivalent to the RPA results obtained 
for  a similar model with an  electron-electron interaction characterized by a small momentum transfer, without
applying the bosonization procedure.

\subsection{Dispersion of the charge excitations }
The charge excitation
spectrum is given by equation 
\begin{equation}
\label{charge-spectrum}
\frac{1}{\chi_{\rho\rho}({\bf q},\omega)}=0,
\end{equation}
which implicitly defines the dispersion relation $\omega=\omega ({\bf q})$.
In the following we analyze
the dispersion of the charge excitations: 
We want to see how the linearized dispersion of a
clean one-dimensional system is modified due to the presence of the 
three-dimensional electrons. In this case Eq.\ (\ref{charge-spectrum}) 
is leads  to 
\begin{equation}
\label{dispeq}
1-\chi^{1d}_{\rho\rho}(\omega,q_{\parallel})
   V\chi_{\pi^*}(\omega,{\bf q})V/N^2=0,
\end{equation}
where $\chi^{1d}_{\rho\rho}(\omega,q_{\parallel})$ is the dynamical susceptibility 
of the Luttinger model model, compare Eq.\ (\ref{chi-charge}),
\begin{eqnarray}
\label{chi1}
\chi^{1d}_{\rho\rho}(\omega,q_{\parallel})&=&
   2Na{\cal N}_{1d}\frac{g_cv_F}{v_c}\;
   \frac{-(q_{\parallel}v_c)^2}{\omega^2-(q_{\parallel}v_c)^2},
\end{eqnarray}  
and $\chi_{\pi*}^0(\omega,{\bf q})$
is the dynamical susceptibility of  the free three-dimensional fermionic system,
\begin{eqnarray}
\label{chi3}
\chi_{\pi*}(\omega,{\bf q})&=&
   2{\cal N}_{3d}{\cal V}\left\langle 
   \frac{{\bf v}_F^{3d} \cdot {\bf q}}{\omega+{\bf v}_F^{3d} \cdot {\bf q} }\right\rangle_{FS},
\end{eqnarray}
where ${\cal N}_{3d}$, ${\bf v}_F^{3d}$ are the 
density of states per spin and  per volume and the  velocity of the
$\pi^*$ electrons at the Fermi surface and $\langle ...\rangle_{FS}$ is a Fermi surface average.
We separate the three-dimensional momentum 
${\bf q}$ in two components, one
parallel to the one-dimensional chain, $q_{\parallel}$, and the other 
perpendicular to it, ${\bf q}_{\perp}$. If we insert Eqs.\ (\ref{chi1}) and (\ref{chi3})
 into 
Eq.\ (\ref{dispeq}), 
the dispersion is found as the solution of the following equation:
\begin{equation}
\omega^2-(q_{\parallel}v_c)^2
\left[1-4 V^2 {\cal N}_{1d}{\cal N}_{3d}\frac{a{\cal V}}{N}\frac{g_c v_F}{v_c}
    \left\langle\frac{{\bf v}_F^{3d}\cdot {\bf q}}
      {\omega+{\bf v}_F^{3d}\cdot {\bf q}}\right\rangle_{FS}\right]=0.
\end{equation}
In the limit ${\omega \ll }{\bf v}_F^{3d}\cdot {\bf q}$ the three dimensional charge susceptibility can be approximated
by its static value
\begin{eqnarray}
\nonumber
 \chi_{\pi*}(\omega,{\bf q})& \approx & \chi_{\pi*}(0,{\bf q})=2{\cal N}_{3d} {\cal V}
 \end{eqnarray}
and the dispersion of the one-dimensional charge excitations is given by 
\begin{eqnarray} 
 \omega &=&\pm q_{\parallel}v_c \sqrt{1-4 V^2{\cal N}_{1d}{\cal N}_{3d}\frac{a{\cal V}}{N}\frac{g_c v_F}{v_c}}.
 \end{eqnarray}
In this case the dispersion maintains its one-dimensional linear characteristic. The only effect of the
coupling to the three-dimensional environment is a  renormalization
of the Luttinger parameters
\begin{equation}
v_c \quad \longrightarrow \quad \tilde{v}_c=
    v_c \sqrt{1-4 V^2{\cal N}_{1d}{\cal N}_{3d}\frac{a{\cal V}}{N}\frac{g_c v_F}{v_c}}\, ,
\end{equation}
\begin{equation}
g_c \quad \longrightarrow \quad \tilde{g}_c=
    g_c \left(\sqrt{1-4 V^2{\cal N}_{1d}{\cal N}_{3d}\frac{a{\cal V}}{N}\frac{g_c v_F}{v_c}}\right)^{-1},
\end{equation}
which is equivalent to a decrease of the interaction strength in the charge
channel.\\
In the opposite limit, $\omega \gg {\bf v}_F^{3d}\cdot {\bf q}$ we can  expand
 in powers of 
${\bf v}_F^{3d}\cdot {\bf q}/\omega$, and obtain 
\begin{equation}
\left\langle\frac{{\bf v}_F^{3d}\cdot{\bf q}}{\omega+{\bf v}_F^{3d}\cdot {\bf q}}\right\rangle_{FS}
    \approx
    \frac{1}{3}\frac {(v_F^{3d})^2(q_{\parallel}^2+{\bf q}_\perp^2)}{\omega^2} .
\end{equation}
In this situation  the dispersion is given by the equation    
\begin{equation}    
\omega^4-\omega^2(q_{\parallel}v_c)^2+(q_{\parallel}v_c)^2V^2 {\cal N}_{1d}{\cal N}_{3d}\frac{a{\cal V}}{N}\frac{4g_c v_F}{3v_c}
    (v_F^{3d})^2(q_{\parallel}^2+{\bf q}_\perp^2)=0,
\end{equation}
with the following solutions:
\begin{equation}    
\omega^2_{+}=(q_{\parallel}v_c)^2\left[1-V^2 {\cal N}_{1d}{\cal N}_{3d}\frac{a{\cal V}}{N}\frac{4g_c v_F}{3v_c}
                     \frac{(v_F^{3d})^2(q_{\parallel}^2+{\bf q}_\perp^2)}
		     {(q_{\parallel}v_c)^2}\right],
\end{equation} 
\begin{equation}    
\omega^2_{-}=V^2 {\cal N}_{1d}{\cal N}_{3d}\frac{a{\cal V}}{N}\frac{4g_c v_F}{3v_c}
                     (v_F^{3d})^2(q_{\parallel}^2+{\bf q}_\perp^2).
 \end{equation}  
In the weak coupling limit the
solution $\omega^2_{-}$  is not compatible with the assumöption   $\omega \gg {\bf v}_F^{3d}\cdot {\bf q}$.  
To analyze the solution $\omega^2_{+}$ we must  distinguish between
 two limiting  cases. 
If the parallel component of the momentum is much larger
than the perpendicular component $q_{\parallel}\gg  q_{\perp} $,
then  again the effect of the tree-dimensional 
electrons is only to renormalize the Luttinger parameters and  
the dispersion remains approximately,
linear
\begin{equation}    
 \omega_+ 
	\approx \pm q_{\parallel}v_c
	\sqrt{1-
	\left(\frac{v_{F \parallel }^{3d}}{v_c}\right)^2
	\frac{4g_cv_F}{3v_c}\frac{a{\cal V}}{N}V^2{\cal N}_{1d}{\cal N}_{3d}\;}.
\end{equation}
A more  interesting situation arises  in the opposite  limit,  
$q_{\parallel}\ll q_{\perp} $, when the charge
dispersion looses its linearized characteristic and 
the one dimensional modes are  mixed with the three-dimensional
modes through the $q_{\perp}$ component, i.e.
\begin{equation}    
 \omega_+ \approx \pm q_{\parallel}v_c
	\sqrt{1-\frac{(v_{F \perp }^{3d}q_{\perp})^2}
	{(q_{\parallel}v)^2}
	\frac{4g_cv_F}{3v_c}\frac{a{\cal V}}{N}V^2{\cal N}_{1d}{\cal N}_{3d}}= 
	\pm v_c \sqrt{q_{\parallel}^2-\alpha q_{\perp}^2\;},
\end{equation}
with 
$$\alpha=\frac{(v_{F \perp }^{3d})^2}
	{v_c^2}
	\frac{4g_cv_F}{3v_c}\frac{a{\cal V}}{N}V^2{\cal N}_{1d}{\cal N}_{3d}
	\approx \frac{V^2}{\epsilon_{F}^{1d}\epsilon_{F}^{3d}}
	\approx 10^{-2}...10^{-1}$$
where we have assumed that $V$ is roughly one order of magnitude smaller than
the Fermi energies.
\section{Discussion}
\label{discussionVO2}
\begin{figure}
\label{cosima-picture}
\centerline{\psfig{figure=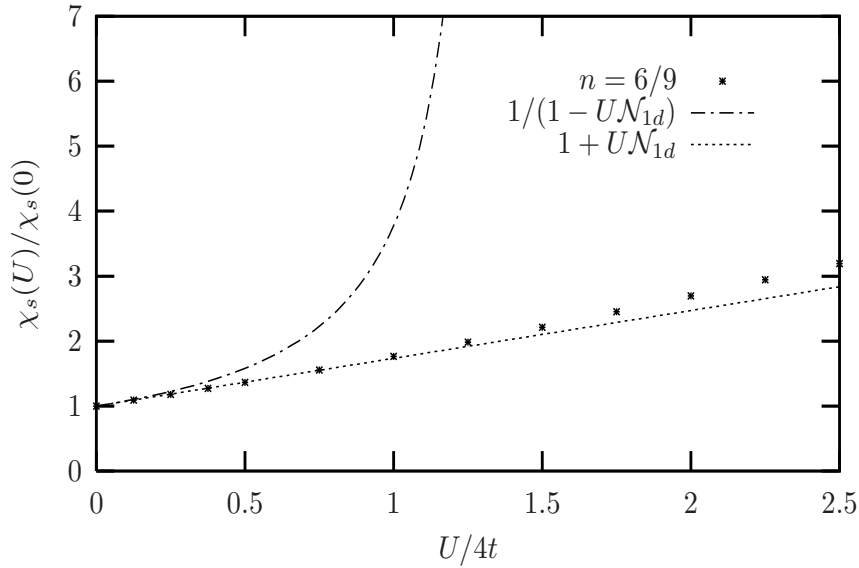,width=12cm,height=7.5cm}}
\caption{The  spin susceptibility of the one-dimensional Hubbard model. 
Within RPA one finds $\chi_{s} (U)=\chi_{s} (0)/(1-U{\cal N}_{1d}) $, where ${\cal N}_{1d}=(\pi v_F)^{-1}$.
 The data points are numerical
results \cite{cosimap} obtained with exact diagonalization of a Hubbard chain with $6$ electrons on $9$ lattice
sites ($n$ is the electron density).}
\end{figure}

In order to see how   the three-dimensional electrons 
affect  the  electrons   in the one-dimensional $d_{\parallel}$ band,
 we have  calculated the 
charge and the spin susceptibility of the system;  the coupling is 
introduced as a local Coulomb interaction of  strength $V$. 
We found  that the spin susceptibility of  the $d_{\parallel}$ electrons
 remains unchanged  in  the presence of the $\pi^{*}$ electrons.
This seems to contradict \cite{Mott75}, where from the spin susceptibiliy in the metallic phase a
smaller $U_{\textrm{eff}}\approx 0.13$ eV has been estimated. The apparent contradiction is
easily solved, however, when calculating the spin susceptibility in a $1d$ Hubbard model numerically, and
comparing with the RPA estimate. Fig.\ \ref{cosima-picture} cleary demonstrates that for a Stoner
enhancement of about $\chi/\chi_0 \approx 3$ the RPA drastically underestimates the interaction
strenght which is necessary to obtain such an enhencement. 

\nop{Numerically, 
 the spin susceptibility has been determinated through 
\begin{equation}
\chi^{-1}_{\sigma}=N[E(n,M)-E(n,M-1)]=
N[E_{\textrm{pbc}}(n,M)-E_{\textrm{abc}}(n,M)],
\end{equation}
where $N$ is the numer of lattice sites, $n$ is the electron density,
 $M$ is the magnetization and pbc(abc) stands for periodic (antiperiodic)
boundary conditions.}

Concerning  the charge susceptibility, we found that it  is changed in  
 the presence of the $\pi^*$ electrons. 
To make the effect more visible, we  analyzed the dispersion of the 
collective charge excitations.
When the dynamics of the $\pi^*$ electrons is fast compared to the $d_{\parallel}$
electrons, 
we found that the major effect of the $\pi^*$ electrons is to 
renormalize the charge velocity. 
When the Fermi velocity of the $d_{\parallel}$ and $\pi^*$ electrons are comparable in size, the
dynamics of the  $\pi^*$ electrons is on comparable time scales. In this limit  collective modes
can mix, so that the  $d_{\parallel}$ electrons loose their one-dimensional character.


\chapter{Summary and outlook}
\label{summary}
In this work we have investigated the low energy properties of different 
one-dimensional fermionic lattice models using the bosonization technique.
We have attached much importance to a proper consideration of the Klein factors
which are  neglected or inaccurately treated in the majority of the
literature.

In order to deal with the nonlinear terms,
arising from the dimerization or similar scattering processes,
we used the self-consistent harmonic
approximation (SCHA), a variational method where the nonlinearities are replaced
by a  quadratic potential. 
We have constructed an extension of the SCHA in order to account for the existence
of Klein factors in bosonized Hamiltonians.

As a first application,  
we have investigated  a model of spinless fermions with nearest-neighbor
interaction and a modulated
hopping. For the infinite system, both, the value of the Luttinger parameter $g_c = 2$ where
the transition from a gapless to a gapped phase takes places,
and the exponent $1/(2-g)$ that characterizes the opening of the gap for small dimerization, are correctly
obtained within the SCHA. 
In the thermodynamic limit  replacing the Klein
operators by a c-number is justified. When considering a finite system
a more careful treatment is necessary. In this case the Klein Hamiltonian 
 can be mapped onto a Mathieu equation which can be solved analytically in
certain limits.
Within our approach it turns out that
the effect of the Klein factors is negligible 
as long as the dimerization gap is larger than the
finite-size gap, $v_F/L$.

In addition, the finite-size formalism allows  calculation of  the size-dependence
of the Drude weight.
The Drude weight   reflects the sensitivity of the system
with respect to a change of boundary conditions and is related to the properties of
the current operator $J$ in the bosonized version of the Hamiltonian.
In a finite system with an energy gap $\Delta$ the Drude weight is expected to be nonzero
but exponentially small, $D \sim \exp(- {\rm const} \cdot L \Delta)$.
In our extended version of the SCHA  we confirm the exponential behavior.

As an application to fermions with spin, we studied 
the role of Klein factors for the Peierls-Hubbard  and the  ionic Hubbard
model. In the bosonized version of these models two kinds of bosonic fields have
to be introduced which are associated with the spin and the charge degrees of
freedom, respectively. Nonlinear terms arising from backward and Umklapp
scattering lead to the opening of spin  or charge gaps depending on the model
parameters. Using the SCHA we have calculated the corresponding gap parameters 
$\Delta_c$ and $\Delta_s$ as a function of the strength of the perturbation 
(the dimerization $u$ in the case of the Peierls-Hubbard model, and the ionic
potential $\Delta$ in the case of the ionic Hubbard model). In both cases we
found a transition from  a Mott insulator phase with vanishing or very small spin
gap to a band insulator phase with almost equal spin and charge gaps. The
opening of the spin gap with increasing perturbation strength occurs according
to a power law with universal exponents, $2/3$ in the case of the
Peierls-Hubbard model, and $2$ in the case of the ionic Hubbard model.
The main difference between the two models, however, is that in the
Peierls-Hubbard model the transition between the Mott and the Peierls phase is
continuous,  while for the ionic Hubbard model a first order transition is
observed, with a jump in $\Delta_c$ and  $\Delta_s$. As to the ionic Hubbard
model we could not confirm the scenario proposed by Fabrizio et al.\ \cite{Fabrizio99} who argued in favor
of two phase transitions.   However, in our bosonization approach we are strictly
limited to weak coupling with regard to the Hubbard interaction $U$ and to the
ionic potential $\Delta$, while Fabrizio   et al.\ \cite{Fabrizio99} used a strong coupling approach in order
to obtain the second transition. 
While there is some evidence for the two phase transitions scenario from DMRG
calculations in the case of a strong ionic potential ($\Delta/t=0.5$ in Ref.
\cite{Kampf03}  and $\Delta/t=20$ in Ref.
\cite{Manmana03}), it is not clear whether this scenario persists to small values of 
$\Delta$, e.g.\ $\Delta/t=0.1$ that we have used in our calculations.

Concerning  the role of Klein factors one has to 
recall that they  
do not commute with the total spin and charge currents,
$J_{c,s}$, and therefore Klein factors and current operators 
 cannot acquire a fixed value at the same time.
In the thermodynamic limit  the ground state of the gapped
systems is a superposition of many states with different $J_{c}$ and $J_{s}$. 
In this situation it
is possible  to choose a fixed phase for the Klein operators. The bosonized
Hamiltonian is then the conventional sine-Gordon-like Hamiltonian. 
In the framework of our variational scheme we decoupled the bosonic Hamiltonian
 from the Klein Hamiltonian.
For finite size systems the eigenvalues of
the Klein factors  are replaced by their  expectation values with
respect to the ground state of the  trial Hamiltonian, and the resulting bosonic
Hamiltonian is also of a sine-Gordon type.
The Klein Hamiltonian   is of the form of a
tight-binding Hamiltonian for a particle moving on a $2d$ lattice in a harmonic
potential. The strength of the potential is inversely  proportional to the length of
the systems, i.e.\ for small systems the potential  contribution to the
ground state energy is relevant.
 On the other hand, 
 when the size of the system approaches the thermodynamic limit, the
 confining potential becomes irrelevant.
  
In an attempt to describe more realistic systems  we considered
a model where
 one-dimensional electrons are coupled to a three-dimensional 
conduction band via a local interaction, having in mind an application to 
embedded Peierls systems like vanadium dioxide, VO$_2$.
In order to see how  the one-dimensional electron system
is modified in  the presence of the  three-dimensional environment, 
we calculated the charge and
the spin susceptibility, and analyzed the collective excitations within RPA.
We found that
the spin excitations remain 
unaffected.
In the charge channel, the collective modes of the $1d$ and $3d$
subsystems are coupled. If the Fermi velocity of the $3d$ system is 
much larger than in the $1d$ system, the $3d$ electrons react practically 
instantaneously to changes in the $1d$ chains. In this case the $3d$ electrons 
simply renormalize or ``screen" the Luttinger parameters in the chains.
When $v_F^{3d}\approx v_F^{1d}$ the situation becomes more complex,
 since the dynamics of screening becomes
relevant.
In this case the charge excitations loose their 
one-dimensional
character and become three-dimensional. 

The next step beyond the current work would be  to
derive an effective model of one-dimensional chains coupled via the $3d$
environment, and to use the methods developed and tested in the first chapters 
to calculate the phase diagram. A major difficulty to overcome is the fact that dynamic
screening leads to retarded interactions which in general cannot be cast into a
Hamiltonian formalism, such  as the one used in the simple model calculations, but
rather requires a path integral formulation for the bosonized systems. While it
is more or less standard to represent a bosonic theory in the path integral
language, it is not clear how the Klein factors in this
formulation can be incorporated.



\appendix

\chapter{Derivation of the gap equations}
\label{finSCHA}
In  this appendix we derive the SCHA equations for 
 the Peierls-Hubbard model. The derivation for the ionic Hubbard model is
similar.
 For a finite  system the Klein factors cannot be replaced by 
 their eigenvalues, but we can 
 choose a trial Hamiltonian where the bosonic fields are decoupled from the Klein factors as
 follows
\begin{equation}
\label{A1}
H_{\rm tr}=H_{\rm tr}^{\Delta_c}+H_{\rm tr}^{\Delta_s}+H_{\rm tr}^{B_{cs}B_cB_s},
\end{equation}
with the usual bosonic part, in which  the nonlinear terms 
are replaced by a quadratic form
\begin{equation}
\label{finHtrial}
H_{\rm tr}^{\Delta_c}+H_{\rm tr}^{\Delta_s}= \sum_{\alpha=c,s}\int_0^L
    \frac{dx}{2\pi}\left\{\frac{v_\alpha}{g_\alpha}(\partial_x\phi_\alpha)^2 +
      v_\alpha g_\alpha(\partial_x\theta_\alpha)^2 +
      \frac{\Delta_\alpha^2}{v_\alpha g_\alpha} \phi_\alpha^2 \right\}
\end{equation}
and a Hamiltonian containing the Klein factors
\begin{eqnarray}
\label{finHKlein}
H_{\rm tr}^{B_{cs}B_cB_s}  & = & i B_{cs}L(F_{R\uparrow}^+ F_{L\uparrow} + F_{R\downarrow}^+
 F_{L\downarrow})
+ B_cL F_{R\uparrow}^+F_{R\downarrow}^+F_{L\downarrow}F_{L\uparrow} \nonumber \\
 &   &  + B_sL F_{R\uparrow}^+F_{L\downarrow}^+F_{R\downarrow}F_{L\uparrow} +{\rm h.c.}
  \nonumber \\   &   &+ \frac{\pi}{4L}(v_c g_c J^2_c + v_s g_s J^2_s).
\end{eqnarray}
Taking the ground state of $H_{\rm tr}$ as a variational state for the full
Hamiltonian we obtain an upper bond $\tilde{E}$ for the ground state 
\begin{eqnarray}
\label{finvarenergy}
\tilde E & = &
      \langle H_0 \rangle_{\rm tr} + \langle H_1 \rangle_{\rm tr}
     +\langle H_2 \rangle_{\rm tr} + \langle H_{\rm{Peierls}} \rangle_{\rm tr}, 
     \nonumber\\
\frac{\tilde E}{L} & = & \frac{E_{\rm tr}}{L} - \sum_{\alpha=c,s}
\frac{\Delta_\alpha^2}{2\pi v_\alpha g_\alpha}
  \langle\phi_\alpha^2 \rangle_{\rm tr} -
B_c\frac{\partial e^0_{\rm{tr}}}{\partial B_c}-
L\frac{\partial e^0_{\rm{tr}}}{\partial B_s}-
B_{cs}\frac{\partial e^0_{\rm{tr}}}{\partial B_{cs}}
\nonumber\\
 & & +\tilde{U}\; {\rm e}^{-4 \langle \phi^2_c \rangle_{\rm tr}}
      \frac{\partial e^0_{\rm{tr}}}{\partial B_c}
     +\tilde{U}\; {\rm e}^{-4 \langle \phi^2_s \rangle_{\rm tr}}
      \frac{\partial e^0_{\rm{tr}}}{\partial B_s}+
      \tilde{u} \; {\rm e}^{- \langle \phi^2_c \rangle_{\rm tr}- 
                \langle \phi^2_s \rangle_{\rm tr}}
      \frac{\partial e^0_{\rm{tr}}}{\partial B_{cs}}.
\end{eqnarray}
Here $E_{\rm{tr}}$ is the ground state energy of $H_{\rm{tr}}$ given by Eq.\ (\ref{A1}), and
 $e^0_{\rm{tr}}=E^0_{\rm{tr}}(B_{cs},B_c,B_s)/L$,  where 
 $E^0_{\rm{tr}}(B_{cs},B_c,B_s)$ is the ground state energy of the Klein Hamiltonian 
 $H_{\rm tr}^{B_{cs}B_cB_s}$ given by Eq.\ (\ref{finHKlein}) . In addition, 
 $\tilde{u}=tu/\pi a$, and $\tilde{U}=UL/[(2\pi a)^2N]$.
We have used  the Feynmann-Hellmann theorem \cite{Feynmann,Hellmann}
 to express the expectation values of the Klein factors in terms of the derivatives of $e^0_{\textrm{tr}}$,
   the ground state
energy $E^0_{\textrm{tr}}$  devided by system length, with respect to the $B$'s:
 \begin{eqnarray}
 \frac{\partial e^0_{\textrm{tr}}}{\partial B_c}&=&
 \langle F_{R\uparrow}^+F_{R\downarrow}^+F_{L\downarrow}F_{L\uparrow} 
  +\rm{h.c.}\rangle,\\
 \frac{\partial e^0_{\textrm{tr}}}{\partial B_s}&=&
 \langle  F_{R\uparrow}^+F_{L\downarrow}^+
 F_{R\downarrow}F_{L\uparrow} +\rm{h.c.}\rangle,\\
 \frac{\partial e^0_{\textrm{tr}}}{\partial B_{cs}}
   &=&i\langle  F_{R\uparrow}^+ F_{L\uparrow} + F_{R\downarrow}^+
 F_{L\downarrow} \rangle +\rm{h.c.}.
 \end{eqnarray}
The  parameters $\Delta_c$, $\Delta_s$, $B_c$, $B_c$ and  $ B_{cs}$ 
are obtained from the condition that $\tilde{E}$
 should be minimized.
For the parameters $B_c$, $B_s$ and $B_{cs}$ the minimum condition
 for $\tilde{E}$
yields the following equations:
\begin{eqnarray}
\label{DerivBcs}
    \lefteqn{  \left[ \tilde{u}
     {\textrm{e}}^{-\langle \phi^2_c \rangle -\langle \phi^2_s\rangle}-
       B_{cs} \right]
                \frac{\partial^2 {\textrm{e}}^0_{\rm{tr}}}{\partial B^2_{cs}}+
        \left[\tilde{U}
     {\textrm{e}}^{-4\langle \phi^2_c\rangle}-B_c\right]
        \frac{\partial^2  e^0_{\rm{tr}}}{\partial B_{cs}\partial B_c}
    }\nonumber\\
     & &+\left[ \tilde{U}
    {\textrm{e}}^{-4\langle \phi^2_s\rangle}-B_s\right]
           \frac{\partial^2  e^0_{\rm{tr}}}{\partial B_{cs}\partial B_s}=0,
\end{eqnarray}  

\begin{eqnarray}
\label{DerivBc}
    \lefteqn{  \left[\tilde{u}
     {\textrm{e}}^{-\langle \phi^2_c \rangle -\langle \phi^2_s\rangle}-
              B_{cs}\right]
                \frac{\partial^2  e^0_{\rm{tr}}}{\partial B_c\partial B_{cs}}+
        \left[\tilde{U}
    {\textrm{e}}^{-4\langle \phi^2_c\rangle}-B_c\right]
              \frac{\partial^2  e^0_{\rm{tr}}}{\partial B^2_c}
    }\nonumber\\
     & &+\left[\tilde{U}
    {\textrm{e}}^{-4\langle \phi^2_s\rangle}-B_s\right]
            \frac{\partial^2  e^0_{\rm{tr}}}{\partial B_{s}\partial B_c}=0,
\end{eqnarray}    

\begin{eqnarray}
\label{DerivBs}
    \lefteqn{\left[\tilde{u}
     {\textrm{e}}^{-\langle \phi^2_c \rangle -\langle \phi^2_s\rangle}-
     B_{cs}\right]
          \frac{\partial^2  e^0_{\rm{tr}}}{\partial B_{s}\partial B_{cs}}+
        \left[\tilde{U}
    {\textrm{e}}^{-4\langle \phi^2_c\rangle}-B_c\right]
                \frac{\partial^2  e^0_{\rm{tr}}}{\partial B_{s}\partial B_c}
    }\nonumber\\
     & &+\left[\tilde{U}
    {\textrm{e}}^{-4\langle \phi^2_s\rangle}-B_s\right]
      \frac{\partial^2  e^0_{\rm{tr}}}{\partial B^2_s}=0,
\end{eqnarray}
with the solutions
\begin{eqnarray}
B_{cs}&=&\tilde{u}
     {\textrm{e}}^{-\langle \phi^2_c \rangle -\langle \phi^2_s\rangle},\\
B_c&=&\tilde{U}
    {\textrm{e}}^{-4\langle \phi^2_c\rangle},\\
B_s&=&\tilde{U}
    {\textrm{e}}^{-4\langle \phi^2_s\rangle}.
\end{eqnarray} 
Minimizing $\tilde{E}$ with respect to 
 $\Delta_c$ and $\Delta_s$ yields
\begin{eqnarray}
\label{gequation1}
  \frac{\Delta_c^2}{2\pi v_c g_c } & = & - 4 B_c \frac{\partial e^0_{\rm{tr}}}{\partial B_c}
 - B_{cs} \frac{\partial e^0_{\rm{tr}}}{\partial B_{cs}}, \\
\label{gequation2}
  \frac{\Delta_s^2}{2\pi v_s g_s } & = & - 4 B_s \frac{\partial e^0_{\rm{tr}}}{\partial B_s}
 - B_{cs} \frac{\partial e^0_{\rm{tr}}}{\partial B_{cs}},
\end{eqnarray}
where we have used that 
\begin{equation}
\frac{\partial e^0_{\rm{tr}}}{\partial (\Delta^2_{\alpha})}=
  \frac{\langle \phi^2_{\alpha} \rangle_{\rm{tr}}}{2\pi v_{\alpha}g_{\alpha}}.
\end{equation}
These equations are identical  to the equations  obtained
 in the thermodynamic limit, see Eqs.\ (\ref{gap-equation1})-(\ref{gap-equation2}),
 if  the minimum of $e(k_{\uparrow},k_{\downarrow})$, given by Eq.\ (\ref{energy}), 
is  replaced with $e^0_{\rm{tr}}$, the ground state energy  $E^0_{\rm{tr}}$ of the Klein
Hamiltonian divided by system length.

\chapter{The trial Hamiltonian in terms of bosonic operators}
\label{H-bosop}
The bosonic part of the trial Hamiltonian for spinless
fermions with static dimerization in the thermodynamic limit reads 
\begin{eqnarray}
  \label{eq:HtrialApp}
      H_{\rm{tr}}^\Delta&=&\int_0^L \frac{\rm{d}x}{2\pi}
      \left\{\frac{v}{g}(\partial_x\phi)^2 + vg(\partial_x\theta)^2
      + \frac{\Delta^2}{v g} \phi^2(x)\right\}.
\end{eqnarray}
After  Fourier transformation
$$\phi(x)=\frac{1}{\sqrt{L}} \sum_{q\ne 0}\textrm{e}^{iqx}\phi (q),\qquad
  \theta(x)=\frac{1}{\sqrt{L}} \sum_{q\ne 0}\textrm{e}^{iqx}\theta (q),$$
it can be rewritten as
\begin{equation}
H_{\rm{tr}}^\Delta=\frac{1}{2\pi}\sum_{q> 0}
  \left[ \left(\frac{v}{g}q^2+\frac{1}{vg}\Delta^2\right)\phi(q)\phi(-q)
  +vgq^2\theta(q)\theta(-q)\right].
\end{equation}
If we express  $\phi(\pm q)$ and $\theta (\pm q)$ in terms of the boson operators
\begin{eqnarray}
\phi(q)&=&
\sqrt{\frac{\pi}{2q}}(b^+_{Lq}-b_{Rq})\textrm{e}^{-aq/2},\\
\theta(q)&=&
\sqrt{\frac{\pi}{2q}}(b^+_{Lq}+b_{Rq})\textrm{e}^{-aq/2},\\
\phi(-q)&=&
\sqrt{\frac{\pi}{2q}}(b_{Lq}-b^+_{Rq})\textrm{e}^{-aq/2},\\
\theta(-q)&=&
\sqrt{\frac{\pi}{2q}}(b_{Lq}+b^+_{Rq})\textrm{e}^{-aq/2},
\end{eqnarray} 
where $q>0$, the trial Hamiltonian becomes
\begin{eqnarray}
\label{k:HtrialApp}
H_{\rm{tr}}^\Delta
 &=&\sum_{q>0}\left[
   \left(  v_F+\frac{g_4}{2\pi }\right)q+\frac{1}{2q}\Delta^2\right]
   \left[b^+_{Lq}b_{Lq}+b^+_{Rq}b_{Rq}\right]\textrm{e}^{-aq}\nonumber\\ 
 & &-\sum_{q>0}\left[
    \frac{g_2}{2\pi}q+\frac{1}{2q}\Delta^2\right]
   \left[b^+_{Lq}b^+_{Lq}+b^+_{Rq}b^+_{Rq}\right]\textrm{e}^{-aq} 
   +\sum_{q>0}\left[\frac{\textrm{e}^{-aq}}{2q}\Delta^2 \right].
\end{eqnarray}
This Hamiltonian can be diagonalized with a Bogoliubov transformation of the 
following form:
\begin{eqnarray}
B_1(q)&=&\alpha b_{Rq}+\beta b^+_{Lq},\\
B_2(q)&=&\beta b_{Rq}+\alpha b^+_{Lq},
\end{eqnarray}
with $\alpha$ and $\beta$ parameters to be determined. 
The operators $B_1$ and $B_2$ are also bosonic operators, i.e.\ 
$$[B_1,B^+_1]=[B_2, B^+_2]=1 \quad \Rightarrow \quad \alpha^2-\beta^2 =1.$$
The $b_{\eta, q}$ operators in the Hamiltonian (\ref{k:HtrialApp}) 
are replaced by the 
new operators $B_1$ and $B_2$. The restriction that the Hamiltonian in
terms of  $B_1$ and $B_2$ has to be diagonal yields the second equation for the
parameters $\alpha$ and $\beta$ 
$$2\alpha \beta ( \gamma_4 q^2+\Delta^2/2)+
(\gamma_2 q^2 +\Delta^2/2)(\alpha^2+\beta^2)=1,$$
with $\gamma_4=v_F+g_4/2\pi v_F$ and $\gamma_2=g_2/2\pi v_F$. 
The diagonalized Hamiltonian is 
\begin{eqnarray}
H_{\rm{tr}}^\Delta
 &=&\sum_{q>0}\sqrt{v^2q^2+\Delta^2}\,\left(B^+_1B_1+B^+_2B_2\right)\textrm{e}^{-aq}
  \nonumber\\
  &+ &\sum_{q>0} \left[ \frac{\textrm{e}^{-aq}}{q}\Delta^2(1+\beta^2)+
  2\beta(\beta \gamma_4+\alpha \gamma_2)q \right]  
\end{eqnarray}
The last  sum gives a infinite constant which is removed  by the
normal ordering operation. At the end we obtain in the limit $a\rightarrow 0$
\begin{equation}
:H_{\rm{tr}}^\Delta :\; = 
  \sum_{q>0}\epsilon(q)\left(B^+_1B_1+B^+_2B_2\right)\quad \textrm{with} \quad
  \epsilon(q)=\sqrt{v^2q^2+\Delta^2}.
\end{equation}
 

\chapter{Quantum theory of Josephson junctions}
\label{josephson}
Here we wish to point out that the study of the Klein factors within 
the SCHA, for the spinless-fermions model,leads to a trial Hamiltonian 
which is well known from the quantum theory of Josephson junctions in the 
zero-damping limit \cite{tinkham}. Considering again Eq.\ (\ref{eq:HB}) and putting
$2p_J\rightarrow x$, we have 
\begin{equation}
H_{\textrm{tr}}^{B}\rightarrow -\frac{2\pi vg }{L}\frac{d^2}{dx^2}+2B\cos x
\end{equation} 
with periodic boundary conditions for the wave function, $\Psi(x+2\pi)=\Psi(x)$.
On the other hand, the Josephson junction Hamiltonian in standard notation 
\cite{tinkham} reads
\begin{equation}
\label{jos-eq}
H^{JJ}=-4E_c\frac{d^2}{d\varphi^2}-E_J\cos \varphi,
\end{equation}
\begin{equation}
\Psi(\varphi+2\pi)=\Psi(\varphi),
\end{equation}
where $\varphi$ is the order parameter phase difference across the junction, 
$E_J$ the Josephson coupling energy, and $E_c=e^2/2C$ ($C$ is the capacitance)
the charging energy. The correspondence is  obvious.

Note that the ``kinetic energy" in (\ref{jos-eq}) corresponds to $Q^2/2C$, 
where $Q=2eN$ is the $2e$ times the Cooper pair number, such that $N$
and $\varphi$ are conjugate variables, $[N,\varphi]=-i$. For large junctions,
$E_c\ll E_J$ (corresponding to large $L$), the variable 
$\varphi$ can be treated classically. 
On the other hand, for small junctions,  
$E_c\gg E_J$ (corresponding to small $L$), 
the appropriate eigenstates are particle, i.e.\ Cooper pair,  number eigenstates. These two limits correspond to two representations of the BCS wave functions, namely with a definite phase and with a definite number of Cooper pairs,
respectively \cite{tinkham}.

\chapter{Mathieu equation}
\label{mathieu}
The Mathieu equation in its canonical form \cite{Abramowitz} reads 
\begin{equation}
\frac{d^2 y}{dx^2}+\left(a-2q\cos (2x)\right)y=0.
\end{equation}
This is exactly equation (\ref{eq:HB}) if we identify
\begin{equation}
y(x)\longrightarrow \Psi (p_J),
\end{equation}
\begin{equation}
a \longrightarrow \frac{2LE}{\pi vg},
\end{equation}
\begin{equation}
q \longrightarrow \frac{2LB}{\pi vg}.
\end{equation}
It can be shown that there exists a  set of {\it
characteristic values } $a_r (q)$ 
which yield even periodic solutions and a  set of {\it
characteristic values } $b_r (q)$ which yield odd periodic
solutions, where $r=0,1,2,...$. If $q$ is real then the sets of {\it characteristic values } 
$a_r$ and $b_r$ have the following properties:
\begin{itemize}
\item[a)] The  characteristic values $a_r$ and $b_r$ are real
and distinct, if $q\ne 0$; $a_0<b_1<a_1<b_2<....., \; q>0$, and $a_r(q)$,
 $b_r(q)$  approaches  $r^2$ as $q$ approaches zero.
 \item[b)] A solution of the Mathieu equation associated with $a_r$ or $b_r$ 
 has $r$ zeros in the interval $0\leq x<\pi$.
\end{itemize}
In Ref. \cite{Meixner54} the asymptotic behavior of the characteristic values $a_r$ and
$b_r$ is derived. In particular for $q\gg 1$
\begin{equation}
b_{r+1}(q)-a_r(q)\approx 2^{4r+5}\sqrt{\frac{2}{\pi}}
   q^{r/2+3/4}{\textrm{e}}^{-4\sqrt{q}}\frac{1}{r!}
\end{equation}   
If we consider the case $r=0$, we can estimate the 
difference $E(\pi)-E(0)$ needed in Eq.\ (\ref{difD}). 
We identify $E(\pi)=b_1(q)q/B$ and $E(0)=a_0(q)q/B$ to obtain Eq.\ (\ref{Drude2}).
\begin{figure}[ht]
   \centerline{\psfig{figure=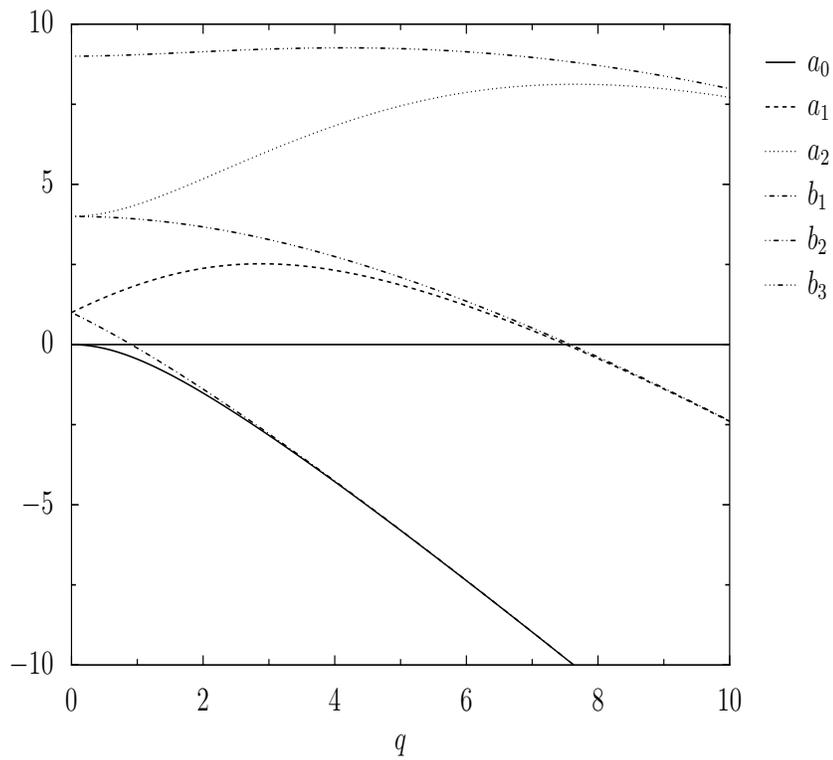,width=11cm,height=10cm}}
   \caption{The lowest eigenvalues  of the Mathieu equation  as function of 
    $q$. The $a$'s correspond to periodic and the $b$'s to anti-periodic solutions. With
    increasing $q$ the difference $b_r-a_{r-1}$ goes to zero.}
\end{figure} 

\chapter{The Klein Hamiltonian as a $2d$ tight binding model}
\label{KH-tbm}
\begin{figure}
   \centerline{\psfig{figure=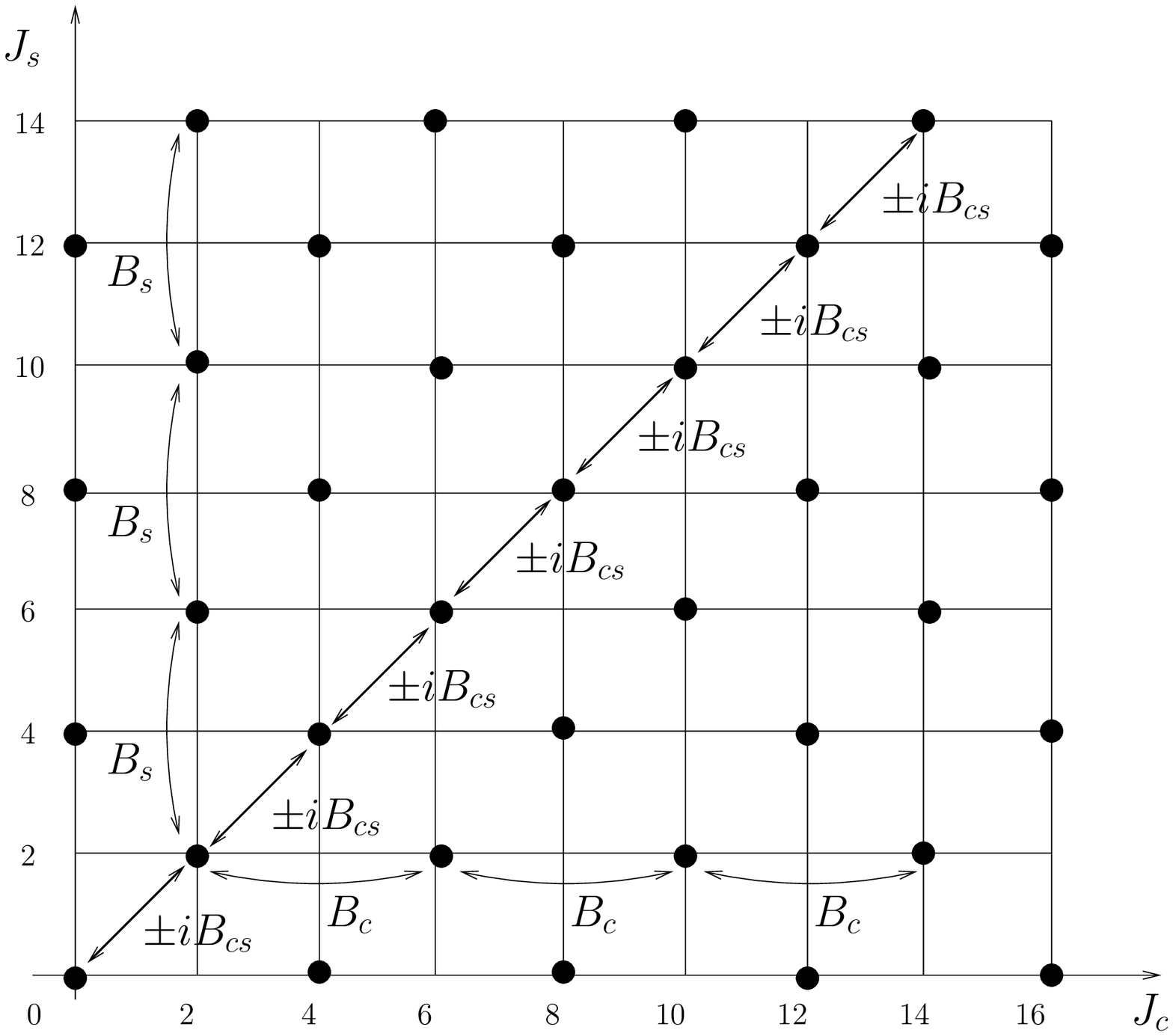,width=13cm,height=11cm}}
   \caption{Pictural representation of the hopping Hamiltonian (\ref{hop-ham}).}
\label{JcJs}
\end{figure}

In the following we 
represent  the Klein Hamiltonian of  the Peierls-Hubbard model,
\begin{eqnarray}
\label{HKleinC}
H_{\rm tr}^{B_{cs}B_cB_s}  & = & i B_{cs}L(F_{R\uparrow}^+ F_{L\uparrow} + F_{R\downarrow}^+
 F_{L\downarrow})
+ B_cL F_{R\uparrow}^+F_{R\downarrow}^+F_{L\downarrow}F_{L\uparrow} \nonumber \\
 &   &  + B_sL F_{R\uparrow}^+F_{L\downarrow}^+F_{R\downarrow}F_{L\uparrow}+{\rm h.c.}
 \nonumber \\
  &   &+ 
 \frac{\pi}{4L}(v_c g_c J^2_c + v_s g_s J^2_s),
\end{eqnarray}
in the basis of $J_c$ and $J_s$ in order to show explicitly that it has the form of a
$2d$ tight binding model for a particle in a harmonic potential. Using the sign convention for the Klein factors that we have chosen in 
section\ \ref{kleinfactors} one obtains 
\begin{eqnarray}
\label{hop-ham}
 H_{\rm tr}^{B_{cs}B_cB_s}  & = &iB_{cs}L
   \sum_{J_c,J_s}(-1)^{\frac{J_c}{2}}|J_c,J_s\rangle\langle J_c+2,J_s+2| \nonumber\\
   & +&iB_{cs}L\sum_{J_c,J_s}(-1)^{\frac{J_s}{2}}|J_c+2,J_s\rangle\langle J_c,J_s+2| \nonumber\\
   & +& B_sL \sum_{J_c,J_s}|J_c,J_s\rangle\langle J_c,J_s+4|
     +   B_cL \sum_{J_c,J_s}|J_c,J_s\rangle\langle J_c+4,J_s|+{\rm h.c.} \nonumber\\
  & +&\frac{\pi v_c g_c}{4L}\sum_{J_c,J_s}J_c^2|J_c,J_s\rangle\langle J_c,J_s|+
    	\frac{\pi v_s g_s}{4L}\sum_{J_c,J_s}J_s^2|J_c,J_s\rangle\langle J_c,J_s|. 
\end{eqnarray}
For $N_c=0$ and $N_s=0$, according to Eqs.\ (\ref{jc}) and (\ref{js})
the quantum numbers $J_c$ and  $J_s$ are even and $J_c\pm J_s$
are multiples of $4$. In  Fig.\ \ref{JcJs} the allowed hoppings
are displayed schematically.

\chapter{Analytical solution of the gap equations}
\label{an-solution}
In this appendix we present the solutions of the gap equations, both for the Peierls-Hubbard
model, Eqs.\ (\ref{gap-equation1})-(\ref{gap-equation2}), and for the ionic Hubbard model,
Eqs.\ (\ref{gapequation12})-(\ref{gapequation22}).
\section*{Peierls-Hubbard model}
In the case of the Peierls-Hubbard model, for non-zero dimerization the charge and spin mean field
parameters are given by 

\begin{equation}
\label{Ap-scha-eq1}
\frac{\Delta_c^2}{2\pi v_cg_c }=4B_{cs}+8B_c=
          \frac{4tu}{\pi} \; {\rm e}^{- \langle \phi^2_c \rangle_{\rm tr}}+
	  \frac{2U}{\pi^2} \; {\rm e}^{-4 \langle \phi^2_c \rangle_{\rm tr}},
\end{equation}
\begin{equation}
\label{Ap-scha-eq2}	 	 
\frac{\Delta_s^2}{2\pi v_sg_s }=4B_{cs}-8B_s=
           \frac{4tu}{\pi} \; {\rm e}^{- \langle \phi^2_c \rangle_{\rm tr}}
	  - \frac{2U}{\pi^2}\; {\rm e}^{-4 \langle \phi^2_s \rangle_{\rm tr}},
\end{equation}
where we set $a$ equal to one. If we insert   the analytical values for  
$\langle \phi^2_c \rangle_{\rm tr} $ and $\langle \phi^2_s \rangle_{\rm tr} $, given by 
Eqs.\ (\ref{phi2c})-(\ref{phi2c}), 
we obtain  the following set of equations:
\begin{equation}
\frac{\Delta^2_c}{2\pi v_cg_c}=
     \frac{4tu}{\pi }\left( \frac{\Delta_c}{\Delta_{0c}}\right)^{g_c/2}
                     \left( \frac{\Delta_s}{\Delta_{0s}}  \right)^{1/2}+
	    \frac{2U}{\pi^2}\left( \frac{\Delta_c}{\Delta_{0c}}  \right)^{2g_c},
\end{equation}

\begin{equation}
\frac{\Delta^2_c}{2\pi v_cg_c}=\frac{4tu}{\pi }\left( \frac{\Delta_s}{\Delta_{0s}}  \right)^{1/2}
                             \left( \frac{\Delta_c}{\Delta_{0c}}\right)^{g_c/2}-
			     \frac{2U}{\pi^2}\left( \frac{\Delta_s}{\Delta_{0s}}\right)^{2}.
\end{equation}
Introducing the following notation:
$\Delta_c/\Delta_{0c}=x$,  $\;\Delta_s/\Delta_{0s}=y$, 
$\;\Delta_{0c}/2\pi v_c g_c =A_c$,
$\;\Delta_{0s}/2\pi v_s g_s =A_s$,
$4tu/\pi  =\tilde{u}$ and $2U/\pi^2=\tilde{U}$ yields 
\begin{eqnarray}
\label{eqx1}
A_cx^2&=&\tilde{u}x^{g_c/2}y^{1/2}+\tilde{U}x^{2g_c},\\
\label{eqy1}
A_sy^2&=&\tilde{u}x^{g_c/2}y^{1/2}-\tilde{U}y^{2}.
\end{eqnarray}
Using  Eq.\ (\ref{eqy1}), we may  express $y$ as a function of $x$
\begin{equation}
\label{eqy2}
y=\tilde{u}^{2/3}x^{g_c/3}\left(\frac{1}{A_s+\tilde{U}}\right)^{2/3},
\end{equation}
which  we insert in  (\ref{eqx1}) to obtain 
\begin{equation}
\label{eqx2}
A_cx^2=\tilde{u}^{4/3}x^{2g_c/3}\left(\frac{1}{A_s+\tilde{U}}\right)^{1/3}+\tilde{U}x^{2g_c}.
\end{equation} 
We solve  this equation in two limits. For $\tilde{u}\rightarrow 0$ the solution
of Eq.\ (\ref{eqx2}) yields a constant 
\begin{equation}
\label{xconst}
x_0=\left(\frac{\tilde{U}}{A_c}\right)^{1/(2-2g_c)}.
\end{equation}
For $\tilde{u}/\tilde{U}\ll 1$ we make the ansatz 
\begin{equation}
\label{xu1}
x=x_0+C\tilde{u}^{\alpha}, 
\end{equation}
with $C$ and $\alpha$ to be determined. 
Since $C\tilde{u}^{\alpha}$ is only a small correction we keep it only in linear
order in the following and obtain
\begin{eqnarray}
\label{xu2}
&&A_c x_0^2 (1-\tilde{U} x_0^{2g_c-2})+
2C\tilde{u}^{\alpha}(A_cx_0-\tilde{U}g_cx_0^{2g_c-1})\nonumber\\
&&\qquad =\tilde{u}^{4/3}
    x_0^{2g_c/3}\left[1+\frac{2Cg_c\tilde{u}^{\alpha}}{x_0}
    \right]
    \left(\frac{1}{A_s+\tilde{U}}\right)^{1/3}.
\end{eqnarray} 
We consider this equation  as a polynomial equation in $\tilde{u}$ 
which has to be fulfilled for every $\tilde{u}$.  Neglecting the term 
$2Cg_c\tilde{u}^{\alpha}/x_0 \ll 1$  in the square brackets we obtain 
the following expressions:
\begin{equation}
\alpha=\frac{4}{3},
\end{equation}
\begin{equation}
C=\frac{x_0^{(3-4g_c)/3}}{2\tilde{U}(1-g_c)(A_s+\tilde{U})^{1/3}},
\end{equation}
Finally  we reinsert the solution for $x$ into equation (\ref{eqy2})
and obtain the result
\begin{eqnarray}
\frac{\Delta_c(u)-\Delta_c(0)}{\Delta_{0c}}&\approx&
    \left(\frac{4u}{\pi}\right)^{4/3}
   \left(\frac{\Delta_c(0)}{\Delta_{0c}}\right)^{(3-4g_c)/3}\nonumber\\
   && \quad \times \frac{\pi^2}{4(1-g_c)U}
    \left(\frac{\Delta_{0s}^2}{2\pi v_sg_s}+\frac{2U}{\pi^2}\right)^{-1/3},\\
\frac{\Delta_s(u)}{\Delta_{0s}}&\approx&
    \left(\frac{4u}{\pi}\right)^{2/3}
   \left(\frac{\Delta_c(0)}{\Delta_{0c}}\right)^{g_c/3}
    \left(\frac{\Delta_{0s}^2}{2\pi v_sg_s}+\frac{2U}{\pi^2}\right)^{-2/3}.
\end{eqnarray}
In the other limit, $\tilde{u}/\tilde{U}\gg 1$ we 
 take the logarithm of  Eq.\ (\ref{eqx2}),
\begin{equation}
\label{xU1}
\ln (A_c x^{2-2g_c})=\ln\left[ \tilde{U}
\left(1+\frac{\tilde{u}^{4/3}x^{-4g_c/3}}{(A_s+\tilde{U})^{1/3}\tilde{U}},
\right)\right]
\end{equation}
and make the ansatz
\begin{equation}
\label{xU2}
 x=B \tilde{u}^{\beta}.
\end{equation}  
In the limit $\tilde{u}\gg \tilde{U}$ we may neglect the $1$ in the 
logarithm of
(\ref{xU1}) compared to the second term and obtain 
\begin{equation}
B=\frac{1}{A_c(A_s+\tilde{U})^{1/3}}^{3/(6-2g_c)},
\end{equation} 
\begin{equation}
\beta=\frac{2}{3-g_c}.
\end{equation}
Finally  the charge and the spin gap are
\begin{eqnarray}
\frac{\Delta_c(u)}{\Delta_{0c}}&\approx&
    \left(\frac{4u}{\pi}\right)^{2/(3-g_c)}
    \left(\frac{2\pi v_c g_c}{\Delta_{0c}^2}\right)^{3/(6-2g_c)}
    \left(\frac{\Delta_{0s}^2}{2\pi v_sg_s}+\frac{2U}{\pi^2}\right)^{-1/(6-2g_c)},\\
\label{betap}
\frac{\Delta_s(u)}{\Delta_{0s}}&\approx&
    \left(\frac{4u}{\pi}\right)^{2/(3-g_c)}
    \left(\frac{2\pi v_c g_c}{\Delta_{0c}^2}\right)^{g_c/2(3-g_c)}
    \left(\frac{\Delta_{0s}^2}{2\pi v_sg_s}+\frac{2U}{\pi^2}\right)^{(4-g_c)/(6-2g_c)}.
\end{eqnarray} 
\section*{Ionic Hubbard model}
For the ionic Hubbard model in the case 
 $2B_c<B_{cs}$ the mean field parameters are given by the equations 
\begin{eqnarray}
\frac{\Delta_c^2}{2\pi v_cg_c}&=&4B_{cs}-8B_c=\frac{2\Delta}{\pi }
    \textrm{e}^{-\langle \phi^2_c \rangle-\langle \phi^2_s \rangle}+
    \frac{2U}{\pi^2}\textrm{e}^{-4\langle \phi^2_c \rangle},\\
\frac{\Delta_s^2}{2\pi v_sg_s}&=&4B_{cs}-8B_s=\frac{2\Delta}{\pi }
    \textrm{e}^{-\langle \phi^2_c \rangle-\langle \phi^2_s \rangle}+
    \frac{2U}{\pi^2}\textrm{e}^{-4\langle \phi^2_s \rangle}.
\end{eqnarray}
As in the previous case
we insert the analytical values of $\langle \phi^2_c \rangle $ and 
$\langle \phi^2_s \rangle $
and  introduce the  notation
 $\Delta_c/\Delta_{0c}=x$,  $\;\Delta_s/\Delta_{0s}=y$,
  $\;\Delta_{0c}/2\pi v_c g_c =A_c$,
$\;\Delta_{0s}/2\pi v_s g_s =A_s$, 
$2\Delta/\pi  =\tilde{u}$ and $2U/\pi^2=\tilde{U}$. We obtain
\begin{eqnarray}
\label{Ix1}
A_cx^2&=&\tilde{\Delta}x^{g_c/2}y^{1/2}+\tilde{U}x^{2g_c},\\
\label{Iy1}
A_sy^2&=&\tilde{\Delta}x^{g_c/2}y^{1/2}+\tilde{U}y^{2}.
\end{eqnarray}
Eqs.\ (\ref{Ix1})-(\ref{Iy1}) are identical 
to Eqs.\ (\ref{eqx1})-(\ref{eqy1}), if  $\tilde{u}$ is replaced with $\tilde{\Delta}$ and
$A_s+\tilde{U}$ with $A_s-\tilde{U}$.
In  the limit 
 $\tilde{\Delta}/\tilde{U}\gg 1$ we can use the results given 
by  Eqs.\ (\ref{xU1})-(\ref{betap})
and obtain 
\begin{eqnarray}
\frac{\Delta_c(\Delta)}{\Delta_{0c}}&\approx&
    \left(\frac{2\Delta}{\pi}\right)^{2/(3-g_c)}
    \left(\frac{2\pi v_c g_c}{\Delta_{0c}^2}\right)^{3/(6-2g_c)}
    \left(\frac{\Delta_{0s}^2}{2\pi v_sg_s}-\frac{2U}{\pi^2}\right)^{-1/(6-2g_c)},\\
\frac{\Delta_s(\Delta)}{\Delta_{0s}}&\approx&
    \left(\frac{2\Delta}{\pi}\right)^{2/(3-g_c)}
    \left(\frac{2\pi v_c g_c}{\Delta_{0c}^2}\right)^{g_c/2(3-g_c)}
    \left(\frac{\Delta_{0s}^2}{2\pi v_sg_s}-\frac{2U}{\pi^2}\right)^{(4-g_c)/(6-2g_c)}.
\end{eqnarray}
For $2B_c>B_{cs}$ the gap equations read   
\begin{eqnarray}
\frac{\Delta_c^2}{2\pi v_cg_c}&=&8B_c-2\frac{B_{cs}^2}{B_c},\\
\frac{\Delta_s^2}{2\pi v_sg_s}&=&-8B_s+2\frac{B_{cs}^2}{B_c},
\end{eqnarray}
which is equivalent  to 
\begin{equation}
\label{IUx1}
A_c x^2=-\frac{2\Delta^2}{U}x^{-g_c}y+\frac{2U}{\pi^2}x^{2g_c},
\end{equation}
\begin{equation}
\label{IUy1}
A_s y^2=\frac{2\Delta^2}{U}x^{-g_c}y-\frac{2U}{\pi^2}y^{2}.
\end{equation}
We assume for $x$ a dependence on $\Delta$ of the form
\begin{equation}
\label{IUx2}
x(\Delta)=x_0+B\Delta^{\beta},
\end{equation} 
with $x_0/B\tilde{\Delta}^{\beta} \gg 1$. If we insert  Eqs.\ (\ref{IUy1})  
 and (\ref{IUx2}) in Eq.\ (\ref{IUx1}) and compare the prefactors of same powers
 of $\Delta$
 we
 obtain for $x_0$, $\beta$ and  $B$  the following expressions:
\begin{equation} 
x_0=\left (\frac{2U}{A_c\pi^2})\right)^{1/(2-2g_c)}, 
\end{equation} 
\begin{equation} 
 \beta=4,
\end{equation} 
\begin{equation} 
 B=-\frac{2}{A_c(1-g_c)U^2(A_s+2U/\pi^2)}x_0^{-2g_c-1}.
\end{equation} 
Finally the charge and spin parameter are given by 
\begin{eqnarray}
\frac{\Delta_c(\Delta)-\Delta_c(0)}{\Delta_{0c}}&\approx&
    -\Delta^4
    \left(\frac{\Delta_{0s}^2}{2\pi v_sg_s}+\frac{2 U}{\pi^2} \right)^{-1}
    \nonumber\\ && \quad \times\frac{2\pi v_cg_c}{(1-g_c)U^2\Delta_{0c}}
   \left(\frac{\Delta_c(0)}{\Delta_{0c}}\right)^{(-1-2g_c)},
   \\
\frac{\Delta_s(\Delta)}{\Delta_{0s}}&\approx&
    \Delta^2
    \left(\frac{\Delta_{0s}^2}{2\pi v_sg_s}+\frac{2 U}{\pi^2} \right)^{-1}
    \frac{2}{U}
   \left(\frac{\Delta_c(0)}{\Delta_{0c}}\right)^{-g_c}.
\end{eqnarray}


%

\end{document}